\pdfoutput=1
\documentclass[11pt]{article}
\usepackage[margin=1in]{geometry}
\usepackage{microtype}
\usepackage{amsmath,amssymb,amsfonts,mathtools,bm}
\usepackage{graphicx,booktabs,tabularx,longtable,array}
\usepackage{xcolor}
\usepackage[numbers,sort&compress]{natbib}
\usepackage[hidelinks]{hyperref}
\usepackage{enumitem}
\usepackage{caption}
\hypersetup{
  pdftitle={Trainable sparse dendritic networks reveal regime-dependent roles for shunting inhibition and morphology, with Supplementary Information},
  pdfauthor={},
  pdfsubject={Sparse dendritic models, shunting inhibition, synaptic allocation, and multilayer networks},
  pdfkeywords={dendrites, shunting inhibition, sparse synapses, neural networks}
}

\IfFileExists{placeins.sty}{\usepackage{placeins}}{}
\providecommand{\FloatBarrier}{}


\graphicspath{{figures_empirical/}{./figures_empirical/}{./}}

\usepackage{manuscript/supplement}

\title{Trainable sparse dendritic networks reveal regime-dependent roles for shunting inhibition and morphology}
\author{%
Houman Safaai$^{1,*,\dagger}$,
Maceo Richards$^{1,*}$,
Ryan Whalen$^{1}$,\\
Adithya Madduri$^{1}$,
Naeem Khoshnevis$^{1}$,
and Bernardo L. Sabatini$^{1,2,\dagger}$\\[5pt]
\begin{minipage}{\dimexpr\textwidth-2\tabcolsep\relax}\centering\footnotesize
$^{1}$Kempner Institute for the Study of Natural and Artificial Intelligence at Harvard University, Cambridge, MA, USA\\
$^{2}$Department of Neurobiology, Howard Hughes Medical Institute, Harvard Medical School, Boston, MA, USA\\[3pt]
$^{*}$These authors contributed equally.\\
$^{\dagger}$Correspondence: \texttt{houman\_safaai@harvard.edu}; \texttt{bernardo\_sabatini@hms.harvard.edu}
\end{minipage}%
}
\date{}

\begin{document}
\maketitle

\begin{abstract}
Neural circuits must process and perform computations with inputs that exhibit irrelevant variation and whose properties may change across contexts.  How such challenges are overcome to ensure correct computation is unclear.  Dendritic properties and the process of synaptic integration, including pooling of multiple noisy inputs, divisive normalization via shunting inhibition, active dendritic conductances, and dynamic routing of signals via inhibitory control over dendritic segments likely contribute.  To examine these possibilities, we introduce DendriNet, a framework for training scalable rate-based networks with sparse excitatory and inhibitory contacts, controllable dendritic architecture and multiple modes of synaptic integration.  We find that, when trained in image classification tasks, networks with conductance based synaptic interactions transmit and rely on distal dendritic signals more than those utilizing linear, additive integration.  Furthermore,  distributing limited contacts across compartments improved classification in small populations, whereas greater serial dendritic depth provided no benefit on clean images within resource-matched trees. Corruption of images by pixel level noise or variable global scaling reveals that no single configuration is best across all conditions: deeper shunting based neurons better preserve classification accuracy under noise, whereas additive integration better tolerates strong scaling.  On a contextual task with variable gain and noise, the primary shunting tree is more accurate than the separately tuned additive tree under the training corruption, yet relies less on cue-specific inhibition. Learning dendritic leak conductances partly restores shunting accuracy on dim, noise-free inputs outside that distribution. Lastly, at AlexNet width, shunting-based integration and learned sparse connectivity support smaller deployed networks with performance comparable to that of the retrained dense layers they replace. Our studies identify features of biological dendrites and synaptic interactions that reduce the sensitivity of network computation to input corruption and suggest tests of dendritic function and design principles in brain-inspired networks.
\end{abstract}

\section{Introduction}
\label{sec:intro}

Neural circuits must recognize stimuli despite noise and correlated changes in input strength, yet also be able to alter their responses when behavioral context demands it. Dendrites provide multiple sites for local interactions between excitation and inhibition before signals reach the soma. Their nonlinear branches can support multistage computation, but under what conditions this organization improves performance, suppresses irrelevant variation or enables contextual control remains incompletely understood. These questions require distinguishing the distribution of inputs across compartments from the rule by which each compartment integrates them.

Dendritic nonlinearities and spikes can shape neuronal responses and sensory selectivity~\citep{Mel1992,PolskyEtAl2004,LondonHausser2005,StuartSpruston2015,SmithEtAl2013}, while different input pathways target distinct dendritic domains~\citep{PetreanuEtAl2009}. Excitatory contacts substantially outnumber inhibitory contacts in cortical pyramidal neurons, with relative densities varying across dendritic regions~\citep{IasconeEtAl2020}. Nevertheless, dendrite-targeting inhibition can regulate calcium signals within individual spines~\citep{ChiuEtAl2013}. Contact number, location and strength therefore need not have equivalent effects on computation. Input variability also takes different forms: stochastic transmitter release introduces variation locally~\citep{AllenStevens1994}, whereas fluctuations in excitability can modulate responses across neurons and time~\citep{GorisEtAl2014}. Pooling inputs may reduce independent fluctuations, shared gain changes require tolerance to coherent variation, and contextual cues require selective changes in the response itself.

Models have established several computational benefits of dendritic structure~\citep{ChavlisPoirazi2021,AcharyaEtAl2022}. Nonlinear subunits can give a pyramidal neuron the computational structure of a two-layer network~\citep{Poirazi2003}, and deep artificial networks can approximate the input--output function of detailed cortical neuron models~\citep{Beniaguev2021}. Explicit dendritic architectures can increase expressive capacity and support image classification~\citep{WuEtAl2018,JonesKording2021}. Normalizing afferent weights by contact count improves sparse-network learning~\citep{BirdEtAl2021}, and active dendrites with inhibitory sparsification support continual and multitask learning~\citep{IyerEtAl2022}. Structured dendritic networks further show that restricted input sampling and dendrite--soma connectivity can contribute separately to accuracy, robustness and parameter efficiency~\citep{ChavlisPoirazi2025}. We therefore asked how the local integration rule and input statistics alter these benefits.

Biological synapses change membrane conductance as well as current, unlike the weighted summation used in conventional artificial neurons. Inhibition with a reversal potential near rest can reduce the voltage response to coactive excitation through shunting, contributing to gain control~\citep{ChanceAbbottReyes2002,MitchellSilver2003,Silver2010}. Its effect depends on operating point and the placement of excitation and inhibition~\citep{KochPoggioTorre1983,GidonSegev2012,JadiEtAl2012,ZhangEtAl2013}; shunting at the membrane need not produce divisive modulation of firing rate~\citep{HoltKoch1997}. Shunting classifiers establish this mechanism as a trainable computational building block~\citep{ArulampalamBouzerdoum2003,TiviveBouzerdoum2005}. Divisive normalization, proposed as a canonical neural computation~\citep{Heeger1992,CarandiniHeeger2012}, has also been incorporated into trainable network models~\citep{RenEtAl2017,GiraldoSchwartz2019,BurgEtAl2021,MillerChungMiller2022}. Models of dendritic disinhibition also show how context-dependent inhibition can regulate selected input pathways~\citep{YangMurrayWang2016}. Whether these local and circuit-level uses of inhibition coincide in a learned network is a separate question.

We developed DendriNet to examine these relationships within one trainable, rate-based framework. Each compartment combines sparse excitatory and inhibitory inputs with outputs from more distal branches, using either additive integration or conductance-inspired shunting inhibition, followed by a learned bounded activation. We vary contact allocation, branching, input sampling and network organization, and compare trees with soma-only neurons under explicit resource and training controls. Image-classification tasks provide controlled tests of the resulting computation rather than models of a particular biological circuit. The framework abstracts away spikes, cable dynamics and biological learning rules, allowing us to ask which signals trained networks use and how that dependence changes with input statistics.

We find that distributing inputs across compartments can improve classification, whereas greater serial depth provides no clean-image benefit within resource-matched trees. Inhibitory contacts help both integration rules, but their allocation changes which rule performs better. Similar accuracy can conceal different computations: shunting trees transmit distal changes more strongly and depend more on their image-specific activity. Robustness depends on input statistics and on how many inputs each compartment pools. Multilayer and AlexNet tests extend the comparisons across network scales. In the contextual task, the primary shunting tree attenuates gain variation more strongly and is more accurate than the separately tuned additive tree, yet relies less on cue-specific inhibition. The local action of shunting alone therefore does not determine how strongly a trained circuit uses inhibition to select a response.

\section{Results}
\label{sec:results}

We ask how excitation and inhibition, sparse connectivity, dendritic morphology and the local integration rule contribute to computation. Classification accuracy measures task performance; performance under noise and gain measures sensitivity to irrelevant input variation; contextual classification tests whether a network can select different responses to the same stimulus according to a cue. Signal associations, transmission measurements, decoding and targeted interventions then relate performance to activity within branches and E/I populations.

Below, accuracy $A$ is the fraction of evaluated images assigned to the correct class; differences are expressed in percentage points. Unless stated otherwise, results summarize independently trained networks with 95\% bootstrap confidence intervals. Training, model selection and evaluation procedures are specified in Methods.

\subsection{A common model separates dendritic structure from synaptic integration}
\label{sec:model_overview}

Each model neuron contains a soma and a tree of dendritic compartments (Fig.~\ref{fig:fig1}A,B). A compartment receives local excitatory (E) and inhibitory (I) inputs and the outputs of its more distal child branches. Sparse connectivity is specified by the number of contacts retained from an eligible source pool. We distinguish this count and its E/I allocation from the spatial placement of inputs and from whether contact locations are fixed or learned.

Synaptic weights are learned in both fixed- and learned-connectivity models. In the latter, each compartment uses a fixed number of its strongest candidate connections; their identities can change during training and are frozen for evaluation. We call these the selected, or retained, synapses. Their source activity can vary from image to image even when the connections are unchanged. Retained contacts describe the deployed network, whereas learned selection can require a much larger candidate weight bank during training.

For nonnegative source activities $x_p$, the inputs to a compartment are
\begin{equation}
\begin{aligned}
E&=\sum_{p\in\mathcal S_E}w_p^E x_p,
& I&=\sum_{p\in\mathcal S_I}w_p^I x_p,\\
C&=\sum_{k\in\mathrm{children}}c_k a_k,
& G&=\sum_{k\in\mathrm{children}}c_k.
\end{aligned}
\label{eq:branch_inputs}
\end{equation}
Here $\mathcal S_E$ and $\mathcal S_I$ are the selected synapses, $w_p^E$ and $w_p^I$ their nonnegative weights, and $c_k$ the nonnegative branch-coupling weights. The local synaptic drives are $E$ and $I$; $C$ is the weighted input from child branches, also called inherited input, and $G$ is their total coupling weight. Terminal branches have $C=G=0$. In the dendritic-tree models of Figures~\ref{fig:fig1}--\ref{fig:fig5}, the soma receives only branch outputs; the contextual networks additionally receive somatic inhibition (Methods).

We compare two integration rules:
\begin{equation}
u_{\rm sh}=\frac{E+C}{1+E+I+G},
\qquad
u_{\rm add}=E-I+C.
\label{eq:integration_overview}
\end{equation}
The shunting rule has a steady-state voltage interpretation with unit leak conductance and reversal levels of 1 for excitation and 0 for inhibition and leak. Under this interpretation, $E$ and $I$ are conductances, $C$ is a weighted child-output source term, and $G$ is coupling conductance; the numerator contains source terms rather than net membrane currents (Methods). Excitation contributes to both numerator and denominator, whereas inhibition increases only the denominator. Raw addition instead subtracts inhibition from the sum of local excitation and child input. Changing the rule therefore changes the integration of all three inputs, not inhibition alone.

In the initial experiments, separate E and I afferent pathways carry image-derived signals directly. Later networks learn inhibition through explicit I-neuron populations. After integration, both rules use a bounded activation with a learned slope $m$ and midpoint $b$:
\begin{equation}
a=\phi(u)=\frac{1+\tanh[m(u-b)]}{2},
\qquad m>0.
\label{eq:activation_overview}
\end{equation}
These parameters let each compartment learn its responsive range and degree of saturation. The model computes a dimensionless integration state $u$ and activity $a$, not a membrane-voltage trajectory. Signals pass from children to parents without reciprocal cable interactions, carrying the results of successive integrations toward the soma.

Morphology specifies branching topology rather than cable length, diameter or a reconstructed neuron. We write it as $[b_1,\ldots,b_D]$, where $D$ is the number of dendritic levels and $b_d$ the branching factor at depth $d$ from the soma. For example, $[2,2]$ has two proximal and four distal compartments. A wider tree distributes inputs across more parallel branches; a deeper tree adds successive integration stages along the path to the soma.

\subsection{A ten-neuron classifier exposes learned dendritic computation}
\label{sec:fig1_results}

\begin{figure}[htbp]
\centering
\IfFileExists{figures_empirical/fig1_model_v16.pdf}{%
  \includegraphics[width=\textwidth]{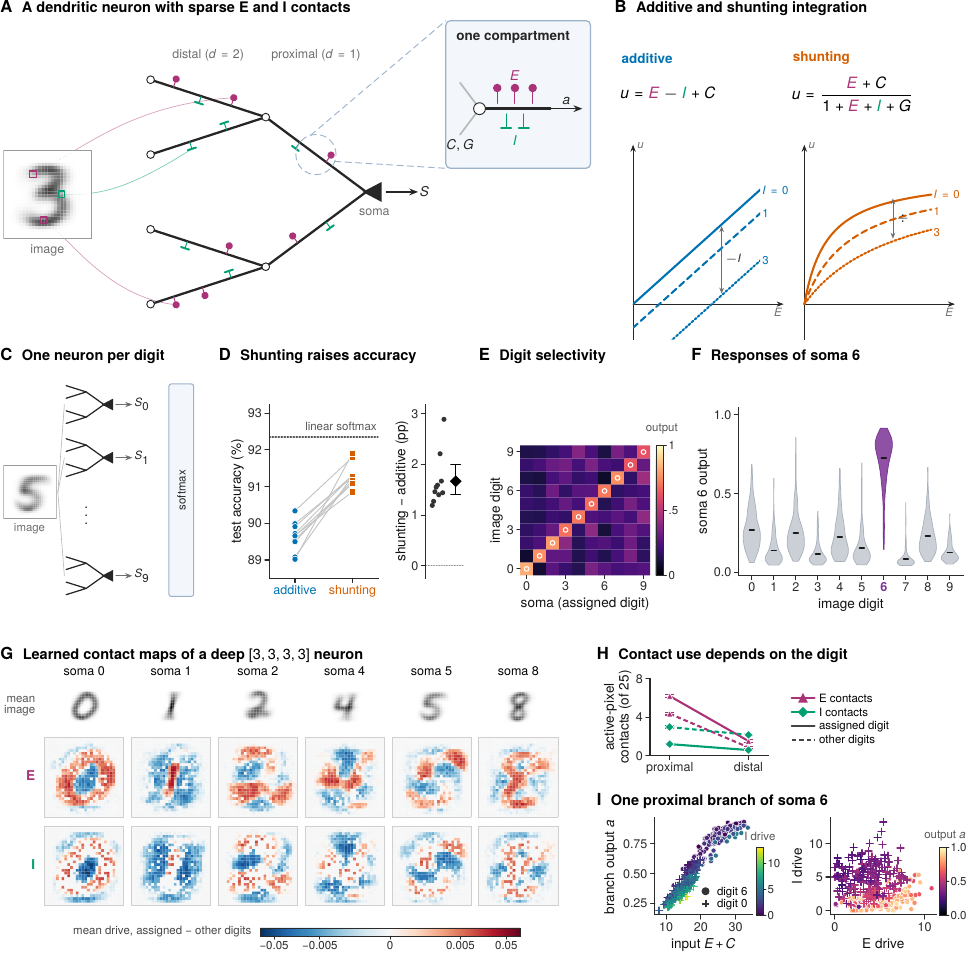}%
}{%
  \PackageError{empirical-manuscript}%
    {Missing required figure `figures_empirical/fig1_model_v16.pdf'}%
    {Restore the figure listed in the manuscript export manifest before compiling.}%
}%
\caption{\textbf{Sparse dendritic inputs support selective somatic responses.} (A) A dendritic neuron: E (boutons) and I (bars) contacts on every compartment receive image pixels; the magnified compartment combines its E and I drives with its children's weighted output $C$ and coupling conductance $G$ and passes $a=\varphi(u)$ to its parent (Eqs.~\ref{eq:branch_inputs}--\ref{eq:activation_overview}). The primary $[2,2]$ tree (25 E/25 I contacts per compartment) supplies C--F, H and I. (B) The two integration rules, with idealized responses of a terminal compartment ($C=G=0$) at three inhibitory drives: inhibition enters subtractively under addition and changes the conductance-dependent denominator under shunting (equations, not data). (C) One single-neuron classifier per digit, without I neurons; a shared learned scale $T$ maps soma outputs $S$ to logits. (D) Paired test accuracies and shunting-minus-additive differences for ten seeds (95\% paired seed-bootstrap interval); the dotted linear softmax reference has a different parameter budget (Methods). (E) Mean soma output per image digit; circles mark assigned digits. (F) Soma-6 response distributions and means. (G) Separate dense-sampling example, not resource matched to the primary tree: assigned-minus-other mean synaptic drive of six somas in one $[3,3,3,3]$ model (120 compartments, 25 E/10 I contacts each), from retained weights and class-mean inputs (Eq.~\ref{eq:retained_weight_map}); color is logarithmic in magnitude and linear near zero; gray images are assigned-class means. (H) Mean contacts per branch on active pixels ($x_p>0.1$) for assigned and other digits. (I) One proximal branch of soma 6: input $E+C$ against output (left; color, I drive) and E against I drive (right; color, output) for the same images of digits 6 (circles) and 0 (plus signs). E, F and I use one primary checkpoint (200 images per digit in F and I), G one deep checkpoint, and H ten primary shunting networks (95\% seed-bootstrap intervals); checkpoint identities are specified in Methods.}
\label{fig:fig1}
\end{figure}

We first examined how sparse E and I input selection and local integration produce stimulus-selective somatic responses, using a ten-neuron MNIST digit classifier.  Each neuron is assigned one digit and its somatic output, multiplied by a shared positive scale, supplies the score or logit of that class (Fig.~\ref{fig:fig1}C).  The neuron with the largest output determines the predicted digit.  The primary example uses $[2,2]$ trees with inputs from the image directly to the dendritic compartments and no direct somatic synapses. Because each neuron provides one class score, its inputs and branch responses can be related directly to the digit it must recognize. The linear softmax reference in Fig.~\ref{fig:fig1}D provides task context and performs better under the same data split and preprocessing.

Both integration rules learn selective somatic responses, with a modest accuracy advantage for shunting across paired training seeds (Fig.~\ref{fig:fig1}D). In the illustrated shunting network, each neuron's mean response is largest for its assigned digit (Fig.~\ref{fig:fig1}E).  The response distributions for soma 6 illustrate this selectivity: its assigned digit generally evokes larger outputs, with some overlap between classes (Fig.~\ref{fig:fig1}F).

The learned contacts also reflect digit selectivity. Maps of the difference in mean synaptic drive between assigned and other digits reveal where each pathway samples discriminating pixels (Fig.~\ref{fig:fig1}G). Inhibitory contacts can contribute strongly on pixels that are dark in the assigned digit but active in other digits. We use a separate, deeper tree to make this spatial organization visible; the quantitative comparisons use the primary $[2,2]$ networks, as specified in the caption.

In those primary networks, E synapses sample more active pixels for the assigned digit, whereas I synapses sample fewer (Fig.~\ref{fig:fig1}H). The connections remain fixed during evaluation; it is their input activity that changes across images. Both integration rules learn this organization, with a larger assigned-versus-other contrast in proximal branches under shunting (\suppfig{figS_mnist_components_source_activity}).

Local excitation and child input alone do not determine the output of a shunting branch. In the illustrated branch, responses at similar $E+C$ vary with inhibitory drive (Fig.~\ref{fig:fig1}I), as expected from Eq.~\ref{eq:integration_overview}. The learned activation then shapes how this integrated input becomes branch activity. Thus, the synaptic maps identify selective input sampling, while the branch responses show how excitation, inhibition and distal signals jointly influence its output.

\subsection{Integration, contact allocation, and readout shape CIFAR-10 performance}
\label{sec:fig2_results}

We next asked how contact number and the division of contacts between E and I pathways affect classification under each integration rule. A population of 64 dendritic neurons supplies ten class scores through a jointly trained readout (Fig.~\ref{fig:fig2}A). The primary readout is an MLP with two hidden layers; a linear readout tests how much the result depends on downstream nonlinear computation. Both combine signals across neurons, unlike the MNIST identity readout. We varied E and I contacts per compartment in four-level $[3,3,3,3]$ trees while holding the remaining architecture fixed. E/I allocation refers to contact counts, not the balance of learned synaptic drives. Within each rule, training settings are fixed across the grid (Methods, Section~\ref{sec:optimization}); sensitivity to these choices is reported in \suppfig{figS13_allocation} and Supplementary Note~\ref{note:recipe_policy}.

Relative performance depends on E/I contact allocation (Fig.~\ref{fig:fig2}B).  The paired difference, $\Delta A=A_{\rm sh}-A_{\rm add}$, compares test accuracy for the two rules at the same allocation and training seed; positive values favor shunting.  Addition is more accurate across most tested allocations, whereas shunting gains are concentrated in a subset of excitation-dominated allocations with few I contacts.  At 40E/10I in this allocation-grid cohort, the two rules perform comparably ($\Delta A=0.3$ percentage points; 95\% interval, $-0.03$ to 0.6).  In the full tested grid, shunting is least accurate with a single E contact and raw addition with a single I contact.

Allocation-dependent differences persist when training settings are allowed to vary, although the average gap between the rules narrows (Supplementary Note~\ref{note:recipe_policy}). Thus, allocation shapes relative performance, but the size of the difference also depends on how the networks are trained.
Holding $N_E+N_I$ fixed distinguishes reallocation between E and I from an increase in contact number.
  At a fixed total of 50 contacts per compartment, the four-level tree favors addition at inhibition-rich allocations and shunting at the most excitation-rich allocation (Fig.~\ref{fig:fig2}C).  The three-level tree shows a similar but weaker trend, whereas the two-level tree favors addition or neither rule across the slice (\suppfig{figS_allocation_slices}A).

To test the benefit of allocating contacts to inhibition, we compared E-only and E/I pathways while holding total retained contacts and trainable parameters constant (Fig.~\ref{fig:fig2}D). These controls keep contact locations fixed. Inhibitory contacts improve performance under both integration rules, with a larger benefit under raw addition. With fixed contacts, shunting remains more accurate both with and without direct I contacts, unlike the learned-contact grid at the same 40E/10I allocation. Its advantage therefore cannot be attributed to inhibition alone, and it depends on how contacts are chosen. The primary contrast changes the complete integration rule, including the normalization of excitation and inherited input.

Replacing the nonlinear readout with a jointly trained linear one generally lowers additive accuracy more than shunting accuracy (Fig.~\ref{fig:fig2}E), bringing the two rules closer (\suppfig{figS_readout_allocation}). The dependence on contact allocation remains: both readouts favor addition when only one E contact is available, and addition is more accurate at every tested equal-E/I budget in the four-level tree (\suppfig{figS_readout_allocation}A,B; \suppfig{figS_allocation_slices}B,E). Across the 50-contact slice, the change in readout effect from inhibition-rich to excitation-rich allocations does not differ clearly between the rules.

The representation and readout learn together, so changing the readout can also change the dendritic computation. To examine class structure in a fixed representation, we fitted auxiliary classifiers, or probes, to the frozen somatic outputs of the MLP-trained networks. Linear decoding favors addition at inhibition-rich allocations and shunting at excitation-rich allocations (\suppfig{figS15_allocation_probes}). Nonlinear probes improve decoding under both rules; their additional benefit is larger for shunting over most of the fixed-budget slice, but not at its E-rich end. At that end, the probes favor shunting more strongly than the jointly trained classifier does. Thus, the class structure expressed by a dendritic population and the accuracy achieved by its trained readout are related but distinct. Contact counts alone also do not specify the resulting E/I drives, which depend on synaptic weights and source activity (\suppfig{figS28_allocation_learned_utilization}).

\subsection{Compartmentalization and spatial organization improve classification without a benefit from greater serial depth}
\label{sec:morphology_results}

\begin{figure}[htbp]
\centering
\IfFileExists{figures_empirical/fig2_allocation_v16.pdf}{%
  \includegraphics[width=0.98\textwidth]{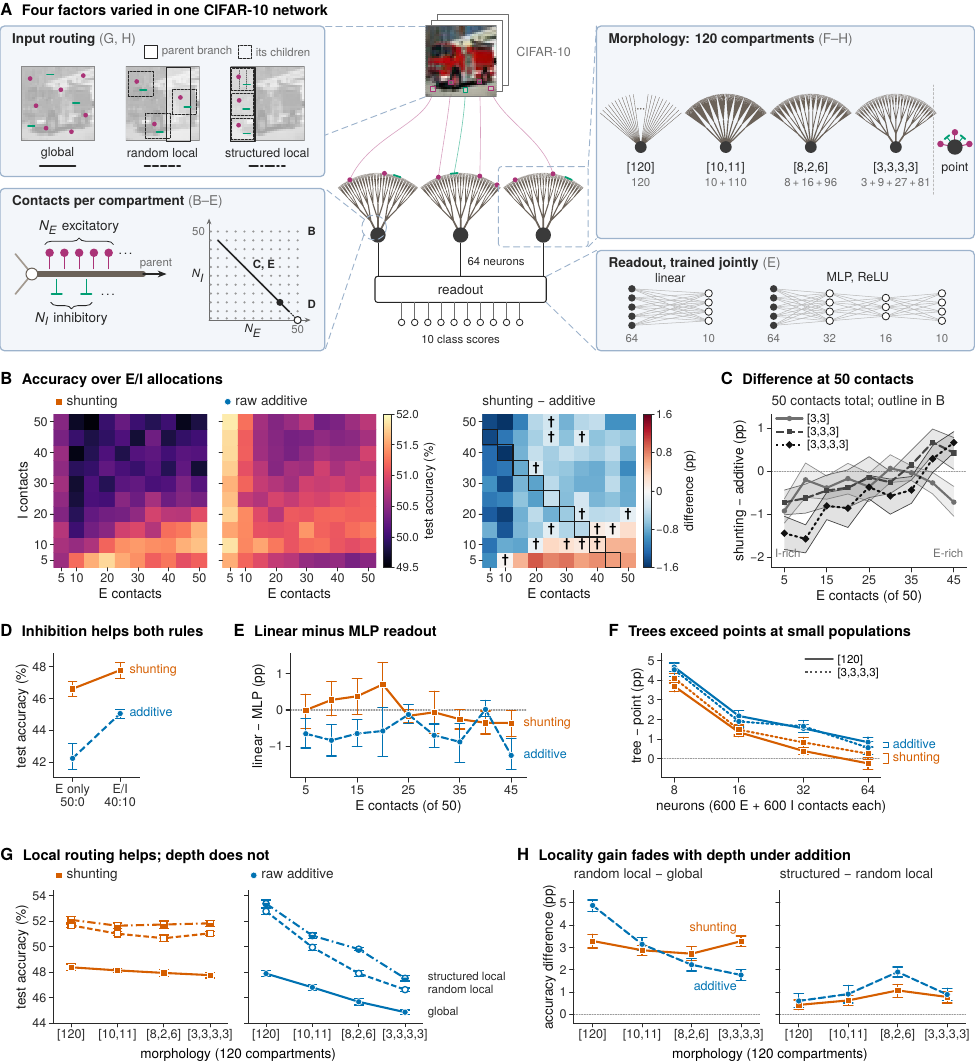}%
}{%
  \PackageError{empirical-manuscript}%
    {Missing required figure `figures_empirical/fig2_allocation_v16.pdf'}%
    {Restore the figure listed in the manuscript export manifest before compiling.}%
}%
\caption{\textbf{E/I allocation changes relative performance.} CIFAR-10, 64 somas except F; training procedures in Methods. (A) Network, morphologies and sampling geometries. (B) Shunting and raw additive accuracy and their paired difference per allocation (5--50 E and I contacts; top-$k$ contacts from a candidate bank); outlined cells form the slice of C; $\dagger$ marks intervals that include zero. (C) Shunting-minus-additive accuracy at $N_E+N_I=50$; shade, marker and line style identify morphology. (D) Fixed-index E-only and E/I models at matched contact and parameter counts. (E) Linear-minus-MLP accuracy on that slice. (F) Shallow $[120]$ and deep $[3,3,3,3]$ tree-minus-point accuracy at 600 E plus 600 I contacts per neuron; trees use 5E/5I per compartment; candidate counts differ. (G) Fixed global, random-local and structured-local routing at matched resources; line style and marker fill identify routing. (H) Local-minus-global and structured-minus-random-local contrasts. B,D,E use $[3,3,3,3]$; B--D,F--H use MLP readouts. B: twenty paired seeds; C: twenty at depth four and thirteen shallower; D: ten paired seeds from the inhibitory-contact control; E: paired seeds 10--19; F--H: twenty paired seeds (ten original and ten added after inspection; Supplementary Table~\ref{tab:seed_extension}). Intervals are pointwise 95\% seed bootstraps, not adjusted across the grid. Differences are percentage points of test accuracy.}
\label{fig:fig2}
\end{figure}

Distributing inputs across compartments can improve classification without a consistent benefit from greater serial depth. With limited contact budgets, both shallow $[120]$ and deep $[3,3,3,3]$ trees outperform soma-only neurons in small populations when E and I contact counts are matched (Fig.~\ref{fig:fig2}F). This benefit declines as the population grows: at 64 neurons and 600 E plus 600 I contacts, the shallow tree retains a small benefit under addition and is not clearly different from the point under shunting. The same model family supplies Fig.~\ref{fig:point_neurons}E. Neither tree is consistently more accurate: the deep tree slightly exceeds the shallow tree under shunting, and the two are similar under addition. These contact-matched comparisons motivate a stricter test in which compartment and parameter counts are also held fixed (\suppfig{figS38_low_capacity_morphology}).

To examine morphology under stricter resource control, we held the numbers of neurons, compartments, E/I contacts, stored weights, and total trainable parameters fixed while varying dendritic depth.  We tested four morphologies---$[120]$, $[10,11]$, $[8,2,6]$ and $[3,3,3,3]$---each with 120 dendritic compartments (Fig.~\ref{fig:fig2}A,G).   We also varied the spatial regions of the images sampled by each branch: \emph{global} routing draws contacts from the whole image, \emph{random-local} routing draws them from a local window positioned independently of branch ancestry, and \emph{structured local} routing places child windows within their parent's region.
  The two local conditions use the same window sizes, and contact locations remain fixed while their weights are learned.  Random-local minus global accuracy therefore measures the contribution of local sampling, while structured-local minus random-local accuracy measures the effect of structured spatial placement, including changes in coverage and overlap (Fig.~\ref{fig:fig2}H).

Local input sampling improves accuracy under both rules and enlarges the shunting advantage in deeper trees. Structured placement gives further accuracy gains in most conditions while leaving the difference between rules similar or slightly smaller (Fig.~\ref{fig:fig2}H). Its benefit is also present in one-level trees, which have no successive dendritic stages to align. Structured placement changes input coverage and overlap as well as nesting (Supplementary Table~\ref{tab:routing_window_coverage}).

Greater depth provides no clean-image accuracy benefit within this resource-matched family: accuracy declines modestly under shunting and more strongly under raw addition (Fig.~\ref{fig:fig2}G). Under global routing, the shunting advantage reaches 2.9 percentage points in the four-level tree (95\% interval, 2.7--3.2), while the one-level difference is small; local routing favors addition in the one-level tree. These models keep contact locations fixed, unlike the learned-contact allocation grid. Together, the comparisons distinguish the benefit of distributing and spatially organizing inputs from the cost of passing them through additional stages.

\subsection{The synaptic integration rule alters signal associations and classifier dependence}
\label{sec:fig3_results}

\begin{figure}[htbp]
\centering
\IfFileExists{figures_empirical/fig3_mechanisms_v13.pdf}{%
  \includegraphics[width=\textwidth]{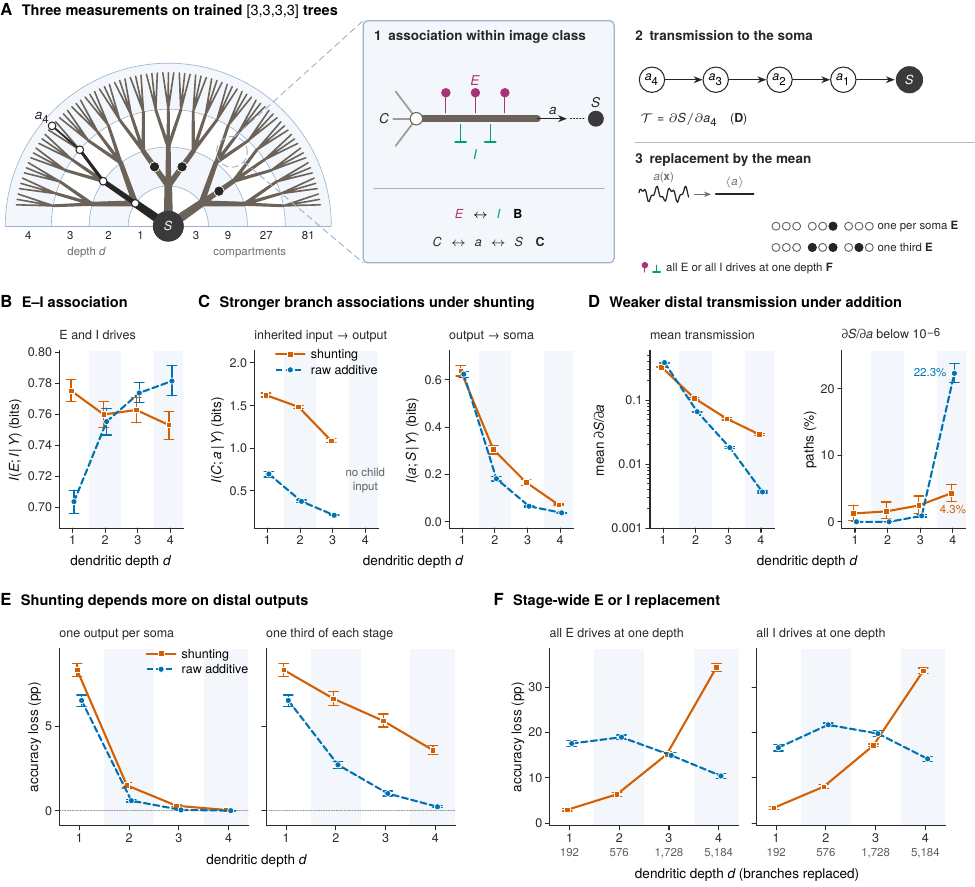}%
}{%
  \PackageError{empirical-manuscript}%
    {Missing required figure `figures_empirical/fig3_mechanisms_v13.pdf'}%
    {Restore the figure listed in the manuscript export manifest before compiling.}%
}%
\caption{\textbf{Shunting supports stronger distal transmission and classifier dependence.} CIFAR-10 models have 64 somas, $[3,3,3,3]$ trees, 40E/10I contacts per compartment, and MLP readouts. (A) Schematic of the three measurements on the complete tree, with soma-relative depth $d$, compartments per tree and illustrative contacts: 1, within-class association of the listed signal pairs; 2, transmission along one path to the soma, a product of edge derivatives; 3, replacement of branch outputs (filled markers indicate replaced outputs) or of all E or all I drives at one depth by mean responses over training images. (B) Mean single-branch E/I association within image class $Y$ (corrected conditional mutual information). (C) Within-class association of branch output with inherited input (left) and soma output (right). (D) Mean transmission $\mathcal T=\partial S/\partial a_d$ on a logarithmic ordinate (left) and fraction of paths below $10^{-6}$ (right). Derivatives include every coupling and activation along the path, on the same 2,048 label-independently selected test images for all checkpoints; branches and images are averaged within seed. The threshold identifies very weak transmission, not a significance test. (E) Accuracy loss after replacing one output per soma (left) or one third of each stage (right); common branch orderings couple the two interventions, and three draws are averaged within seed. (F) Loss after replacing all E (left) or all I (right) drives at one depth; lower ticks give affected branch counts. Squares and solid lines, shunting; circles and dashed lines, raw addition. B--F use twenty paired networks per rule (seeds 0--19, of which 10--19 were added after inspection; Supplementary Table~\ref{tab:seed_extension}); bars are pointwise 95\% seed-bootstrap intervals.}
\label{fig:fig3}
\end{figure}

The integration rule can change how distal signals contribute even when classification accuracy is similar. We therefore asked whether distal activity covaries with somatic output, whether small changes reach the soma, and whether removing image-specific distal activity impairs classification. These measurements distinguish statistical association, signal transmission and functional contribution. We used a separate primary cohort with 40E/10I contacts per compartment and MLP readouts, whose clean accuracies are comparable with a small advantage for shunting (Supplementary Table~\ref{tab:seed_extension}). Figure~\ref{fig:fig3}A identifies branch depth relative to the soma. We denote somatic output by $S$, with $S_s=a_{s,0}$ for neuron $s$.

We first measured within-class associations: whether signals covary among images of the same class. Conditional mutual information, $I(E;I\mid Y)$, measures this dependence for E and I drives after accounting for class $Y$; it is distinct from information about class identity. E/I association was greater under shunting in proximal branches and under addition at the most distal level, with overlapping intervals at intermediate depths (Fig.~\ref{fig:fig3}B). Shunting also produced stronger associations between child input and branch output, $I(C;a\mid Y)$, and, beyond the proximal level, between branch and somatic output, $I(a;S\mid Y)$ (Fig.~\ref{fig:fig3}C). Estimation and shuffled-baseline correction are described in Methods.

To measure transmission, we asked how a small change in a branch's output affects its soma. For branch $b$ with output $a_b$, the transmission derivative is
\begin{equation}
\mathcal T_{b\to s}(x)=\frac{\partial S_s}{\partial a_b},
\label{eq:path_transmission}
\end{equation}
evaluated at the response to image $x$, with other inputs held fixed. It includes all learned couplings and activations along the path. Measuring transmission between bounded outputs allows comparison without the arbitrary scaling of additive input weights (Methods).

The same transmission also enters learning. In these single-layer classifiers, a branch affects the task loss $\mathcal L$ through its owning soma, so the chain rule gives
\begin{equation}
\frac{\partial\mathcal L}{\partial a_b}
=\frac{\partial\mathcal L}{\partial S_s}\,\mathcal T_{b\to s}.
\label{eq:transmission_learning}
\end{equation}
For one child-to-parent connection, the local factor is $\phi'(u)c_k/Q$ under shunting, where $Q=1+E+I+G$, and $\phi'(u)c_k$ under addition. Inhibition therefore influences learning through the activation state under both rules and through total conductance under shunting. Division by $Q$ does not itself imply stronger transmission: the couplings and activation functions also adapt. The measurements below describe fitted networks, not their gradients throughout training. The synaptic and coupling gradients are given in Methods.

Mean transmission decreases with dendritic depth under both rules, but more steeply under raw addition (Fig.~\ref{fig:fig3}D). Proximal transmission is similar, whereas distal transmission is substantially greater under shunting. Very weak transmission is more frequent among the most distal additive paths, but rare at other depths. Thus, small distal changes reach the soma less effectively in the fitted additive trees.

We next tested whether these signals matter for classification. Replacing selected branch outputs with their mean responses over training images removes their image-specific variation while leaving the branches and their connections in place. The accuracy loss, $\Delta A_{\rm replace}=A_{\rm intact}-A_{\rm replaced}$, measures the classifier's dependence on that variation. Replacing one output per soma is most disruptive proximally. Replacing the same fraction of outputs at each depth retains this proximal concentration, but reveals much stronger dependence on intermediate and distal signals under shunting (Fig.~\ref{fig:fig3}E). Replacing all terminal outputs reduces accuracy by 26.1 percentage points under shunting (95\% interval, 25.2--27.1), compared with 1.2 under addition (1.0--1.3). Larger replaced fractions produce larger losses (\suppfig{figS40_whole_layer_ablation}E,F).

The strong proximal effect also holds with a linear readout. Replacing one proximal output per soma costs shunting more than addition under both readouts, without a resolved change in this contrast between readouts (\suppfig{figS_readout_mechanisms}D; Supplementary Note~\ref{note:readout_mechanisms}). Replacing every proximal output makes the soma independent of the image because it has no direct input.

To distinguish local synaptic contributions from the complete branch output, we replaced E or I drive at one depth with its training-image mean while leaving child input intact. Accuracy loss increased with depth under shunting but was smallest at the terminal level under raw addition (Fig.~\ref{fig:fig3}F). This intervention affects more branches at greater depth and measures their collective contribution. Initialization, replacement and estimator controls support the main distinction (\suppfig{figS40_whole_layer_ablation}, \suppfig{figS_tree_factorial_v5}, \suppfig{figS34_tree_scalar_information}, \suppfig{figS41_information_estimator_sensitivity}). The association profiles also recur with a linear readout (\suppfig{figS_readout_mechanisms}A--C). Together, the associations, transmission measurements and interventions show greater use of distal signals in the trained shunting trees.

\subsection{Noise and gain reveal different contributions of dendritic organization}
\label{sec:fig4_results}
\label{sec:corruption_results}

\begin{figure}[htbp]
\centering
\IfFileExists{figures_empirical/fig4_regimes_v15.pdf}{%
  \includegraphics[width=\textwidth]{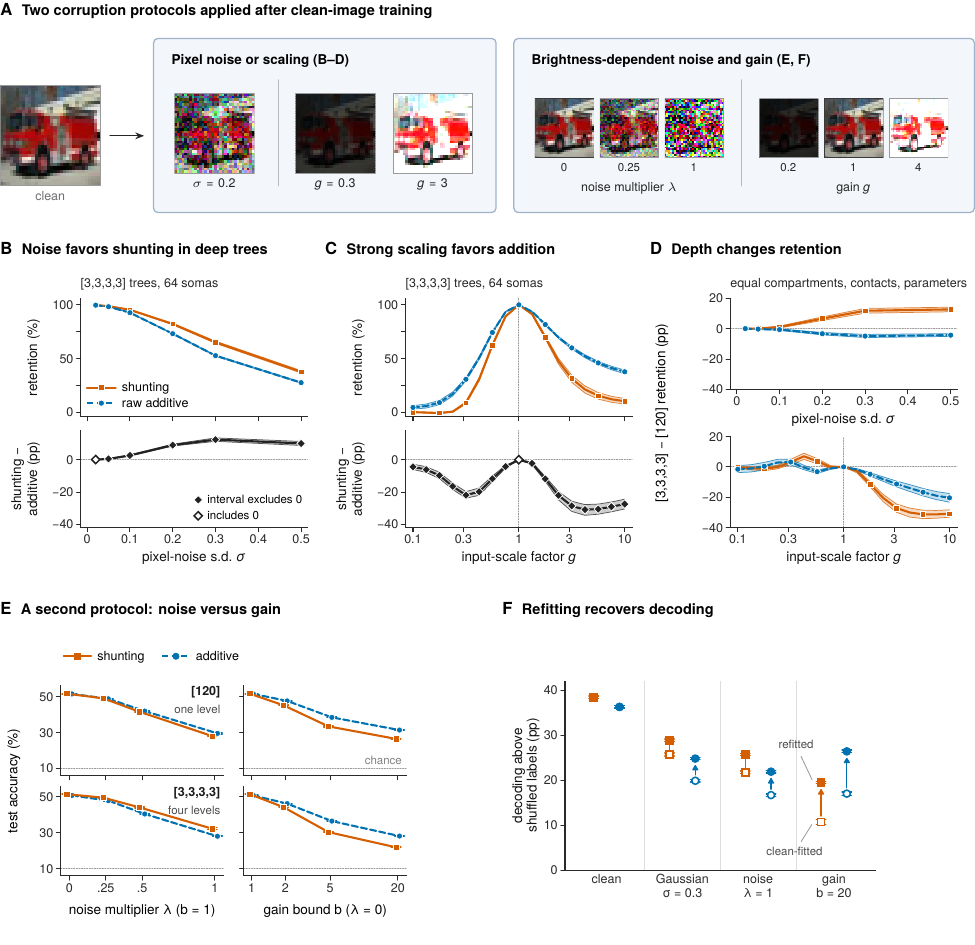}%
}{%
  \PackageError{empirical-manuscript}%
    {Missing required figure `figures_empirical/fig4_regimes_v15.pdf'}%
    {Restore the figure listed in the manuscript export manifest before compiling.}%
}%
\caption{\textbf{Noise and gain reveal different contributions of dendritic organization.} Clean-trained CIFAR-10 networks have 64 somas, MLP readouts and 40E/10I contacts per compartment. (A) The two corruption protocols on an illustrative image: independent pixel noise or deterministic scaling, and brightness-dependent noise or image gain, shown at fixed levels. B--D use the first protocol, E the second, and F compares conditions from both. Images are clipped to $[0,1]$ for display. Both protocols are applied without retraining. (B,C) Above-chance retention (upper) and paired shunting-minus-additive retention differences (lower) in four-level trees under Gaussian noise and deterministic scaling; open diamonds mark intervals including zero. (D) Four-level-minus-one-level retention at matched compartment count, contacts and parameters under noise (upper) and scaling (lower). B--D use twenty paired seeds, ten of them added after inspection (Supplementary Table~\ref{tab:seed_extension}); stochastic perturbations average five common draws. (E) Absolute test accuracy under brightness-dependent noise without gain spread (left; noise multiplier $\lambda$) and log-uniform gain without noise (right; gain bound $b$, with gains drawn between $1/b$ and $b$), for one-level and four-level trees; twenty seeds and five draws. (F) Linear probes of four-level soma outputs under clean input, Gaussian noise ($\sigma=0.3$), brightness-dependent noise ($\lambda=1$, $b=1$), or random gain ($b=20$, $\lambda=0$). Probes are fitted on clean validation images (open) or refitted under the test corruption (filled), then evaluated on separate test images above shuffled labels; twenty seeds and one shared corruption draw. Squares and solid lines identify shunting; circles and dashed lines addition. Intervals are pointwise 95\% seed bootstraps. Full factorial surfaces and information estimates appear in \suppfig{figS_corruption_protocols}.}
\label{fig:fig4}
\end{figure}

We next asked how integration and morphology affect tolerance to independent fluctuations and shared changes in input strength. In the clean-trained networks, the first perturbation protocol separates independent pixel noise from coherent changes in image strength (Fig.~\ref{fig:fig4}A). Noise changes each pixel independently, with values clipped to the image range; scaling multiplies every pixel by the same factor without clipping. We evaluated both perturbations without retraining. To compare performance relative to each model's clean accuracy, we use above-chance retention:
\[
R=\frac{A_{\rm perturbed}-A_{\rm chance}}{A_{\rm clean}-A_{\rm chance}},
\qquad A_{\rm chance}=\frac{1}{10}.
\]
Here $R=1$ denotes unchanged above-chance performance and $R=0$ chance performance. Noise distributions and scaling ranges are specified in Methods.

In four-level trees, shunting retained more performance under strong pixel noise, whereas strong global scaling favored addition (Fig.~\ref{fig:fig4}B,C). Uniform noise gave the same ordering (\suppfig{figS_morphology_noise}), as did linear-readout models trained on the same seeds (\suppfig{figS_readout_robustness}).

Morphology changes this comparison even at matched compartment counts, contacts and trainable parameters (Fig.~\ref{fig:fig4}D). Under Gaussian noise, the deeper tree retained more performance than the one-level tree under shunting and less under addition. At $\sigma=0.3$, the four-level shunting-minus-additive retention difference was 12.4 percentage points (95\% interval, 11.1--13.6), whereas the one-level tree favored addition (\suppfig{figS_morphology_noise}G). Strong brightening penalized the deeper tree under both rules; moderate dimming favored the deeper shunting tree.

The second protocol combines random image gain with brightness-dependent Gaussian noise: its variance grows with the mean brightness of the scaled image. We vary the gain range and noise strength independently: the gain bound $b$ sets the range $1/b$ to $b$ from which each image's gain is drawn, and the noise multiplier $\lambda$ scales the brightness-dependent noise amplitude. Negative pixels are set to zero without an upper cap (Methods). This connects noise amplitude to input strength while allowing their effects to be examined separately.

Brightness-dependent noise again favored shunting in the serial trees and addition in the one-level tree; random gain favored addition in every morphology (Fig.~\ref{fig:fig4}E). The full factorial comparisons retain both this protocol and the separate mean-one log-normal gain protocol with independent pixel noise (\suppfig{figS_corruption_protocols}A,B; \suppfig{figS_gain_load_morphology}). Their different gain distributions and noise laws should not be treated as interchangeable. Under log-normal shared gain, independent noise can reverse the ordering in serial trees; under the strongest log-uniform gain variation, addition remains ahead even with brightness-dependent noise.

Probes refitted on corrupted responses decoded class labels better than probes fitted on clean responses under both rules (Fig.~\ref{fig:fig4}F). Thus, some class structure remains decodable after corruption even when a readout fitted on clean responses no longer uses it effectively. Estimated class-information retention showed the same noise-versus-gain ordering (\suppfig{figS_corruption_protocols}C).

The contrast between noise and gain shows that a divisive local rule does not confer general gain invariance. Fixed leak and coupling conductances, learned activations and readouts all influence how input scaling affects the network; the next controls examine the contribution of local input pooling.

\subsection{Input pooling shapes robustness in trees and point neurons}
\label{sec:point_results}

\begin{figure}[htbp]
\centering
\IfFileExists{figures_empirical/fig6_points_v15.pdf}{%
  \includegraphics[width=\textwidth]{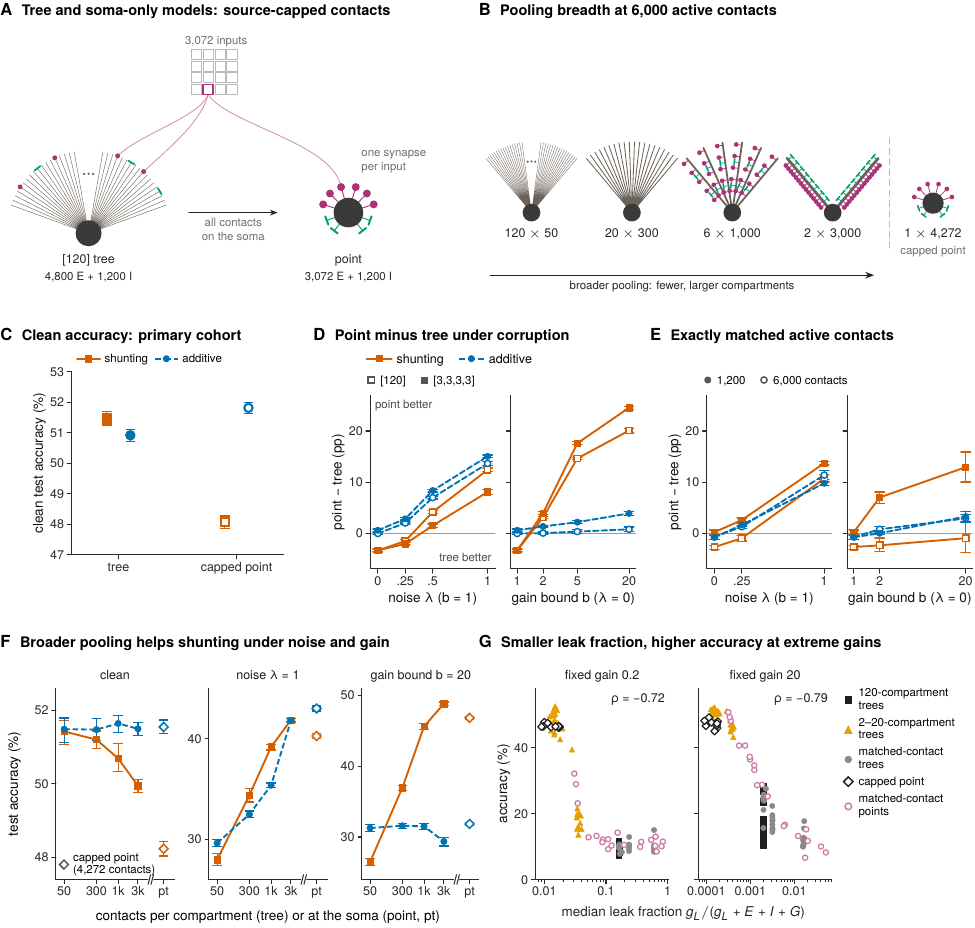}%
}{%
  \PackageError{empirical-manuscript}%
    {Missing required figure `figures_empirical/fig6_points_v15.pdf'}%
    {Restore the figure listed in the manuscript export manifest before compiling.}%
}%
\caption{\textbf{Input pooling shapes robustness in trees and point neurons.} Soma-only points have E and I contacts capped by the number of sources per pathway; their construction and training are specified in Methods. (A) A $[120]$ tree with 4,800 E and 1,200 I contacts and its point neuron (schematic): one synapse per input caps the point's E contacts at the 3,072 inputs. (B) The breadth control retains the same E/I totals as A (schematic). (C) Absolute clean test accuracy of the four primary CIFAR-10 arms: trees (filled) and capped points (open), shunting (squares) and addition (circles); twenty seeds. Cross-cohort rule contrasts appear in Supplementary Table~\ref{tab:point_interactions}. (D) Point minus tree accuracy for the $[120]$ (open) and $[3,3,3,3]$ (filled) trees of the resource-matched morphology cohort under the corruption of Fig.~\ref{fig:fig4}: brightness-dependent noise without gain variation (left) and random gain without noise (right); 20 seeds. (E) Point minus $[120]$ tree at exactly matched active contacts (filled: 600 E + 600 I; open: 3,000 E + 3,000 I), 64 neurons, ten seeds. (F) Accuracy with the 6,000 active contacts on 120, 20, 6 or 2 compartments (50 to 3,000 each), on clean images, at noise multiplier 1 and at gain bound 20; open diamonds (pt) mark the capped point (4,272 active contacts), a reference rather than a member of the series; ten seeds. (G) Median leak fraction of the most distal compartments (the soma of a point) against accuracy on the same 1,000 test images at fixed gains 0.2 and 20 for 130 trained shunting models; trees filled, points open; Spearman correlations across all models. C--F show means with pointwise 95\% seed-bootstrap intervals; differences are percentage points. Panel G shows individual models.}
\label{fig:point_neurons}
\end{figure}

The preceding robustness comparisons raise a further question about morphology: how much does tolerance to noise and gain depend on the number of inputs pooled before each compartment's nonlinearity? We compared trees with soma-only point neurons under both integration rules (Fig.~\ref{fig:point_neurons}A). Points receive their tree counterpart's active contacts at the soma, capped at one connection per available source in each pathway. We compare the shunting-minus-additive accuracy difference in trees with the same difference in points; control construction and training are specified in Methods. This tests how the effect of the integration rule depends on compartmentalization.

In the primary and resource-matched morphology comparisons, compartmentalization benefits shunting more than addition on clean images. In the primary and resource-matched CIFAR-10 cohorts, every shunting point is less accurate than its corresponding tree, whereas additive points have similar or higher mean accuracy than additive trees (Fig.~\ref{fig:point_neurons}C; \suppfig{figS_three_arms}A,B). The difference between the trees' and points' shunting-minus-additive contrasts is 4.3 percentage points in the primary cohort (95\% interval, 3.9--4.7) and is positive in every seed of the resource-matched family (Supplementary Note~\ref{note:three_arms}). The same interaction appears in MNIST and the fixed-contact controls with and without direct inhibition, but is not general across E/I allocations (Supplementary Table~\ref{tab:allocation_points}). This stronger benefit of compartments under shunting does not make shunting trees universally more accurate: their own advantage over additive trees is small in the resource-matched family, and the additive point slightly exceeds the shunting tree in the primary cohort.

Under the corruption of Fig.~\ref{fig:fig4}, in the resource-matched morphology cohort, the ordering depended on the integration rule and corruption strength (Fig.~\ref{fig:point_neurons}D). Shunting trees led on clean images and under weak noise; points led under stronger noise across all four morphologies. Under random gain without noise, shunting points also led at every nontrivial gain bound, with a much larger separation at the strongest variation. With raw addition, points and trees were similarly accurate on clean images, points were more accurate under noise, and gain variation alone separated them comparatively little. The strong-noise advantage of points therefore held under both rules, whereas their large advantage under random gain was specific to shunting in these cohorts.

Points also differ from trees in how many contacts each nonlinearity pools. A first-layer point neuron sums 4,272 contacts before one activation and one shunting denominator, whereas each compartment of a $[120]$ tree pools 50. To vary this pooling while retaining dendrites, we spread the tree's 6,000 active contacts per neuron over 20, 6 or 2 compartments (Fig.~\ref{fig:point_neurons}B,F). Under shunting, broader pooling slightly lowered clean accuracy but improved accuracy under strong noise and random gain: two compartments rather than 120 gave a gain of 13.9 percentage points under noise (95\% interval, 13.2--14.5), with a larger improvement under gain variation. Under raw addition, broader pooling improved noise tolerance but did not improve tolerance to random gain. The capped point is a separate reference because it has fewer active contacts; the two-compartment shunting tree outperformed it in all three displayed conditions, in every seed. Thus, substantial robustness can be recovered while retaining dendritic compartments.

The capped point also has fewer active contacts than the trees. We therefore compared the 64-neuron populations of the contact-bottleneck experiment with points having exactly the same E and I contact counts, at budgets of 1,200 and 6,000 contacts per neuron (Fig.~\ref{fig:point_neurons}E). These use the contact-matched models of Fig.~\ref{fig:fig2}F. Under strong noise, point neurons were substantially more accurate under both rules and at both budgets. Under random gain, shunting point neurons were more accurate than the corresponding trees at the lower contact budget, but the higher-budget pair was not clearly separated. Additive point neurons showed a smaller advantage at both budgets. On clean images, trees were more accurate at the larger budget under shunting and at both budgets under addition, whereas the low-budget shunting pair was similarly accurate. Exactly matching active contacts therefore preserves the points' noise advantage but does not produce a uniform advantage under gain variation.

A possible contributor to gain tolerance is the balance between synaptic and leak conductance. In a terminal compartment or soma-only neuron, common scaling of excitation and inhibition is increasingly cancelled as their summed conductance dominates leak. Scaling these inputs by $g$ gives $u(g)=gE_0/[g_L+g(E_0+I_0)]$, where $E_0$ and $I_0$ are the unscaled drives and $g_L=1$ is the leak conductance. For nonzero excitation, the relative gain sensitivity is
\begin{equation}
\frac{\partial\log u}{\partial\log g}
=\frac{g_L}{g_L+g(E_0+I_0)}.
\label{eq:leaf_gain_sensitivity}
\end{equation}
The right-hand side is leak's fraction of the total conductance. A smaller fraction therefore predicts less sensitivity before the activation function. Across trained shunting trees and points, a smaller median terminal leak fraction accompanied higher accuracy at extreme fixed gains (Fig.~\ref{fig:point_neurons}G). This association links the learned operating point to gain tolerance; it does not isolate leak from the other model differences. Together, these controls show that broader input pools can recover robustness to strong corruption while retaining dendritic compartments.

\subsection{Shunting reduces the performance cost of repeated dendritic E/I layers}
\label{sec:fig5_results}

\begin{figure}[htbp]
\centering
\IfFileExists{figures_empirical/fig7_networks_v14.pdf}{%
  \includegraphics[width=\textwidth]{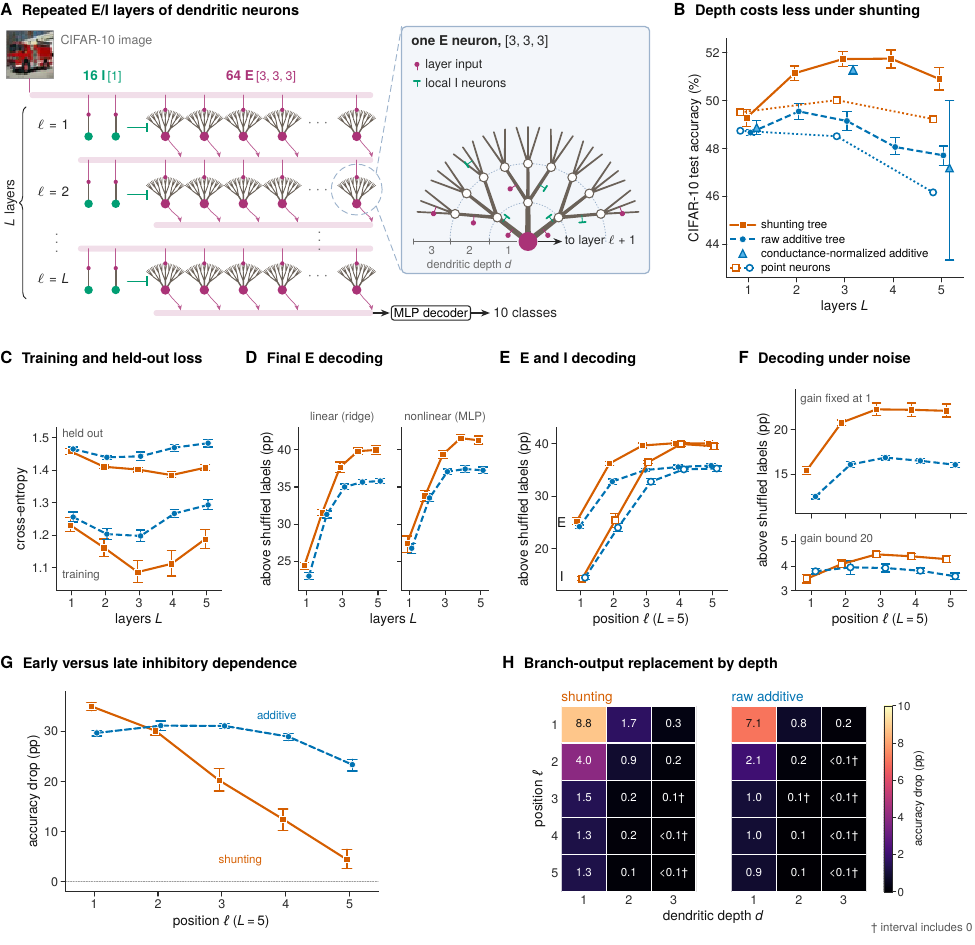}%
}{%
  \PackageError{empirical-manuscript}%
    {Missing required figure `figures_empirical/fig7_networks_v14.pdf'}%
    {Restore the figure listed in the manuscript export manifest before compiling.}%
}%
\caption{\textbf{Shunting reduces the performance cost of repeated dendritic E/I layers.} Each layer contains 64 E neurons with $[3,3,3]$ trees and 16 I neurons with $[1]$ trees. (A) Local I inhibits E; only E proceeds onward. Total network depth $L$, position $\ell$ and dendritic depth $d$ (magnified; contacts illustrative) are distinct. (B) Classifier accuracy versus $L$; triangles show conductance-normalized addition at $L=1,3,5$. Dotted lines with open markers, offset left, show the point-neuron reference, with a different contact and candidate-parameter budget (Supplementary Note~\ref{note:multilayer_point_reference}). (C) Logged training and held-out cross-entropy at the validation-selected epoch; training loss averages the epoch and held-out loss is measured at its end, so they diagnose fitting rather than a generalization gap. (D) Final-E linear and nonlinear decoding above shuffled-label baselines. (E) Decoding of E (filled) and I (open) populations at matched dimensionality within five-layer networks; information estimates are in \suppfig{figS29_multilayer_network_tree_atlas}E. (F) E-population decoding within five-layer networks under brightness-dependent noise (noise multiplier 1), with gain fixed at one (top) or drawn between 1/20 and 20 (gain bound 20; bottom); probes are refitted on corrupted validation images and scored on separately corrupted test images. (G) Accuracy loss after replacing local I output with a response to another image of the same class. (H) Accuracy loss after replacing one selected branch output per E neuron with mean responses over training images; somas are not replaced. Ten seeds per condition; intervals are pointwise 95\% seed bootstraps. In B and D--F coincident series are offset horizontally; $\dagger$ marks the eight intervals of H that include zero.}
\label{fig:fig5}
\end{figure}

In the preceding models, inhibitory pathways receive image-derived signals directly. In neural circuits, excitation can instead recruit local inhibitory neurons, as in feedforward inhibition. We therefore asked how repeated dendritic integration and population-mediated inhibition affect task performance and the use of signals across a network. Each layer contains a dendritic E population and a smaller I population driven by the same input pool; local I neurons inhibit E neurons, and only E output reaches the next layer (Fig.~\ref{fig:fig5}A). A single-layer control shows that population-mediated inhibition leaves shunting accuracy nearly unchanged but lowers additive accuracy (Supplementary Table~\ref{tab:inhibitory_source}; perturbation responses of the same networks in \suppfig{figS22_input_regime_routing}). We distinguish total network depth $L$, layer position $\ell$ and dendritic depth $d$ within a neuron. Networks contain one to five E/I layers, with fixed population sizes and morphology; an MLP reads out the final E population.

Classification accuracy initially improves with additional network layers, then declines under both integration rules. The decline is smaller under shunting: at $L=5$, it exceeds raw addition by 3.2 percentage points (paired 95\% interval, 2.6--3.8; Fig.~\ref{fig:fig5}B). Training and held-out losses rise more clearly in deeper additive networks. Under shunting, held-out loss falls up to four layers and rises slightly at five, while training loss rises beyond three (Fig.~\ref{fig:fig5}C). These logs indicate poorer fitting more clearly under addition, but do not distinguish optimization difficulty from limitations of the fitted architecture.

A conductance-normalized additive control divides the signed additive input by the same total conductance as shunting (Methods). It nearly matches shunting at three layers but is less accurate and less stable at five (Fig.~\ref{fig:fig5}B; Supplementary Note~\ref{note:multilayer_confirmatory}). The dotted point reference assigns each soma the contacts of one tree compartment and also declines in the longest networks. A second point control combines more contacts at the soma. Its additive networks can match or exceed the trees, whereas its shunting networks remain less accurate (\suppfig{figS_three_arms}C). This outcome depends on training conditions, so the poorer additive point performance in the primary comparison is not an architectural limit. The two point controls and their training settings are specified in Methods. Linear-readout retraining at the two depth endpoints retains the larger shunting advantage at five layers (\suppfig{figS_readout_depth}). The representation analyses below use MLP-trained networks.

To examine the representations underlying these predictions, we fitted linear and nonlinear auxiliary classifiers, or probes, to the final E population after network training. These probes predict class from fixed population responses. We report their test accuracy above a shuffled-label baseline, which measures how readily a specified readout can extract class identity (Methods). Both probe families show the same dependence on total network depth: decoding improves up to four layers and then levels off, and is higher under shunting from three layers onward (Fig.~\ref{fig:fig5}D). The shunting advantage therefore extends beyond linear decoding and persists in the training-sensitivity control (\suppfig{figS17_multilayer_policy_sensitivity}).

We then followed E and I representations within the five-layer networks.  Because the populations differ in size, we projected E outputs, without using labels, to the dimensionality of the I population.  Decoding improves from early to late populations, with a clearer late advantage under shunting (Fig.~\ref{fig:fig5}E). Population-information estimates provide a descriptive complement and are shown with their estimator controls in \suppfig{figS29_multilayer_network_tree_atlas}E,F.

Decoding under input corruption extends the within-network comparison (Fig.~\ref{fig:fig5}F). Probes were refitted to the corrupted responses, while the networks remained fixed after clean training (Methods). Under brightness-dependent noise without gain variation, the shunting advantage grew toward the output, reaching a between-rule difference of 6.1 percentage points in baseline-corrected decoding at the final layer (95\% interval, 5.3--6.9). Wide random gain greatly reduced decoding under both rules and left little difference between them. Thus, the late-layer decoding advantage widens under this noise alone but nearly disappears when gain also varies widely.

To test dependence on example-specific inhibition, we replaced one local I population's output with activity evoked by another image of the same class.  This preserves the empirical distribution of I responses within each class while disrupting their pairing with the current E response.  The layerwise comparison reveals opposite profiles: disrupting local inhibition costs shunting networks most near the input, whereas additive networks remain more dependent on it near the output. The accuracy loss, $\Delta A_{\rm replace}=A_{\rm intact}-A_{\rm replaced}$, is larger for shunting in the first layer and for addition in later layers (Fig.~\ref{fig:fig5}G; \suppfig{figS21_multilayer_subset_geometry_local_i}C adds a training-sensitivity control).  Thus, the contribution of example-specific E/I coordination depends on both integration rule and network position.  The intervention uses evaluation labels as a mechanistic control.  Removing I output altogether has a much larger effect, bringing both networks close to chance (Supplementary Note~\ref{note:multilayer_subset_geometry_local_i}).

This population-level intervention differs from replacing outputs inside individual dendritic trees. We separately tested how classifier dependence varies with network position $\ell$ and dendritic depth $d$. One soma-to-distal path was selected in every E neuron without using labels. Replacing the selected branch outputs with their mean responses over training images produces the greatest accuracy loss proximally (Fig.~\ref{fig:fig5}H). This intervention replaces one branch output per E neuron at each tested depth, leaving the other branches available; somas themselves are not replaced.

Improved decoding means that class labels are more readily extracted by the tested readout; it need not imply an increase in the total class information carried by the population. The data-processing constraint and finite-sample estimator checks are given in \suppfig{figS29_multilayer_network_tree_atlas} and its accompanying note.

Across these analyses, shunting better preserves decodable class structure in longer E/I networks, while informative signals and classifier dependence remain distinct. The concentration of replacement effects near the soma persists under alternative replacement references and training conditions (\suppfig{figS32_multilayer_post_gate_output_intervention}). The inhibitory donor intervention in Fig.~\ref{fig:fig5}G establishes dependence on example-specific coordination, but ordinary classification does not specify an alternative answer that a donor signal should produce. The contextual task below provides that prediction by changing the required response with a cue.

\subsection{Gain sensitivity and inhibitory control of contextual responses}
\label{sec:contextual_remapping_results}

\begin{figure}[htbp]
\centering
\IfFileExists{figures_empirical/fig7_contextual_v16.pdf}{%
  \includegraphics[width=\textwidth]{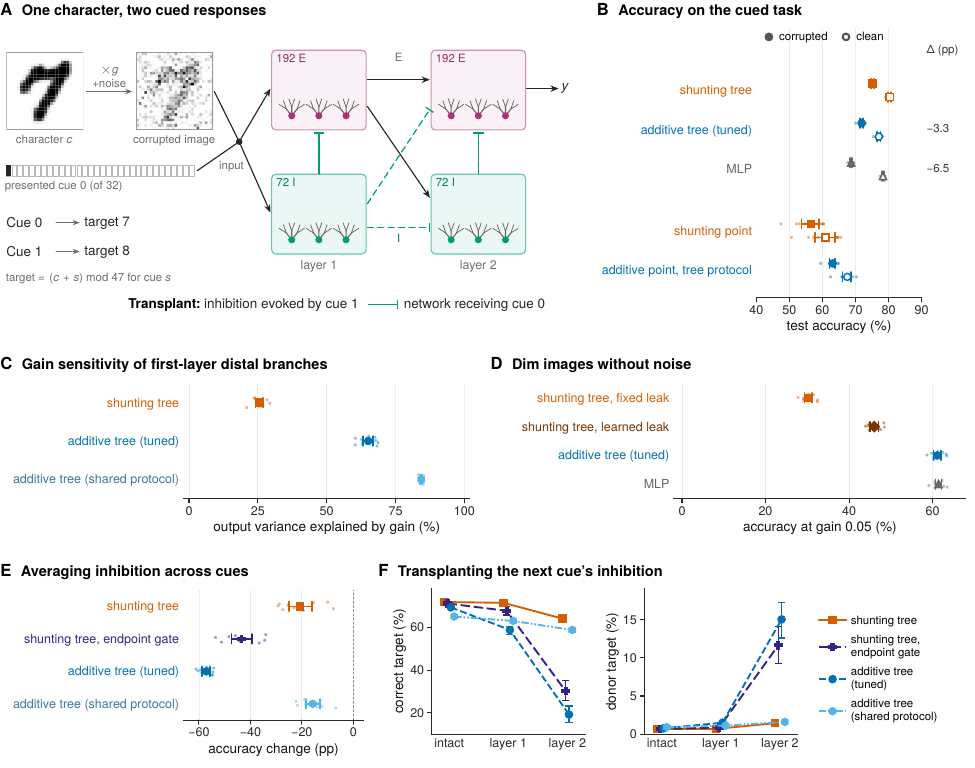}%
}{%
  \PackageError{empirical-manuscript}%
    {Missing required figure `figures_empirical/fig7_contextual_v16.pdf'}%
    {Restore the figure listed in the manuscript export manifest before compiling.}%
}%
\caption{%
\textbf{Gain sensitivity and inhibitory control of contextual responses.}
(A) The same corrupted character requires different outputs under different cues. For character 7 (gain 0.25, noise divisor 2), cue 0 requires target 7 and cue 1 requires target 8. The transplant supplies inhibition from cue 1 while the network receives cue 0. Each layer has 192 E and 72 I neurons with $[3,2]$ trees. Arrows denote excitation and bars inhibition.
(B) Accuracy on clean and corrupted test images for trees, soma-only neurons and the retained-parameter-matched MLP. Filled symbols show accuracy under the gain and noise used in training and open symbols show clean accuracy. Values at right give the indicated model's accuracy minus the shunting tree's on corrupted images, in percentage points.
(C) Fraction of output variance explained by image gain in the distal compartments of first-layer E neurons, measured on held-out training images.
(D) Accuracy at gain 0.05 without noise, with shunting leak fixed at one or learned from one. The tuned additive tree and MLP provide reference values.
(E) Change from intact accuracy after replacing the inhibitory drive to each second-layer E compartment with its mean over all 32 cues for the same corrupted image. Differences are paired by training seed.
(F) Fractions of predictions matching the presented cue's target (left) or the donor cue's target (right). The next cue in cyclic order supplies the inhibitory drive to E compartments in layer 1 or layer 2 in separate tests. Intact networks provide the baseline. E and F use the same fixed grid of gains and noise levels. The endpoint gate is a tanh activation anchored at zero and one. B--F show means and pointwise 95\% seed-bootstrap intervals across ten trained networks per model. Small dots in B--E mark individual networks.}
\label{fig:fig7}
\end{figure}

A change in image brightness should leave the identity of a character intact, but a cue can change the appropriate response to that same character. To test these two demands together, we presented networks with images from the 47 classes of EMNIST Balanced~\citep{CohenEtAl2017EMNIST}. An uncorrupted cue selected one of 32 cyclic shifts of the character-to-output mapping (Fig.~\ref{fig:fig7}A). For character $c$ and shift $s$, the required output was
\begin{equation}
y=(c+s)\bmod 47.
\label{eq:context_target}
\end{equation}
Gain and independent pixel noise varied during training as well as testing. The same corrupted image could require a different answer under each cue.

The image and cue entered the first of two E/I layers. E neurons received inhibition from local I neurons in both layers. I neurons in the first layer also projected to the second. On test images corrupted as during training, the primary shunting tree exceeded the separately tuned additive tree by 3.3 percentage points (95\% interval, 2.7--3.8) and also outperformed the retained-parameter-matched MLP and soma-only controls (Fig.~\ref{fig:fig7}B). The shunting advantage persisted when both trees used the same training protocol (\suppfig{figS_context_task_controls}A). At a fixed noise level, the shunting tree also outperformed the separately tuned additive tree and the MLP at every gain tested (\suppfig{figS_context_task_controls}B).

Image gain explained less of the variation in first-layer distal branch outputs in the shunting tree than in either additive tree (Fig.~\ref{fig:fig7}C). Although their excitatory inputs varied similarly across gains, shunting attenuated this variation more strongly at the branch output (\suppfig{figS_context_controls}B,C). This attenuation had limits. On dim images presented without the noise encountered during training, the fixed-leak shunting tree was less accurate than the additive tree and MLP. Learning the leak partly restored its accuracy (Fig.~\ref{fig:fig7}D). Leak's contribution to total conductance also fell in first-layer distal compartments on dim, noisy inputs (\suppfig{figS_context_controls}F).

In the shunting tree, the cue was readily decoded from first-layer E neurons. The required answer became reliably decodable in second-layer E neurons (\suppfig{figS_context_task_controls}C). A decoder fitted on clean responses from those neurons transferred well across most gains. On the dimmest noise-free images, both decoders lost accuracy, and refitting at that gain recovered about four percentage points (\suppfig{figS_context_task_controls}D).

To ask whether inhibition helped select the cued answer, we first replaced second-layer inhibitory drive with its average over all cues for the same corrupted image. Removing cue-specific inhibition reduced accuracy much more in the separately tuned additive tree than in the shunting tree (Fig.~\ref{fig:fig7}E). An accuracy loss alone leaves open which response the inhibitory signal favored. We therefore held the presented image and cue fixed and supplied the inhibition evoked by the next cue. With this second-layer transplant, the tuned additive tree selected the donor cue's target on 15.0\% of trials, compared with 1.4\% for the more accurate shunting tree (Fig.~\ref{fig:fig7}F).

The degree of inhibitory control also varied within each integration rule. An alternative shunting tree used a learned tanh activation anchored at zero and one at the endpoints of the integration range (the \emph{endpoint gate}; Methods, Eq.~\ref{eq:context_activation}). Its accuracy under the training corruption was slightly lower than the primary shunting tree's (\suppfig{figS_context_controls}A), yet it selected the donor target much more often after the second-layer transplant. The additive tree trained with the shunting protocol rarely selected that target (Fig.~\ref{fig:fig7}F). Second-layer E neurons also selected local inhibitory inputs whose cue tuning was less similar to their own than that of the eligible sources (\suppfig{figS_contextual_wiring}C). The transplant thus distinguishes how strongly a learned circuit uses inhibition to select a response, even among networks with the same local integration rule.

\subsection{Dendritic E/I populations support computation at AlexNet width}
\label{sec:fig6_results}

\begin{figure}[htbp]
\centering
\IfFileExists{figures_empirical/fig8_alexnet_v14.pdf}{%
  \includegraphics[width=\textwidth]{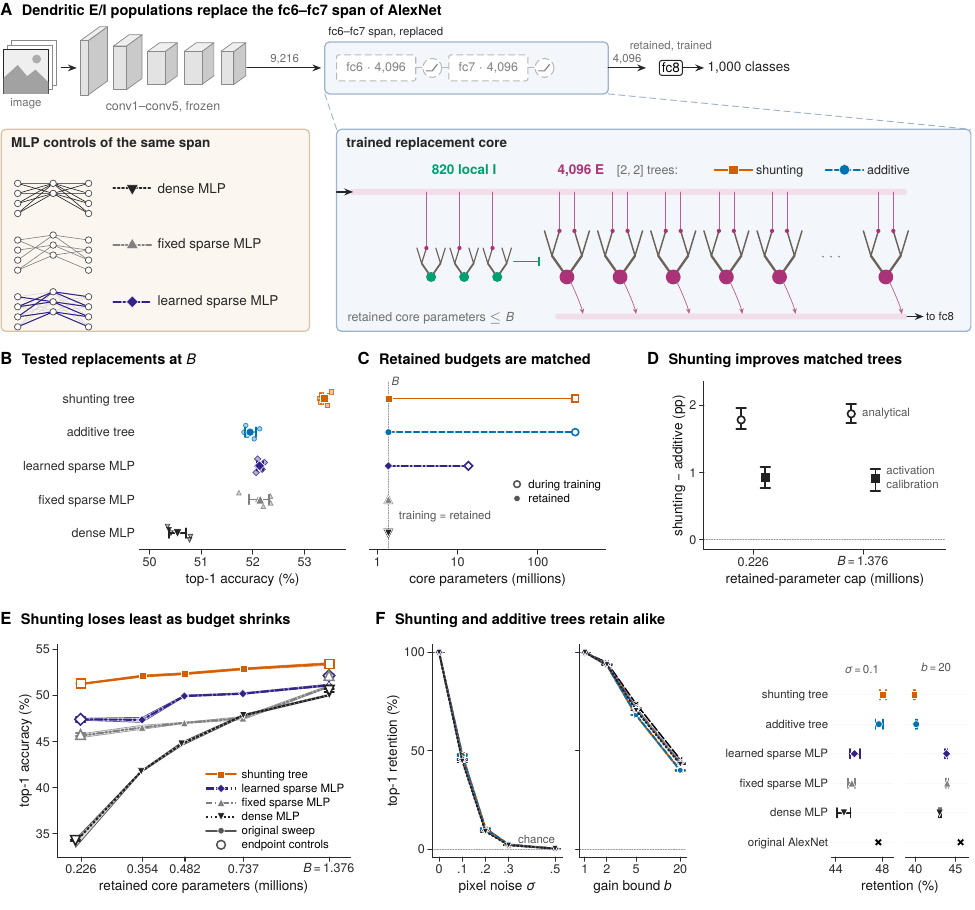}%
}{%
  \PackageError{empirical-manuscript}%
    {Missing required figure `figures_empirical/fig8_alexnet_v14.pdf'}%
    {Restore the figure listed in the manuscript export manifest before compiling.}%
}%
\caption{\textbf{Shunting improves matched trees, and learned trees retain accuracy at smaller deployment budgets.} (A) The fc6--fc7 replacement analyzed in B--F. The frozen convolutional prefix feeds a trained core of 4,096 E neurons with $[2,2]$ trees (shunting or additive) and 820 local I neurons, which replaces fc6--ReLU--fc7--ReLU; only E outputs enter fc8, which is retained and trained. MLP controls replace the same span. (B) Selected fc6--fc7 replacements at $B=1{,}376{,}320$ retained core parameters. Model-selection procedures are specified in Methods. Small points: five paired seeds; large symbols and bars: means and pointwise 95\% seed-bootstrap intervals. (C) Stored trainable core parameters during optimization (open) and retained after training (filled) at $B$. Tree rules coincide; the common classifier and nontrainable indices are excluded. (D) Paired shunting-minus-additive accuracy under each matched initialization procedure, at caps 226,120 and $B$. Analytical initialization uses expected input drives; activation calibration uses training inputs to set the activation range (Methods). Both rules have identical population widths, morphology and parameter counts; calibration fits rule-dependent activation parameters. (E) Original five-budget sweep: connected filled symbols and bands. Isolated open symbols show the endpoint controls; no regularization-tuned intermediate points are implied. (F) Retention of above-chance top-1 accuracy, $(A-0.1\%)/(A_{\rm clean}-0.1\%)$, of endpoint models at $B$ and untouched AlexNet under Gaussian pixel noise and under a gain drawn per image between $1/b$ and $b$ (gain bound $b$), applied to pixels without retraining; means of five corruption draws; dotted lines: chance; right: $\sigma=0.1$, $b=20$. B--D, F and open symbols in E use five confirmation seeds; connected curves use a separate set of five seeds. All are evaluated on the same 50,000 official-validation images; training and model selection are specified in Methods. Tree--MLP comparisons match retained budgets closely, but differ in width and training candidate-bank size.}
\label{fig:fig6}
\end{figure}

We finally asked whether sparse dendritic E/I populations support accurate classification at much larger network width, how accuracy depends on retained contact budgets, and which input-sensitivity differences persist. AlexNet extracts image features through convolutional layers and passes them through fully connected layers named fc6, fc7, and fc8 for classification.  Starting from ImageNet-pretrained AlexNet~\citep{Deng2009,Krizhevsky2012}, we replaced either fc7 alone or the combined fc6--fc7 transformation with dendritic E/I populations.  The preceding layers remained fixed, while the replacement and final classification layer were trained (Fig.~\ref{fig:fig6}A shows the fc6--fc7 replacement analyzed in the figure).  As in the smaller multilayer networks, local I neurons inhibit E neurons, and only E output passes onward.

We compared shunting networks that learn their contact locations with dense MLPs and sparse MLPs whose contacts are either fixed or learned. For the fc6--fc7 replacement, we counted parameters retained after training separately from the full candidate bank used during optimization and from the common final classifier (Methods).

We tested two retained-parameter caps spanning the deployment-budget range; the larger cap, $B$, is the budget of the reference learned-contact shunting model. Shunting remains more accurate than the selected MLPs at both caps. At $B$, its advantage over the learned-contact sparse MLP is 1.3 percentage points (paired 95\% interval, 1.1--1.4; Fig.~\ref{fig:fig6}B). This comparison includes regularization of the sparse MLPs, which narrows the advantage relative to their unregularized counterparts. Learned and fixed sparse MLPs perform similarly at the larger cap, whereas learning contacts helps at the smaller cap; the dense bottleneck loses the most accuracy as its budget shrinks (Supplementary Table~\ref{tab:alexnet_followup}).

These comparisons match deployment budgets closely, with small differences from integer widths and contact counts, but do not match the candidate banks used during training (Fig.~\ref{fig:fig6}C). Shunting keeps the same population widths and large candidate bank as its retained budget falls, whereas the MLPs reduce hidden width. These accuracy differences compare models at similar retained parameter counts, but with different training banks and internal widths.

Within the tree architecture, learned-contact additive networks match the shunting networks in population sizes, morphology and parameter budgets. Shunting is more accurate at both caps, and this ordering persists under both tested initialization procedures (Fig.~\ref{fig:fig6}D; Methods). The additive tree also exceeds the learned sparse MLP at the smallest cap, so the tree--MLP difference cannot be assigned entirely to shunting.

A five-budget sweep shows how accuracy varies across the deployment range (Fig.~\ref{fig:fig6}E; Supplementary Table~\ref{tab:downward_matched}). Shunting retains higher accuracy than the tested MLPs throughout this sweep and loses less accuracy as retained parameters decrease. The regularized endpoint comparisons are shown separately because the intermediate budgets were not retuned (Methods; Supplementary Note~\ref{note:alexnet_followup}).

Fixed-contact comparisons support complementary conclusions: learning contact locations improves shunting at the same retained count, and fixed shunting outperforms raw addition and a sparse point replacement in the original cohort (Supplementary Table~\ref{tab:fig6_fc6fc7_confirmation}). A separate sparse-MLP comparison also shows that the relative accuracy of fixed shunting depends on training duration (Supplementary Table~\ref{tab:alexnet_matched_mlp}; Methods).

All replacements remain below untouched pretrained AlexNet in these comparisons.  Retraining the native dense fc6--fc7 layers also lowers accuracy relative to the untouched reference, showing that the observed deficit includes the effect of retraining.  Separate compression controls also demonstrate useful conventional replacements: after retraining, pruned conventional layers attain accuracy similar to learned-contact dendritic replacements at comparable active-weight budgets (Supplementary Table~\ref{tab:alexnet_conventional_compression}; Supplementary Note~\ref{note:conventional_mlp_controls}). These controls inherit pretrained weights and retain dense masked arrays during optimization, so they address compression rather than learning a replacement from scratch at the same training-storage budget.

Unused candidate weights can be discarded after learning, and both budget comparisons evaluate the compact models directly. This conversion reduces deployment storage without reducing the candidate bank used during training~\citep{HanEtAl2015}. Whole-model savings also depend on connectivity indices and the unchanged AlexNet layers; an output-preservation check and storage accounting are reported in Supplementary Note~\ref{note:scale_compatibility}.

A separate fc7-only analysis locates decodable class structure within the replacement. Both distal E branches and E somas carry class-related signals, but their decoding differences are small relative to between-network variability (\suppfig{figS_alexnet_width_parameter_scaling}B,C). These measurements describe internal representations without assigning a unique causal contribution to either pathway.

Finally, we tested corrupted images without retraining any model (Fig.~\ref{fig:fig6}F). Corruption acts before the frozen convolutional prefix and therefore tests the whole network. All models lost accuracy under Gaussian pixel noise; at the illustrated endpoint, trees retained modestly more above-chance accuracy than the MLPs and about as much as untouched AlexNet. Under wide random gain, the trees retained modestly less than both. Shunting and additive trees had similar retention at the illustrated noise and gain endpoints. Linear decoding of the replacement outputs also favored the MLPs and AlexNet's original fully connected representations under both corruptions (\suppfig{figS_alexnet_corruption_decoding}). The strong corruption-dependent separation between integration rules in the smaller CIFAR-10 networks therefore does not recur in this replacement setting.

\clearpage
\section{Discussion}
\label{sec:discussion}

We used DendriNet to examine how dendritic compartments, conductance-based integration and sparse E/I contacts affect computation. We found that networks providing similar accuracies on different tasks show different representations and computations, suggesting three main hypotheses. First, a branch's contribution depends on which inputs it groups, not just on the number of branches or dendritic depth. Second, integration shapes branch influence and can affect excitatory or inhibitory synapse recruitment during learning. Third, the inhibitory signal provides a mechanism that can regulate gain, sharpen selectivity or influence somatic response, depending on the input or signal that drives it.

\textbf{Dendritic morphology}\par\nobreak

Dendritic branching divides a neuron's inputs into local pools that interact before reaching the soma. Small pools keep different input combinations separate, whereas large pools can average out independent fluctuations, such that the optimal arrangement should depend on what the grouped inputs share. Consistent with this, distributing a limited contact budget across compartments improved classification over soma-only neurons, most clearly in small populations (Fig.~\ref{fig:fig2}F), and restricting each branch to a local image region improved accuracy at matched resources (Fig.~\ref{fig:fig2}G,H). Under strong noise the balance shifted: spreading the same contacts over fewer, broader compartments improved accuracy while retaining dendrites (Fig.~\ref{fig:point_neurons}F). In biological networks, afferents target distinct dendritic domains~\citep{PetreanuEtAl2009}, such that which inputs share a branch may matter as much as the number of branches. Our experiments complement previous studies of sparse connectivity~\citep{MocanuEtAl2018,EvciEtAl2020,JayakumarEtAl2020} by examining where retained contacts terminate and how their inputs are combined.

Aizenbud et al.~\citep{AizenbudEtAl2026} showed that dendritic morphology and synaptic nonlinearities increase the functional complexity of a neuron's input--output transformation. We examined a complementary question, whether deeper dendrites yield better learned performance. In resource-matched trees, additional serial stages failed to improve accuracy on noise-free images (Fig.~\ref{fig:fig2}G). Therefore, a more complex transformation does not automatically help with a task that has to be learned.

\textbf{The mode of synaptic integration}\par\nobreak

Grouping inputs on a branch matters only if that branch influences the somatic output. In networks of similar accuracy, distal branches contribute to performance under shunting: removing the image-specific activity of the terminal branches sharply reduced accuracy in shunting trees but barely changed it in additive trees (Fig.~\ref{fig:fig3}; \suppfig{figS40_whole_layer_ablation}). Under shunting, inhibition changes a compartment's total conductance, and with it the gradients at excitatory synapses and branch couplings (Methods, Eq.~\ref{eq:shunting_task_gradients}). Conductance-based integration may therefore help determine which branches are recruited during learning, a possibility that measurements after training cannot resolve.

Shunting is widely associated with gain control and divisive normalization~\citep{Silver2010,CarandiniHeeger2012}, an association our results qualify. In a terminal compartment, common scaling of excitation and inhibition is increasingly canceled as synaptic conductance dominates leak (Eq.~\ref{eq:leaf_gain_sensitivity}). In networks trained on uncorrupted images, deeper shunting trees tolerated independent noise better than additive trees, but were more affected by strong global scaling (Fig.~\ref{fig:fig4}). Robustness comparisons must therefore specify both the input perturbation and the network receiving it~\citep{HendrycksDietterich2019,ChavlisPoirazi2025,HuangEtAl2025}. Gain tolerance may depend on the learned balance between synaptic and leak conductance: in the contextual networks trained with variable gain, gain explained less of the output variance in first-layer distal branches under shunting than under addition (Fig.~\ref{fig:fig7}C), and learning the leak improved accuracy on dim, noise-free inputs outside the training distribution (Fig.~\ref{fig:fig7}D). Beyond single layers, shunting reduced the accuracy cost of stacking E/I layers (Fig.~\ref{fig:fig5}) and outperformed matched additive trees at AlexNet width (Fig.~\ref{fig:fig6}).

\textbf{Computation by inhibition}\par\nobreak

Distal dendritic inhibition is often considered a gate for plasticity, controlling the induction of calcium signals that drive synaptic change~\citep{Higley2014,LetzkusEtAl2015,WilmesEtAl2016}. In our networks, as in other networks trained with separate excitatory and inhibitory units~\citep{SongYangWang2016,CornfordEtAl2021}, it was also part of the computation, and what it computes should depend on how its drive relates to excitation. Inhibition that covaries with excitation can regulate gain~\citep{LovettBarronEtAl2012}; inhibition tuned away from a neuron's preferred input can sharpen selectivity; and inhibition that is largely uncorrelated with excitation can carry a separate signal, such as a contextual cue, that selects the response~\citep{YangMurrayWang2016}.

We found signs of all three modes. In the first layer of the contextual networks, accounting for gain and noise weakened the correlation between excitatory and inhibitory drive (\suppfig{figS_contextual_wiring}B). Shunting also reduced gain-related variation in distal branch outputs (Fig.~\ref{fig:fig7}C). Inhibitory contacts showed lower alignment with the receiving neuron's preferred inputs: in the digit classifier they fell on fewer active pixels for the assigned digit than for other digits (Fig.~\ref{fig:fig1}H), and in the contextual network they were less aligned with the receiving neuron's context tuning than the available sources (\suppfig{figS_contextual_wiring}C). This is consistent with a function in sharpening selectivity, which these measurements do not test. In that network's second layer, where the two drives were only weakly correlated, the response depended on cue-specific inhibition, although context was also decodable from excitatory populations (\suppfig{figS_context_task_controls}C). Replacing this inhibition with its average over cues reduced accuracy more in the separately tuned additive tree than in the primary shunting tree (Fig.~\ref{fig:fig7}E). When the additive tree received inhibition evoked by another cue, it selected that cue's target far more often than it did with inhibition intact (Fig.~\ref{fig:fig7}F).

The primary shunting tree was slightly more accurate than the endpoint-gate variant on corrupted images (\suppfig{figS_context_controls}A), but averaging or transplanting inhibition had a larger effect in the variant (Fig.~\ref{fig:fig7}E,F). Inhibitory pathways with the same local mechanism can therefore support different computations through their learned signals and coordination with excitation. In biological circuits, this predicts that learning could change which response an inhibitory pathway favors without changing its local synaptic action. For brain-inspired networks, it motivates learning inhibition for contextual response selection alongside gain regulation.

\textbf{Limitations}\par\nobreak

These experiments reveal the solutions that the tested learning procedures reach, not the full computational capacity of each architecture: a computation that training does not find goes untested. The models are steady-state rate units trained by backpropagation, without cable filtering, conductance kinetics, spikes, the timing of feedforward inhibition~\citep{PouilleScanziani2001} or compartment-specific plasticity~\citep{Guerguiev2017,Sacramento2018}, and single-layer networks give excitatory and inhibitory afferents the same image features. Changing compartment number or depth also changes couplings, activation parameters and the weights available during training. Tuning and covariation identify candidate roles for inhibition that only interventions test, and interventions can create activity absent during training. Pixel noise and image scaling do not reproduce the structure of neural variability~\citep{AllenStevens1994,GorisEtAl2014}, and benchmark test sets informed earlier model development.

\textbf{Future directions}\par\nobreak

Each hypothesis suggests an experiment. Varying task-relevant and nuisance correlations among inputs independently, at fixed contact counts and single-input statistics, would test whether the favorable pool size follows them. Measuring branch-to-soma transmission and synaptic gradients in DendriNet from initialization onward would show when the two integration rules begin to differ in their reliance on distal input. Measuring gain sensitivity and response bias in the same circuit, with transplants matched for total inhibition on every trial, would help distinguish the modes an inhibitory pathway serves. The same predictions can be tested in models with cable and synaptic dynamics, and in biological circuits by combining dendritic and somatic recordings with branch- or interneuron-specific perturbations during learning. More broadly, morphology, integration and inhibition should be studied together: which inputs meet on a branch, how strongly that branch influences the output and what its inhibition contributes need not vary in parallel, and none can be read from accuracy alone.

\section{Methods}
\label{sec:methods}

\subsection{Datasets and evaluation splits}

MNIST~\citep{LeCun1998} and CIFAR-10~\citep{Krizhevsky2009} use their canonical training and test sets.  The canonical training set is divided deterministically into an 80\% optimization subset used for weight updates and a 20\% validation subset used for model selection. All single-layer CIFAR-10 cohorts share one split (split seed 61031); MNIST and multilayer CIFAR-10 cohorts use the experiment seed for their split. Paired model comparisons share split membership.  Validation negative log likelihood determines early stopping and model selection; the canonical test set provides the reported predictive accuracy.  MNIST and CIFAR-10 images are represented as flattened vectors in $[0,1]$ without data augmentation.  In direct-input models, the same image is supplied to independently learned E and I synaptic banks.  These E and I input channels represent distinct afferent populations with matched feature tuning, each with a fixed synaptic polarity; the upstream circuit that would produce these afferent rates is not modeled. This is an explicit modeling assumption, not an inhibitory circuit learned by the network.  Because canonical test results were inspected during development of the broader model family, the reported intervals describe variation across trained seeds on these established benchmarks rather than performance on a previously unseen confirmatory dataset.

Figure~\ref{fig:fig1} uses ten MNIST somas and an identity readout.  The baseline CIFAR-10 experiments use a $64\mathbin{\to}32\mathbin{\to}16\mathbin{\to}10$ ReLU MLP, trained jointly with the dendritic network.  Its 2,778 parameters are included in the reported model totals.  The soma-count experiment in Fig.~\ref{fig:fig2}F varies the input width of this readout while retaining its two hidden layers.  Figure~\ref{fig:fig5} supplies the readout with the final 64-dimensional E-population output.  All classifiers minimize categorical negative log likelihood, or cross-entropy: lower values mean that higher probabilities are assigned to the correct classes.

For the allocation readout comparison, both integration rules are retrained under the common protocol with a single affine map from 64 somatic outputs to ten class scores, conventionally termed a linear readout.  This readout has 650 parameters.  E and I counts are independently selected from $\{1,5,10,15,20,25,30,35,40,45,50\}$, yielding 121 allocations for each of two integration rules and two readouts.  The comparison uses seeds 10--19, pairing the linear models with the same ten-seed subset of the twenty-seed MLP grid.  Architecture, the contact-selection procedure, training schedule, the common protocol and each integration rule's initialization are retained; only the readout architecture changes.  The readout comparison therefore has 4,840 trained-model observations.  Post hoc probes of frozen MLP-trained representations are analyzed separately (\suppfig{figS15_allocation_probes}).

The readout effect, $\Delta A_{\rm readout}=A_{\rm linear}-A_{\rm MLP}$, is computed within seed before averaging or bootstrapping; positive values favor linear-readout training.  For the fixed-total interaction, we average each seed's readout effect over the two excitation-dominated allocations $(N_E,N_I)=(40,10),(45,5)$ and separately over the two inhibition-dominated allocations $(5,45),(10,40)$.  We subtract these averages within each integration rule, then compare the resulting changes between integration rules.  Ten seed vectors are resampled jointly across all constituent conditions.  This exploratory summary tests the direction and size of the observed allocation interaction; it does not estimate a universal crossing threshold.  Pointwise grid intervals are not adjusted for multiple comparisons.  Later readout checks include primary scalar-information, intervention and perturbation analyses, an integration-rule comparison, and multilayer endpoint comparisons; their coverage is specified in \suppfig{figS_readout_mechanisms}, \suppfig{figS_readout_robustness}, and \suppfig{figS_readout_depth}.  The primary checks retrain the networks of Figure~\ref{fig:fig3}B--F with a linear readout on the same seeds 0--9.  The integration-rule comparison reads the shunting and raw-additive models of the 40E/10I cells of the Figure~\ref{fig:fig2} allocation and readout grids (seeds 10--19) and a separately trained conductance-normalized cohort with the same seeds; the multilayer endpoints use seeds 10--19.

\subsection{Dendritic geometry and network depth}

Neuronal activity is represented by continuous rates rather than spikes~\citep{WilsonCowan1972}. A morphology $[b_1,\ldots,b_D]$ specifies the number of children per branch at each successive soma-relative depth.  For $n$ somas, depth $d$ contains
\[
n\prod_{j=1}^{d}b_j
\]
compartments.  We define the soma as $d=0$, proximal dendrites as $d=1$, and increasingly distal dendrites as larger $d$ (notation in Supplementary Table~\ref{tab:notation}).  A $[3,3,3,3]$ neuron therefore contains $3+9+27+81=120$ nonsomatic compartments.  Network-layer position is denoted $\ell=1,\ldots,L$, where $L$ is the total number of E/I population layers.  The coordinates $d$ and $\ell$ are distinct: $d$ locates a compartment within one tree, whereas $\ell$ locates that tree within a feedforward network.

The dendritic-tree models of Figures~\ref{fig:fig1}--\ref{fig:fig5} contain no direct somatic synapses; point-neuron controls integrate their inputs at the soma, and the contextual networks of Figure~\ref{fig:fig7} add inhibitory contacts at the soma.  Each nonsomatic compartment receives its own E and I drives, and child-branch outputs propagate toward the soma.  The explicit-inhibition controls and multilayer experiments in those figures add a local I population.  At layer $\ell$, the same feedforward source drives the E and I populations, the I output supplies the inhibitory input to the E population, and only the E output proceeds to layer $\ell+1$. In the contextual networks, first-layer I output also inhibits both second-layer populations.

\subsection{Sparse synapses with nonnegative weights and branch coupling}

Unless stated otherwise in the task-specific Methods, each unconstrained synaptic parameter $\theta$ is transformed into a nonnegative weight,
\[
w=\operatorname{softplus}(\theta)=\log(1+e^\theta).
\]
For nonnegative input coordinates $x_j\geq0$ , the local synaptic drives are
\[
E=\sum_{j\in\mathcal S_E}x_jw_j^E,\qquad
I=\sum_{j\in\mathcal S_I}x_jw_j^I,
\]
where $\mathcal S_E$ and $\mathcal S_I$ are the source indices of the retained E and I contacts.  The standard sparse layer optimizes a dense candidate bank but retains only the $N_E$ and $N_I$ largest transformed weights per compartment during the forward pass.  Here, an active contact is a retained connection that contributes to the forward computation; its source may still have zero activity for a particular example.  Thus $N_E$ and $N_I$ count retained contacts, not stored parameters.  Candidate synaptic parameters are the stored weights from which these contacts are selected.  Total trainable parameters additionally include branch-coupling weights, activation-function parameters, and decoder parameters.  The fixed-index dendritic controls in Fig.~\ref{fig:fig2}D,G,H and the earlier AlexNet comparisons instead store exactly the specified weights and integer source indices. The fixed sparse MLP in Fig.~\ref{fig:fig6} also stores only its retained weights, but allows them to be signed.

For a single population with $n$ somas, $N_{\rm comp}$ nonsomatic compartments per soma and direct E/I input widths $M_E,M_I$, learned-contact trees store $nN_{\rm comp}(M_E+M_I)$ candidate synaptic weights, $nN_{\rm comp}$ child couplings and $2n(N_{\rm comp}+1)$ activation parameters, in addition to the readout. Thus the 64-neuron, 120-compartment CIFAR-10 model with two 3,072-input banks and the 2,778-parameter readout trains 47,211,866 parameters. Its soma-only counterpart trains 396,122. These totals do not depend on how many candidate contacts are retained when both banks remain present. Fixed-index routing stores only the retained synaptic weights, giving 409,946 trainable parameters at 40E/10I. The fixed-index routing and learned-contact morphology cohorts consequently match different resources, despite sharing compartment and active-contact counts.

The spatial-routing comparison in Fig.~\ref{fig:fig2}G,H stores fixed integer source indices for every branch.  Each neuron receives the complete $3\times32\times32$ CIFAR-10 image.  In the structured-local condition, successive branch factors place child supports on nested two-dimensional grids; terminal supports are expanded when necessary to a minimum $4\times4$ pixel window, allowing windows from different paths to overlap.  The random-local condition preserves the resulting window-size distribution but relocates each branch window independently, thereby retaining locality without alignment to tree ancestry.  Each branch draws unique source indices within its own window, independently across somas and without shared weights; E and I contacts are sampled independently within the same spatial geometry.  The global control draws the same numbers of fixed indices from the complete flattened image.  All three routes use 64 somas, $N_E=40$, $N_I=10$, 120 nonsomatic compartments per soma, 384,000 active contacts and 409,946 trainable parameters, with twenty paired seeds and otherwise matched optimization for each integration rule and morphology (Fig.~\ref{fig:fig2}A).

Outputs from child branches are combined through nonnegative weights $c_k=\operatorname{softplus}(\theta_k)$, one per child.  Branch $b$ receives inherited input and activity-independent conductance,
\[
C_b=\sum_{k\in\mathrm{ch}(b)}c_{b,k}\,a_k,\qquad
G_b=\sum_{k\in\mathrm{ch}(b)}c_{b,k},
\]
where $a_k$ is the post-activation output of child $k$; $C_{b,d}$ is the same input indexed by branch and depth. In trees without direct somatic contacts, the soma is the root compartment of this recursion, with pre-activation state $C/(1+G)$ under shunting and $C$ under raw addition, followed by the same learned activation. Contextual trees additionally receive somatic inhibition and retain its contribution to the corresponding integration rule.

\subsection{Additive and shunting integration}

The shunting pre-activation state is
\begin{equation}
u_{\rm sh}=\frac{E+C}{1+E+I+G}.
\label{eq:shunting}
\end{equation}
With unit excitatory reversal and zero inhibitory and leak reversals, the steady-state balance underlying the analogy is $0=E(1-u)-Iu-u+C-Gu$. Thus $C$ is a source term, and the net child-coupling current would be $C-Gu$; neither $C$ nor $E$ alone is a voltage-dependent net current. The implementation evaluates the resulting algebraic state rather than simulating these currents or membrane dynamics. For numerical safety it adds a constant $10^{-8}$ to the denominator; because the denominator is at least the leak conductance, this constant is negligible at the precision of every reported quantity, and we omit it from the equations.  Excitation contributes to both numerator and conductance, inhibition contributes to the denominator, and child branches contribute weighted input and coupling conductance.  The additive comparison uses
\begin{equation}
u_{\rm add}=E-I+C.
\label{eq:additive}
\end{equation}
These equations define complete local integration rules; the comparison is not an isolated manipulation of the inhibitory term.  The shunting rule can be interpreted using fixed normalized reversal levels of one for excitation and zero for inhibition and leak.  These implicit reference levels do not model heterogeneity within synapse classes or dynamically changing reversal potentials.

Both rules are followed by a learned compartment-specific activation function,
\begin{equation}
a=\frac{\tanh[m(u-b)]+1}{2},
\label{eq:reactivation}
\end{equation}
where $m=\exp(\theta_m)>0$ is the learned slope and $b$ is the learned midpoint.  We write the post-activation output of branch $b$ at depth $d$ as $a_{b,d}$.  For descriptive operating-state analyses, an output is active when $0.05\leq a\leq0.95$, near its lower saturation limit when $a<0.05$, and near its upper saturation limit when $a>0.95$.

\subsection{Optimization}
\label{sec:optimization}

Gradients are computed by end-to-end backpropagation, using Adam~\citep{KingmaBa2015} with default moment coefficients, zero optimizer weight decay and elementwise gradient clipping at magnitude 5. No dendrite-specific gradient scaling is applied. The MNIST models in Fig.~\ref{fig:fig1}, the CIFAR-10 cohorts of Fig.~\ref{fig:fig2} and Supplementary Figs.~\ref{fig:figS13_allocation}--\ref{fig:figS38_low_capacity_morphology} and \ref{fig:figS22_input_regime_routing}, and the single-layer tree and point models in Fig.~\ref{fig:point_neurons} use the common training protocol adopted for consistency across the linked single-layer comparisons: learning rate 0.002 for all parameter groups, active-weight maintenance coefficient 0.03 and no inactive-candidate boosting. In the allocation, readout and initialization-control cohorts, this protocol replaced, in every allocation, the protocol selected on validation for that rule and allocation; the decision was made after the complete validation screen and after the per-cell test results had been inspected, and before the newly trained replacements were evaluated; per-cell runs already trained under this protocol were reused (Supplementary Note~\ref{note:recipe_policy}). These are the configuration fields \texttt{optimizer.lr}, \texttt{common.param\_groups}, \texttt{common.synapse\_decay\_rate} and \texttt{common.weight\_boosting} within \texttt{training.main}. Batch size is 256. MNIST trains for at most 300 epochs with patience 75. The primary, matched-morphology, allocation and resource-matched routing CIFAR-10 cohorts train for at most 2,000 epochs with patience 150; the matched-contact comparison uses the same horizon with patience 400. Tree and point arms share these settings within each comparison. The multilayer and contextual point arms instead retain their respective task protocols, described below.

Figures~\ref{fig:fig3} and \ref{fig:fig4} and Supplementary Figs.~\ref{fig:figS40_whole_layer_ablation}, \ref{fig:figS41_information_estimator_sensitivity}, \ref{fig:figS_morphology_noise} and \ref{fig:figS_gain_load_morphology} use twenty seeds per rule of the primary or matched-morphology cohorts, trained with the common protocol above: the ten original seeds and ten added after their results had been inspected (Supplementary Table~\ref{tab:seed_extension}). Supplementary Figs.~\ref{fig:figS_tree_factorial_v5} and \ref{fig:figS34_tree_scalar_information} retain ten seeds because the initialization-policy controls were not extended. The linear-readout comparisons (Supplementary Figs.~\ref{fig:figS_readout_mechanisms} and \ref{fig:figS_readout_robustness}) retrain the primary architecture with a linear readout under the common protocol on seeds 0--9 and pair it with the MLP-trained networks of the same seeds.

The single-layer allocation comparisons use the common protocol for both rules; the protocol selected on validation for each rule and allocation, from five tested protocols, is reported separately (Supplementary Note~\ref{note:recipe_policy}).  The multilayer comparisons below instead use training settings selected by validation within each integration rule; comparisons under matched training settings are analyzed separately in the Supplementary Information.

The multilayer networks use batch size 512, clipping at 5, active-weight maintenance coefficient 0.01, inactive-candidate boosting, and zero Adam weight decay.  Shunting uses learning rate $10^{-3}$; validation selected $10^{-4}$ for additive.  A shared-$10^{-3}$ comparison is reported separately.  For conductance-normalized additive, an independent three-seed screen over $10^{-4}$, $3\times10^{-4}$, and $10^{-3}$ selected $3\times10^{-4}$, $10^{-4}$, and $10^{-3}$ at $L=1,3,5$ by mean validation categorical log likelihood; ten new seeds per depth form the confirmation shown in Fig.~\ref{fig:fig5}B.  All multilayer models are trained for at most 5,000 epochs with patience 600 and reload the state with minimum validation loss.  Gradient scaling is disabled for synapses, child coupling, and activation-function parameters. The linear-readout endpoint control in \suppfig{figS_readout_depth} uses the same implementation and maintenance settings, retaining each rule's selected learning rate and pairing seeds 10--19 across readouts.

For standard top-$k$ support maintenance, each forward pass selects the largest effective positive weights per compartment; only retained weights contribute to synaptic drive and receive gradients. Before the Adam update, active weights are multiplied by $1-\eta_{\rm lr}\lambda$ and, when boosting is enabled, inactive eligible weights by $1+\eta_{\rm lr}\lambda$, where $\eta_{\rm lr}$ is the learning rate and $\lambda$ the maintenance coefficient. Values are mapped back to the stored unconstrained parameterization before Adam applies its update. Selection is discrete; it is not differentiated. Candidate parameters and their optimizer states remain stored when their weights leave the selected set. Support is recomputed from the current weights on the next forward pass; fixed-index controls retain their source indices throughout training. Equal-score selection follows the library's top-$k$ operation, without an additional stable tie-breaking rule. This maintenance is distinct from Adam weight decay.

\subsection{Training-setting selection and computational precision}

Training settings were compared using absolute validation accuracy, not the shunting-minus-additive difference. For the allocation study, three screening seeds (61000--61002) tested the five complete protocols in Supplementary Table~\ref{tab:recipe_sensitivity} at every depth-four allocation. The highest mean validation accuracy selected a protocol separately for each integration rule and allocation; ties followed the fixed order A, B, C, D, P. The linear-readout grid reused these selections after a ten-allocation transfer check; both shallower morphologies were screened over their full grids. The common protocol A was subsequently adopted across these related experiments to make allocation and morphology the varying factors. It had the highest screened accuracy for both rules at 40E/10I, but not the highest grid average, and is not claimed to be optimal. The per-cell selected results remain a sensitivity analysis. The matched-contact control retains the same learning rates, maintenance and initialization, with patience 400 rather than 150. Point controls reuse their corresponding tree settings without a separate search.

The main MNIST, CIFAR-10 and contextual training uses float32 arithmetic, with TF32 disabled and four CPU threads per training process. Paired conditions use the same seed and GPU family where specified in the cohort records; exceptions and hardware-restricted checks for the allocation point controls are reported in Supplementary Note~\ref{note:allocation_points}. AlexNet uses mixed precision as specified below. Checkpoint selection and hyperparameter selection use validation data; the canonical test sets had nevertheless informed earlier model development. In addition, the common-protocol amendment followed inspection of the per-cell test results, four cohorts were extended from ten to twenty seeds after inspection of their first ten results, and the AlexNet regularization follow-up and fixed-contact 100-epoch continuation followed inspection of earlier official-validation results. These comparisons therefore describe the tested procedures on established benchmarks, rather than an untouched, prospectively selected benchmark.

\subsection{Initialization policies}

Synaptic scales and bounded-activation operating points are initialized without class labels or validation or test outcomes. For a pathway with $M$ eligible sources and $K$ retained contacts, the analytical initializer draws raw weights $\theta\sim\mathcal N(\mu,s^2)$, with excitatory $\mu_E=0$ and $s_E=\sqrt{2/(M_E+1)}$. It approximates the mean retained weight by the upper-tail expectation $\bar w=\mathbb E[f(\theta)\mid\theta>t]$, where $\Pr(\theta>t)=K/M$ and $f$ is the positive weight transform. For the exponential transform this is the truncated log-normal expectation; for softplus it is a second-order expansion about the truncated-normal mean, $f(\mu_t)+f''(\mu_t)v_t/2$, where $\mu_t$ and $v_t$ are that mean and variance. Using a reference source activity of $1/2$, expected drives are $E_0=K_E\bar w_E/2$ and $I_0=K_I\bar w_I/2$. When both pathways are present, $s_I=\sqrt{2/(M_I+1)}\min(1,K_I/K_E)$ and the inhibitory raw mean is shifted from zero in steps of 0.01 until $K_I\bar w_I$ first crosses $K_E\bar w_E$. Thus approximate equality of expected E and I drives is imposed at initialization. A terminal compartment with excitation alone instead shifts its raw mean downward, if needed, until $\bar w_E\leq2/K_E$. The native neuronal configurations disable additional morphology-dependent rescaling of these synaptic distributions.

For shunting, write $D_0=E_0+I_0+g_L$ and $r=E_0/D_0$. A terminal compartment starts with activation midpoint $b_0=r$. With $n_c$ children whose reference output is $1/2$, the midpoint and total child coupling are
\begin{equation}
 b_0=\begin{cases}\max(0.49,r/2+1/4),&r<1/2,\\r/2+1/4,&r\geq1/2,\end{cases}
 \qquad
 G_0=\begin{cases}(D_0b_0-E_0)/(1/2-b_0),&r<1/2,\\D_0,&r\geq1/2.\end{cases}
 \label{eq:initial_coupling}
\end{equation}
Each child coupling starts at $\max(G_0/n_c,10^{-6})$, and the shunting activation slope starts at $m=1.5$. Additive child couplings instead start at one, with slope $m=0.1$ and midpoint $E_0-I_0$ for nonterminal compartments or zero for terminal compartments. Equation~\ref{eq:initial_coupling} therefore gives rule-dependent initial couplings even when the same analytical procedure is requested.

The occupancy policy starts from this analytical state and calibrates the activation using only optimization images. For each compartment stage, it pools activation inputs $u$ over three batches of 256 images and obtains their 10th and 90th percentiles, $q_{0.1}$ and $q_{0.9}$. With $z_r=\operatorname{arctanh}(2r-1)$, the fitted parameters are
\[
m=\frac{z_{0.9}-z_{0.1}}{q_{0.9}-q_{0.1}},\qquad
b=q_{0.1}-\frac{z_{0.1}}{m}.
\]
This maps the two input quantiles to output levels 0.1 and 0.9. Fits with nonfinite parameters, quantile width below $10^{-3}$ or slope above 50 retain the preceding valid activation parameters. Calibration reuses the same image batches while updating the stages, because changing an upstream activation changes downstream inputs. The archived training implementation repeats this procedure to parameter convergence (maximum absolute change below $10^{-3}$), with a cap of 50 iterations. The cohort summary appears in Supplementary Table~\ref{tab:cohort_manifest}; the separate experiment manifest records configuration and stage-specific source identities. Activation slopes and midpoints are subsequently optimized by backpropagation.

MNIST and the primary single-layer CIFAR-10 comparisons use analytical initialization for shunting and occupancy initialization for raw addition. Multilayer E/I networks and the primary contextual networks use occupancy calibration under both rules; the contextual variants are specified in their task Methods. Shared-policy controls apply each policy to both rules and evaluate sensitivity to this training choice (Supplementary Figs.~\ref{fig:figS13_allocation} and~\ref{fig:figS_tree_factorial_v5}). A shared policy does not imply identical initial synaptic weights, activation parameters or activity distributions.

\subsection{MNIST identity readout and synaptic maps}
\label{sec:mnist_readout}

Figure~\ref{fig:fig1} uses ten primary networks per integration rule, with seeds 100--109. Soma $c$ maps directly to class logit
\[
g_c=Ta_c,\qquad T>0,
\]
where $a_c$ is the soma output and the learned scalar $T$ is shared by all classes.  The scale changes loss calibration but cannot change the predicted class.  A fixed seed-100 $[2,2]$ network supplies the illustrative response and branch panels.  All six synaptic-map columns in G use a seed-100 $[3,3,3,3]$ shunting network from an existing MNIST identity-readout cohort.  The branch in I belongs to the separate primary $[2,2]$ network; the shared soma index denotes the assigned digit, not the same trained neuron across the two models.  Cohort-level conclusions in Fig.~\ref{fig:fig1}H and Supplementary Fig.~\ref{fig:figS_mnist_components_source_activity} use the ten independently trained $[2,2]$ networks per rule.

Both branch views in Fig.~\ref{fig:fig1}I use soma 6, its first proximal branch (global branch index 12, depth 1), and digits 6 and 0 from the fixed seed-100 checkpoint.  For a view with appreciable excitation and inhibition, we compared all twenty proximal branches using their existing, class-balanced sample of 2,000 images.  For each branch, joint E/I coverage is the fraction of images on which both drives exceed 1\% of their respective pooled 95th percentiles.  We selected the branch with the largest coverage, without using class separation or classifier performance.  Digit 0 supplies a fixed comparison to the selected neuron's assigned digit.  The same 200 originally sampled images per digit appear in both views; no image or recorded value was changed.  The left horizontal variable is the recorded depolarizing input $E+C$, not the normalized input $u$ to the activation function.  This selection favors a visible joint-input example and is not an estimate of typical branch behavior or an inference across trained networks.

In the primary example, each soma has morphology $[2,2]$, corresponding to two proximal and four distal compartments.  Each compartment retains 25 direct-image E contacts and 25 direct-image I contacts.  The denser example in Fig.~\ref{fig:fig1}G has morphology $[3,3,3,3]$, with 120 nonsomatic compartments and 25 E plus 10 I contacts per compartment.  Both examples use ten class-aligned neurons, direct pixel inputs to both pathways, an identity readout, and the same training schedule.  The denser example provides an illustration of spatial input coverage under a different contact budget, rather than a resource-matched morphology comparison.  Soma identity fixes the assigned digit.  The independent optimization-split response preference is used only to verify this identity assignment and agrees with it in every analyzed soma.

For soma $s$ and pathway $P\in\{E,I\}$, the assigned-minus-other mean synaptic drive at pixel $p$ is
\begin{equation}
\Delta d_s^P(p)=\bar w_s^P(p)\left(\mathbb E[x_p\mid Y=s]-\mathbb E[x_p\mid Y\ne s]\right),
\label{eq:retained_weight_map}
\end{equation}
where $Y$ is the digit label and $\bar w_s^P(p)$ is the mean selected weight at that pixel across synapse-bearing branches. Positive values indicate more drive for the assigned digit and negative values more drive for other digits. Pixels inactive across all images contribute zero. These maps describe local synaptic input; its influence on the soma additionally depends on dendritic integration.

For the synaptic-drive contrasts in Figure~\ref{fig:fig1}G, effective retained weights are averaged over all 120 nonsomatic branches of the illustrated $[3,3,3,3]$ model, counting absent contacts as zero. The assigned-class mean pixel intensity is compared with the mean over all other canonical test images, pooled in proportion to their class counts. Multiplying this intensity difference by the mean weight gives Eq.~\ref{eq:retained_weight_map}. Direct averaging of the branchwise drive contrasts gives the same result. One signed logarithmic color scale spans all displayed E and I maps, with a linear region within $\pm0.001$; no smoothing, thresholding, clipping or per-map normalization is applied. Separate gray reference images show assigned-class means in $[0,1]$. The original retained-weight maps remain in the source archive. Checkpoints and displayed somas retain their pre-existing selection.

The source-activity statistic uses independent optimization and validation splits.  Soma $s$ has assigned class $c_s=s$ under the identity readout.  As a validation of this assignment rather than an alternative definition, the class producing the largest mean soma response is selected on optimization data; it equals $c_s$ for every soma in all 20 analyzed checkpoints.  Let $\mathcal B_{s,d}$ be the branches of soma $s$ at dendritic depth $d$, and $\mathcal S_b^P$ the 25 retained source pixels on branch $b$ in pathway $P\in\{E,I\}$. We compute
\[
q_{s,d}^{P,+}=\frac{1}{|\mathcal B_{s,d}|}\sum_{b\in\mathcal B_{s,d}}\frac{1}{25}\sum_{p\in\mathcal S_b^P}\mathbb E[\mathbf 1(x_p>0.1)\mid Y=c_s],
\]
with $q_{s,d}^{P,-}$ defined analogously after pooling $Y\ne c_s$.  Figure~\ref{fig:fig1}H reports $25q_{s,d}^{P,+}$ and $25q_{s,d}^{P,-}$, the expected number of ON-pixel contacts per branch.  Supplementary Fig.~\ref{fig:figS_mnist_components_source_activity} reports their within-checkpoint contrast $25(q_{s,d}^{P,+}-q_{s,d}^{P,-})$.  Expectations are evaluated on validation data; somas and branches are averaged within seed before inference across trained seeds.
The threshold includes antialiased stroke pixels while excluding near-zero intensity.  Fixed-checkpoint sensitivity checks at thresholds 0.05, 0.25, and 0.5 preserve the E/I and depth ordering.

\paragraph{Linear MNIST reference.}
An unconstrained linear softmax classifier uses the seed-100 optimization/validation split (48,000/12,000 images) and the same pixel preprocessing as the primary MNIST example. L-BFGS fits each of three inverse regularization strengths, $C\in\{0.01,0.1,1\}$; validation cross-entropy selects $C=0.1$. All fits meet the optimizer's convergence criterion before final evaluation on the 10,000 test images. Its 7,850 weights and biases are a contextual reference, not a contact- or parameter-matched control.

\subsection{Contact allocation and morphology experiments}

The allocation experiment crosses $N_E,N_I\in\{1,5,10,15,20,25,30,35,40,45,50\}$, giving 121 conditions per integration rule.  Figure~\ref{fig:fig2}B displays the 100 conditions with 5--50 contacts in each pathway, showing mean test accuracies on the left and middle and the paired difference between rules on the right. Single-contact conditions remain in the full-grid summaries and supplementary allocation analyses.  Curves connect observed means and interval endpoints without interpolation or smoothing.

The fixed-budget slice sets $N_E+N_I=50$.  The equal-ratio slice sets $N_E=N_I$ and defines total active contacts per compartment as $B=N_E+N_I$.  The inhibitory-tail contrast is calculated within seed before averaging:
\[
\Delta_I=\frac{A_{N_I=45}+A_{N_I=50}}{2}
-\frac{A_{N_I=1}+A_{N_I=5}}{2}.
\]
Figure~\ref{fig:fig2}C forms the shunting-minus-additive accuracy difference within matched morphology and seed; Figure~\ref{fig:fig2}E instead forms the linear-minus-MLP readout effect within each integration rule and seed.  Supplementary Fig.~\ref{fig:figS_allocation_slices}D first computes $\Delta_I$ within seed and then forms $\Delta_I^{\rm shunt}-\Delta_I^{\rm add}$ before bootstrapping seeds.  The two lowest and two highest sampled inhibitory allocations were prespecified to average adjacent conditions rather than use a single endpoint; this contrast is descriptive, not a derivative with respect to $N_I$.  In every morphology, the interaction is positive at $N_E=1$ and negative from $N_E=5$, except that in $[3,3]$ it approaches zero at the excitation-dominated end.  The shallower grids contain thirteen paired seeds per condition.  The depth-four grid and its accuracy slices use twenty, while its prespecified inhibitory-allocation contrast uses ten.

The contact-bottleneck experiment compares a point model with $[120]$ and $[3,3,3,3]$ trees over 8--64 somas.  At $N_E=N_I=5$ per tree compartment, each tree has 600 E and 600 I contacts per neuron; the point model receives the same total numbers.  Candidate parameter counts are not matched.  The fixed-index resource-matched experiment instead compares $[120]$, $[10,11]$, $[8,2,6]$, and $[3,3,3,3]$ at 64 somas and $N_E=40,N_I=10$.  Every morphology has 120 nonsomatic compartments per soma, 307,200 active E contacts, 76,800 active I contacts, 384,000 stored and trainable synaptic weights, 25,946 nonsynaptic parameters, and 409,946 total trainable parameters.

The direct-inhibition control uses the same 64 somas, $[3,3,3,3]$ morphology, optimizer, and ten paired seeds while comparing $40$E:$10$I with $50$E:$0$I.  Fixed source indices are sampled once before training and remain unchanged.  Holding the total at 50 contacts per compartment makes both arms contain exactly 384,000 stored and trainable synaptic weights and 409,946 total trainable parameters.  The E-only arm contains no direct I weights or I drive; it does not introduce an inhibitory population.

The explicit-I routing control contains 64 E neurons with morphology $[3,3,3,3]$ and 50 I neurons with morphology $[1]$.  The image drives both populations, the I output is the only inhibitory input to E, and only E enters the decoder.  The control tests a different inhibitory route and therefore does not match the direct-input model in source dimension or total parameters.

\subsection{Integration controls and readout comparison}

The conductance-normalized additive control uses $u_{\rm norm}=(E-I+C)/(1+E+I+G)$ and retains learned bounded activations. Its selected initialization and training schedule are specified with the corresponding cohort. Readout comparisons hold settings fixed within each integration rule; they compare separately optimized complete classifiers, rather than decoders of one fixed representation. The primary information, replacement and perturbation readout analyses use seeds 0--9 under both readouts; information is compared between integration rules within each readout, and replacement and retention contrasts pair the readouts within seed. These cohorts are identified in the experiment manifest.

\subsection{Branch-resolved information and dendritic-stage interventions}

Figure~\ref{fig:fig3}B--F use twenty paired shunting and additive networks, with MLP readouts and seeds 0--19 (seeds 10--19 added after inspection), at 64 somas, morphology $[3,3,3,3]$, $N_E=40$, and $N_I=10$.  At each depth, one branch is selected from every soma without using labels.  Mutual information measures statistical dependence~\citep{CoverThomas2006} and is estimated separately for each scalar branch pair on all 10,000 validation examples and averaged across the 64 branches only after estimation.

The primary estimator uses ten empirical-quantile bins and Miller--Madow sampling-bias correction~\citep{Miller1955,Paninski2003}.  Within-class information is computed separately within each class and averaged according to class frequency.  For $I(X;Z\mid Y)$, twenty independent within-class permutations of $Z$ preserve its class-conditional marginal while removing example-specific pairing.  Their mean is subtracted as an estimator-specific finite-sample floor:
\[
\widehat I_{\rm corr}(X;Z\mid Y)=
\widehat I_{\rm MM}(X;Z\mid Y)
-\frac{1}{20}\sum_{q=1}^{20}
\widehat I_{\rm MM}(X;\pi_{q,Y}Z\mid Y).
\]
The primary within-class associations are $I(E;I\mid Y)$, $I(C;a\mid Y)$, and $I(a;S\mid Y)$. Observed and shuffled estimates are shown separately in the shared-initialization control. A common-data sensitivity check compares the binned estimator with corrected nearest-neighbor estimates~\citep{Kraskov2004} at $k=10$ for these three quantities and the joint association $I((E,I);S\mid Y)$. Both estimators use identical 3,000-example, 16-soma supports, label-free standardization and five within-class permutations (\suppfig{figS41_information_estimator_sensitivity}).

The local derivatives clarify how each integration rule propagates changes. Let $N=E+C$ and $Q=1+E+I+G$, so $u_{\rm sh}=N/Q$.  Before the activation function, the shunting partial derivatives are
\[
\frac{\partial u}{\partial E}=\frac{Q-N}{Q^2},\quad
\frac{\partial u}{\partial I}=\frac{\partial u}{\partial G}=-\frac{N}{Q^2},\quad
\frac{\partial u}{\partial C}=\frac{1}{Q},
\]
whereas the raw-additive derivatives are $1,-1,1,0$ for $E,I,C,G$, respectively.  The corresponding output derivative is $\partial a/\partial X=\phi'(u)\,\partial u/\partial X$ for $X\in\{E,I,C,G\}$. These local partial derivatives hold the other aggregate inputs fixed. Transmission along an actual branch-to-soma path additionally includes coupling and activation at each successive compartment, as defined in Eq.~\ref{eq:path_transmission}.

An individual coupling $c_k$ changes both $C$ and $G$.  Holding the child output fixed, its shunting derivative is therefore
\[
\frac{\partial a}{\partial c_k}
=\phi'(u)\frac{a_k-u}{Q}.
\]
Here $a_k$ is the child's post-activation output and $u$ the parent's pre-activation integration state; their difference is an effective driving-force term, not a difference between two simulated membrane-voltage trajectories.

To express the corresponding task gradients, define $\delta=(\partial\mathcal L/\partial a)\phi'(u)$. Holding the selected synaptic support fixed, the gradients with respect to effective weights and child outputs are
\begin{equation}
\begin{aligned}
\frac{\partial\mathcal L}{\partial w_p^E}
&=\delta x_p\frac{1-u}{Q},
&\frac{\partial\mathcal L}{\partial w_p^I}
&=-\delta x_p\frac{u}{Q},\\
\frac{\partial\mathcal L}{\partial c_k}
&=\delta\frac{a_k-u}{Q},
&\frac{\partial\mathcal L}{\partial a_k}
&=\delta\frac{c_k}{Q}.
\end{aligned}
\label{eq:shunting_task_gradients}
\end{equation}
For raw addition the corresponding expressions are $\delta_{\rm add}x_p$, $-\delta_{\rm add}x_p$, $\delta_{\rm add}a_k$ and $\delta_{\rm add}c_k$, where $\delta_{\rm add}$ is evaluated in the additive network. Inhibition affects this factor through the activation and downstream task error, even though the local integration is additive. The positive weight parameterization contributes its own chain-rule factor when differentiating with respect to unconstrained parameters: for $w=\operatorname{softplus}(\theta)$, this is the sigmoid $\sigma(\theta)=(1+e^{-\theta})^{-1}$. These derivatives describe backpropagation through the selected connections; topology maintenance and optimizer updates are separate operations. For learned leak, replace the unit leak in $Q$ by $g_L$, giving $\partial\mathcal L/\partial g_L=-\delta u/Q$. At terminal compartments, $C=G=0$ and the recorded aggregate partials remain well defined, but no child-coupling parameter or child transmission pathway is present.

For complete-stage mean replacement, every E or every I drive at one depth is replaced by its own mean over 40,000 optimization examples; accuracy is evaluated on all 10,000 test examples.  For 64 $[3,3,3,3]$ trees, this changes 192, 576, 1,728, or 5,184 branches at $d=1,2,3,4$.  The main output-replacement comparison uses two doses: one branch per soma and one third of all branches at each depth. For each of three draws, branches within each soma and depth are permuted using seed $20260921+100d+\mathrm{draw}$, with draw indices 0--2. The first branch supplies the fixed-number support; prefixes of length $3^{d-1}$, $2\cdot3^{d-1}$ and $3^d$ supply fractions one third, two thirds and one. Supports match across paired integration rules and are nested across doses within a stage; they do not impose a nested anatomical path across stages. Each selected output is replaced by its mean over all 40,000 optimization images, and accuracy is evaluated on all 10,000 test images. Three draws are averaged within trained seed. The earlier readout controls retain their original nested-path construction: one to three complete paths per soma, smaller doses and shallower supports as prefixes, and three draws from seed 271828. The shared-initialization control instead selects one complete path per soma for each of five draws from the same base seed. These analyses are distinguished by their branch selections and number of draws. Structural lesions, zero replacement and joint E/I clamps provide supplementary controls.

\paragraph{Transmission along a dendritic path.}
The chain rule gives
\[
\mathcal T_{b\to s}(x)=\prod_{(k\to j)\in\mathrm{path}(b,s)}\frac{\partial a_j}{\partial a_k}.
\]
For an edge with coupling weight $c$, the local derivative is $c\phi'(u)$ under addition and $c\phi'(u)/Q$ under shunting, with $Q=1+E+I+G$. Transmission therefore depends on learned couplings and activations as well as the integration rule; it is not a passive cable property. For addition, multiplying synaptic and coupling weights and the activation midpoint by $\lambda>0$, while dividing the slope by $\lambda$, preserves branch outputs and these edge derivatives. The transmission measure is consequently unchanged by an equivalent additive reparameterization, unlike derivatives with respect to aggregate input drives.

We computed Eq.~\ref{eq:path_transmission} by automatic differentiation on the same twenty primary four-level checkpoints per rule used for the branch interventions. A fixed random permutation (seed 20260924) selected 2,048 test images without using labels; the support was shared across rules and trained seeds. The derivative of the sum of soma outputs with respect to a branch output equals its derivative to the owning soma because the single-layer somas have disjoint trees. We verified the result against the product of local edge derivatives. As a numerical check, an equivalent twofold additive weight/midpoint rescaling preserved logits and transmission to double-precision accuracy. All branch and image measurements were reduced within trained seed before 10,000 bootstrap draws. The near-zero fraction uses a fixed descriptive threshold of $10^{-6}$ in the bounded output coordinates.

\subsection{Input perturbations}

Gaussian perturbations add independent $\mathcal N(0,\sigma^2)$ pixel noise and uniform perturbations add independent $\mathcal U[-a,a]$ noise, with $\sigma$ or $a\in\{0.02,0.05,0.1,0.2,0.3,0.5\}$.  The resulting pixels are clipped to $[0,1]$.  Five shared random draws are evaluated at every magnitude and averaged within trained seed.  Global input scaling is deterministic: $\mathbf{x}_g=g\mathbf{x}$ for 17 logarithmically spaced values $g\in[0.1,10]$, applied without clipping.  It is an out-of-distribution scale-transfer assay, not stochastic gain noise or an independently controlled inhibitory/conductance load.  No model is retrained for a perturbation.  Figure~\ref{fig:fig4}B,C uses the frozen Fig.~\ref{fig:fig3} cohort: 64 somas, $[3,3,3,3]$ morphology, $N_E=40$, $N_I=10$, and twenty paired seeds under the common single-layer protocol.

The factorial comparison of shared gain and independent pixel noise uses the independently trained four-morphology cohort; its paired differences and rule-specific surfaces appear in Supplementary Figs.~\ref{fig:figS_corruption_protocols}A and \ref{fig:figS_gain_load_morphology}.  At each condition, one scalar $g_n$ is drawn per image and shared across all its pixels, with $\log g_n\sim\mathcal N(-\sigma_g^2/2,\sigma_g^2)$ for $\sigma_g\in\{0,0.2,0.5,0.8\}$.  Independent pixel noise is $\eta_{np}\sim\mathcal N(0,\sigma_{\rm pix}^2)$ for $\sigma_{\rm pix}\in\{0,0.1,0.3,0.5\}$, and the analyzed input is $\max(0,g_nx_{np}+\eta_{np})$: negative values are set to zero and there is no upper bound.  Five condition-stable common-random-number draws are shared across paired checkpoints and averaged within seed.  The twenty independently trained seeds per rule and morphology, not the perturbation draws, are the inferential units.  Chance-corrected retention is computed relative to each checkpoint's own unperturbed accuracy.

The brightness-dependent corruption of Figs.~\ref{fig:fig4}E and \ref{fig:point_neurons}D applies the noise law used for contextual training (Eq.~\ref{eq:context_noise}) to frozen networks trained without it.  For each image, a gain $g$ and a noise divisor $X$ are drawn with $\log g\sim\mathcal U(-\log b,\log b)$ and $\log X\sim\mathcal U(0,\log 4)$, and the analyzed input is $\max\bigl(0,gx_p+\lambda\sqrt{\overline{g\mathbf x}}\,z_p/X\bigr)$, where $\overline{g\mathbf x}$ is the mean scaled intensity over all pixels and color channels of the image and the $z_p$ are independent standard normal variables.  The gain bound $b\in\{1,2,5,20\}$ sets the range of the random gain, not a fixed gain factor, and the noise multiplier $\lambda\in\{0,0.25,0.5,1\}$ scales the noise; $b=1$ and $\lambda=0$ leave the image unchanged.  Inputs are not clipped above or renormalized.  Gains, divisors and Gaussian draws are generated in single precision on the CPU from fixed seeds and are shared across models, training seeds and conditions; five draws are averaged within each trained network.

Chance-corrected retention is
\[
R=\frac{A_{\rm perturbed}-0.1}{A_{\rm clean}-0.1}.
\]
Contrasts between integration rules are paired within seed.

The resource-matched morphology transfer uses the same perturbation grids on 160 frozen CIFAR-10 checkpoints: four shapes, two integration rules, and twenty paired training seeds (21000--21019, of which 21010--21019 were added after inspection).  Every model has 64 somas, 120 nonsomatic compartments per soma, and 40 E plus 10 I contacts per compartment.  Thus each shape contains 384,000 active feedforward contacts and the same trainable parameter count.  Noise draws are averaged within seed; seed-bootstrap intervals and contrasts between integration rules or morphologies are then computed across independently trained seeds.

\paragraph{Decoding and class information under corruption.}
Figure~\ref{fig:fig4}F and \suppfig{figS_corruption_protocols}C use the 64-dimensional soma outputs of the four-level resource-matched trees, with twenty trained seeds per rule. Ridge classifiers with penalty one are fitted on all 10,000 validation images, either clean or corrupted under the evaluation condition, and scored on the separate 10,000 test images. Coordinates are standardized using the probe-fit data. We subtract the mean accuracy of five probes fitted on shuffled validation labels. These representation analyses use one corruption draw shared across models, with independent validation and test realizations, rather than the five-draw average used for network accuracy. For class information, one label-independent subset of 3,000 test images per trained seed is shared across conditions. The full soma vector is standardized without labels and analyzed with the continuous--discrete nearest-neighbor estimator of Ross~\citep{Ross2014} with $k=5$; the mean of twenty shuffled-label estimates is subtracted. Information retention is the corrupted estimate divided by that network's clean estimate, expressed as a percentage. It is distinct from accuracy retention and remains subject to finite-sample estimation error.

\subsection{Point neurons and input-pooling controls}
\label{sec:point_methods}

A point arm keeps a tree configuration and changes only its neurons: every population becomes soma-only, with all synapses on the soma, and each pathway keeps the tree's active contacts per neuron, capped at the pathway's source width because a soma contacts each source at most once. Contact targets are read from the library's synapse accounting of the built tree and verified on the built point model. The caps apply to first-layer CIFAR-10 E pathways (4,800 or 6,000 contacts become 3,072), to the E/I networks (local I pathways of 390 contacts become 16, and later-layer feedforward E pathways of 1,560 contacts become 64) and to the second layer of the contextual network (360 E contacts become 192). MNIST pathways are not capped. Data, readout, optimizer, learning rate, initialization policy, schedule and seeds are those of the tree arm with the same integration rule; point arms were not screened separately. In the resource-matched morphology cohort, one point model per rule and seed stands beside all four morphologies, and in the selected pairing the shunting point uses the analytical and the additive point the occupancy protocol. In the E/I networks, the shunting points enter both learning-rate comparisons, and additive points were trained at $10^{-4}$ and $10^{-3}$. The contextual point arms reuse the protocols of the shunting tree, the tuned additive tree, the additive tree trained with the shunting protocol, and the two rule-independent initializations. Active contacts are not trainable parameters: a top-$k$ compartment trains a candidate weight for every input of its pathways, so a $[120]$ tree trains 737,280 candidate synaptic weights per neuron and a CIFAR-10 point 6,144.

The breadth control places the 4,800 E and 1,200 I active contacts per neuron of the $[120]$ trees on one-level trees with 20, 6 or 2 compartments (240E/60I, 800E/200I or 2,400E/600I per compartment), with the cohort's selected protocols (analytical initialization for shunting, occupancy for raw addition) and seeds 21000--21009. Candidate banks per neuron range from 12,288 weights with two compartments to 737,280 with 120. The matched-contact pairs use the 64-neuron trees of the contact-bottleneck experiment (seeds 52000--52009) with 5E/5I or 25E/25I per compartment and points with exactly their 600E/600I or 3,000E/3,000I active contacts, which need no cap.

All of these models are evaluated under the brightness-dependent corruption with the grid and common draws used for the trees, and paired differences are computed within seed before seed bootstrapping. Before corruption, each evaluation replays the model's clean test accuracy on the evaluating GPU; replays may differ from the recorded score by at most one test image for trees and five for points (observed differences were at most two). Tree and point arms of one seed were in some cases trained on different GPU models (A100, H100 or H200). In the breadth control, the four seed pairs trained on one GPU model gave similar contrasts to the others (two compartments minus 120 under shunting at noise multiplier 1: 14.2 against 13.7 points), and in the contextual task the shunting point was 18.7 points below shunting trees trained with the same protocol on its GPU model, against 18.6 overall.

The terminal relation in Eq.~\ref{eq:leaf_gain_sensitivity} does not describe an internal compartment with child input or the complete network's response to random gain. For the leak-fraction diagnostic, the leak fraction $g_L/(g_L+E+I+G)$ of every excitatory compartment of a shunting model is averaged over the first 1,000 CIFAR-10 test images at fixed gains 0.05, 0.2, 1, 5 and 20 without noise, and at unit gain with noise multiplier 1. Its median over the most distal compartment level (the soma for a point) is compared with the model's accuracy on the same images. The 130 models comprise the resource-matched trees and point and the breadth series (seeds 21000--21009) and the 64-neuron trees and points of the contact-bottleneck cohort (seeds 52000--52009). Spearman correlations across models are descriptive; models of one construction share seeds and protocols and are not independent replicates of an integration rule. Every model keeps the fixed unit leak; a point given the trees' total leak would be a separately defined control and was not trained.

\subsection{Multilayer representation and intervention analyses}

Two point references accompany the multilayer comparisons. The dotted reference in Fig.~\ref{fig:fig5}B assigns each soma the contacts of one tree compartment and is evaluated under the shared learning rate (Supplementary Note~\ref{note:multilayer_point_reference}). The second reference instead combines the tree's contacts at its soma, subject to source-width caps, and includes both the selected and shared learning-rate conditions described under Point neurons and input-pooling controls (\suppfig{figS_three_arms}C). They differ in contact and candidate-parameter budgets and are not interchangeable.

Each multilayer network contains 64 $[3,3,3]$ E neurons and 16 $[1]$ I neurons per network position.  Nonsomatic E compartments retain 40 feedforward-E and 10 local-I contacts; nonsomatic I compartments retain 40 feedforward-E contacts and no inhibitory contacts.  Only the final E output enters the decoder.  Networks with total depth $L=1,\ldots,5$ are trained with ten paired seeds per integration rule.

For Figure~\ref{fig:fig5}C, we extracted logged training and validation cross-entropy at the minimum-validation-loss epoch for each selected-protocol checkpoint. Training loss averages minibatches while parameters change during that epoch; validation loss is evaluated afterward. Thus the two curves diagnose fitting across depth, rather than estimating a fixed-checkpoint generalization gap. Ten trained seeds at each depth supply the intervals.

Population decoding is measured after training using auxiliary decoders, often called probes~\citep{AlainBengio2016}.  Decoding accuracy above the shuffled-label reference is $\Delta A_{\rm dec}=A_{\rm dec}-\langle A_{\rm dec}^{\pi}\rangle_{\pi}$, where $A_{\rm dec}^{\pi}$ is test accuracy after fitting a probe on permuted validation labels. Figures express $100\Delta A_{\rm dec}$ in percentage points; we use accessibility only in this operational sense.  At position $\ell$, the complete E and I output vectors have dimensions 64 and 16.  A balanced signed CountSketch~\citep{Charikar2002} maps E to 16 coordinates; I remains in its native 16-dimensional space.  Transformed coordinates are standardized without labels.  Ridge probes with penalty $\alpha=1$ are fitted on validation representations and evaluated on test representations.  Twenty shuffled-label validation fits estimate the finite-sample floor.  The nonlinear probe is a $16\mathbin{\to}32\mathbin{\to}16\mathbin{\to}10$ ReLU MLP with $\ell_2$ coefficient $10^{-4}$, learning rate $10^{-3}$, at most 1,000 iterations, and internal early stopping.  Projection, initialization, and label-shuffle repeats are averaged within trained network.

For Fig.~\ref{fig:fig5}F, E and I outputs at every position were recorded under the brightness-dependent corruption (Input perturbations) with gain bounds 1 and 20 and noise multipliers 0, 0.25, 0.5 and 1, for corrupted validation images and separately corrupted test images.  The linear probes of panel E, with the same projections, penalty and twenty shuffled-label fits, were refitted on the corrupted validation responses of each condition and draw and scored on the corrupted test responses.  The networks and their readouts were not retrained.  Corruption and projection draws are averaged within trained network before seed-bootstrap intervals and paired differences are computed.

Population information is estimated on all 10,000 validation examples after the same label-independent 16-dimensional transform.  The Ross continuous--discrete estimator~\citep{Ross2014} uses $k=5$, and the mean of 20 label permutations is subtracted.  E and I populations are analyzed separately.

For branch-output interventions, five label-independent supports select one nested soma-to-distal path in every E tree. At each $(\ell,d)$, the 64 selected post-activation values are replaced by their mean responses over training images, and the resulting test-accuracy loss is measured. Zero replacement provides a supplementary control. The separate population-information analysis uses complete E-soma vectors and varies sample size, neighbor count and a fixed coordinate subset (Supplementary Note~\ref{note:population_estimator_audit}). These analyses assess classifier dependence on selected branches and statistical class information in soma populations, respectively.

Complete-I interventions act on the 16-dimensional output of one local I population.  Zero replacement removes that route.  For within-class donor replacement, each I vector is replaced by the vector from another test example of the same class, preserving the class-conditional marginal distribution while disrupting example-specific E--I coordination.  Five donor permutations are averaged within network.  Because donor construction uses test labels, it is a mechanistic diagnostic rather than a deployable manipulation.

\subsection{Contextual remapping with gain-dependent input noise}
\label{sec:contextual_remapping_methods}

The balanced split of the Extended MNIST dataset (EMNIST Balanced) supplies 47 character classes~\citep{CohenEtAl2017EMNIST}. Images are converted to upright orientation and normalized by their own maximum intensity before augmentation. A 32-dimensional one-hot cue with amplitude 8 encodes the shift $s\in\{0,\ldots,31\}$ in Eq.~\ref{eq:context_target} for every model. With independent standard normal pixel variables $z_p$, the corrupted image is
\begin{equation}
\widetilde x_p=\max\left(0,g x_p+\frac{\sqrt{\overline{g\mathbf x}}}{q}z_p\right),
\qquad
\overline{g\mathbf x}=\frac{1}{784}\sum_{p=1}^{784}g x_p.
\label{eq:context_noise}
\end{equation}
During training, gain $g$ and noise divisor $q$ are independently log-uniform over $[0.05,20]$ and $[1,4]$; we call $1/q$ the noise level. The same gain applies to every pixel within an image, whereas the Gaussian draws are independent across pixels. Values are clipped only below zero, with no upper clipping or subsequent normalization. Clean evaluation sets $g=1$ and removes the noise while retaining the sampled contexts.

The dendritic model has two E/I layers, each with 192 E neurons, 72 I neurons and $[3,2]$ morphology. Within each layer, E and I trees receive the feedforward excitatory source and I neurons inhibit E trees; first-layer I output also inhibits both second-layer populations. Each nonsomatic compartment retains 40 excitatory contacts. First-layer E compartments retain four inhibitory contacts per dendritic branch and six at the soma, and first-layer I neurons receive no inhibition. At each stage, including the soma, second-layer E neurons retain four inhibitory contacts jointly from the 72 first-layer and 72 local I neurons; second-layer I neurons retain four contacts from the first-layer I population. Synaptic weights are $w=\exp(\theta)$ and are selected by the standard sparse layer from a dense candidate bank, with active-weight maintenance coefficient 0.01 and no inactive-candidate boosting (Methods, Sparse synapses and Optimization). Compartments use the activation of Eq.~\ref{eq:reactivation}, and the leak conductance is fixed at one, so the state is that of Eq.~\ref{eq:shunting}. Synaptic scales and child couplings follow the analytical policy, which sets the couplings of shunting compartments from their expected E and I drives and sets additive couplings to one; the activation is then calibrated by the occupancy procedure on three batches of 256 corrupted training presentations (at most 50 iterations). Shunting and additive trees with the same seed therefore start from rule-dependent initial parameters. Sharing the optimization settings and initialization procedure does not isolate the integration rule from its resulting initial state. A linear readout maps the final E outputs to the 47 classes. The model retains 233,087 parameters and trains 2,885,951, the difference being the unselected candidate synapses.

The independently tuned tree comparison appears in Fig.~\ref{fig:fig7}B; the shared-protocol comparison and its repeat on a second GPU model appear in \suppfig{figS_context_task_controls}A. The references share this construction except where stated. The additive tree replaces Eq.~\ref{eq:shunting} by Eq.~\ref{eq:additive}; its separately tuned recipe uses learning rate 0.005 without maintenance, selected on validation accuracy of separate screening seeds, and its same-recipe version keeps the shunting tree's recipe. The MLP has widths $816\to235\to145\to47$, ReLU hidden activations and no dropout, and retains and trains 233,077 parameters. Two task-specific variants change initialization or leak: one draws zero-mean normal synaptic parameters and unit child couplings irrespective of integration rule, so that the two rules share their initial synaptic and coupling parameters for a given seed, although each keeps its own recipe and its occupancy-calibrated activation parameters; the other learns leak conductance from one. The endpoint-gate shunting variant in Fig.~\ref{fig:fig7}E,F, closer to the model that motivated the task, retains shunting integration but changes activation, initialization and training settings together. It replaces Eq.~\ref{eq:reactivation} with the endpoint-normalized tanh activation
\begin{equation}
\phi_{\rm end}(u)=\frac{\tanh(mu-b)-\tanh(-b)}{\tanh(m-b)-\tanh(-b)},
\qquad m>0,
\label{eq:context_activation}
\end{equation}
The normalization fixes $\phi_{\rm end}(0)=0$ and $\phi_{\rm end}(1)=1$ as $m$ and $b$ are learned separately in each compartment; the primary activation in Eq.~\ref{eq:reactivation} instead approaches zero and one asymptotically. Here $b$ is the tanh offset, distinct from the midpoint in Eq.~\ref{eq:reactivation}. The variant initializes $m=3$ and $b=1.5$ without calibration, draws log-normal synaptic conductances with mean equal to the reciprocal of the compartment's contact count and log standard deviation 0.5, and starts child couplings at 0.3. It uses learning rate 0.01 and inactive-candidate boosting. Because activation, initialization and training settings change together, this comparison cannot attribute differences in inhibitory context dependence to the activation alone.

For the primary tree comparison, three screening seeds (61000--61002) tested learning rates $0.001,0.002,0.005,0.01$ with analytical or rule-independent synaptic initialization, for 1,200 epochs per candidate. Within each rule and initialization, the learning rate was selected by mean corrupted validation accuracy, retaining maintenance 0.01 for shunting and zero for addition, with boosting disabled. The selected primary rates were 0.002 and 0.005, respectively. The MLP screen tested learning rates $0.0005,0.001,0.002$ and dropout 0 or 0.05; the retained MLP uses 0.001 and no dropout. The common-protocol additive control adopts the primary shunting settings rather than a newly selected setting.

Earlier results labelled ``original implementation'' were obtained by B.L.S. in a separate implementation that motivated the contextual task (unpublished data). The accuracy references are means over three seeds, transcribed from his study reports; the intervention and drive-correlation summaries are also reported values from those analyses. They are shown separately in Supplementary Note~\ref{note:contextual_details} and Table~\ref{tab:original_routing}, not treated as paired replicates or pooled with the DendriNet runs. The endpoint-gate variant is a separately trained DendriNet model and is not that earlier implementation.

All models train for 1,200 epochs of 65,536 presentations, each with a fresh character, context, gain and noise draw, in batches of 512. Adam has zero weight decay, and the learning rate follows a cosine schedule from its initial value (0.002 for the trees except where stated, 0.001 for the MLP) to one percent of it. Of the 200 training images per class held out from optimization, the first 50 per class form the validation set. After each epoch the network is evaluated on corrupted and clean validation images, and the checkpoint with the highest corrupted validation accuracy, ties broken by validation cross-entropy, is kept. Ten seeds (62000--62009) are trained on A100 GPUs; the same-recipe pair is also trained, independently, on H100 GPUs. Test accuracy is the mean over five context and corruption draws on all 18,800 official test images. Recipes were fixed on validation accuracy of separate screening seeds before these seeds were scored on test images, but the task's test images had been examined during earlier development, so the comparison is not an untouched benchmark.

Decoding uses the 9,400 held-out training images (200 per class) for fitting and the official test images for scoring. Ridge probes (standardized features, penalty 1) read the soma outputs of each population and layer, or the MLP's hidden layers, for character, context and target. A probe is either refitted under each condition or fitted once on clean images and applied under every condition, and accuracy is reported above the mean of five label-shuffled probes. Conditions are clean images, gains 0.05, 0.2, 5 and 20 without noise, noise levels 0.25, 0.5 and 1 at unit gain, and the four gains with noise at level 0.5; contexts and noise draws come from fixed generators shared by all models.

The inhibition interventions in Fig.~\ref{fig:fig7}E,F use official test images 8--27 of each class, crossed with gains 0.05, 0.25, 1, 5 and 20, noise divisors 1, 2 and 4 and all 32 contexts, with one noise draw per image and condition shared across contexts. Context-mean replacement substitutes the inhibitory drive arriving at each excitatory compartment with its mean over the 32 contexts for the same corrupted image. The transplant supplies the drive for context $(s+1)\bmod32$; the target of that context defines the donor label. Because every corrupted image is presented under all 32 contexts, this cyclic permutation preserves the empirical distribution of supplied inhibitory drive across contexts at each targeted compartment, including its mean. It changes the pairing with the presented cue and does not constrain the total inhibitory drive received by a neuron on an individual trial. All other signals are recomputed through the unchanged network. Leak-scaling counterfactuals multiply the leak of every first- or second-layer excitatory compartment of a trained fixed-leak network by 0.25 or 4 and evaluate images 28--67 of each class without noise at gains $1/20$, $1/4$, 1, 5 and 20.

Gain profiles and leak shares use held-out training images 50--57 of each class, which neither optimization nor model selection used, at five gains spaced geometrically from 0.05 to 20, the three noise divisors and all 32 contexts. For each excitatory compartment and inhibitory soma, attenuation is $1-\mathrm{CV}(Y)/\mathrm{CV}(N)$, where the coefficients of variation are computed across gains for the mean output $Y$ and numerator $N=E+C$. For Fig.~\ref{fig:fig7}C, mean outputs at each gain on images 50--53 of each class predict the outputs on images 54--57. Each network contributes the median over first-layer distal E compartments with mean output above $10^{-6}$ and output variance above $10^{-10}$ on the second subset. The leak share of a shunting compartment is $g_L/(g_L+T)$ for each input, where $T$ is the remaining denominator ($E+I+G$), recorded by the native branch diagnostics; each compartment contributes its mean over inputs, and each seed contributes the median over compartments of a layer, population and level. Supplementary Fig.~\ref{fig:figS_context_controls}F shows the leak shares at noise divisor $q=1$. The dim-image accuracy test in Fig.~\ref{fig:fig7}D uses noise-free inputs. Figures report means over the ten seeds with 95\% percentile intervals from 10,000 bootstrap resamples of seeds, and paired differences are computed within seeds.

E/I drive correlations use the second half of the gain-profile images (54--57 of each class) under the same gains, noise divisors and contexts. For each excitatory compartment, the Pearson correlation of its excitatory and inhibitory drives is computed over all inputs, after removing the mean of each character--context pair (60 inputs: four images, five gains, three divisors), and after removing the mean of each exact condition of character, context, gain and divisor (four images); each seed contributes the median over the compartments of a layer and level. Contact alignment (\texttt{contextual\_connectivity.py}) uses official test images 68--75 of each class at unit gain and noise divisor 2, each presented with all 32 contexts, and the core library's source-tuning analyzer. For each second-layer excitatory soma, its mean output for each context (or character) on images 68--71 forms its tuning profile; on images 72--75 the same profile is computed for every source the soma could contact through a pathway, and each source's profile is correlated with the soma's. The alignment of selected inputs, $\Delta$, is the mean correlation over the soma's active contacts minus the mean over all eligible sources; each seed contributes the mean over somas.

\subsection{ImageNet-scale replacement}
\label{sec:imagenet_methods}

The ImageNet experiment uses the 1,000-class classification dataset and torchvision's pretrained AlexNet weights, identified by \texttt{AlexNet\_Weights.IMAGENET1K\_V1}~\citep{Deng2009,Krizhevsky2012}.  The fc7-only experiment freezes the convolutional backbone and fc6, replaces fc7 and its ReLU, and trains the replacement with fc8.  The fc6--fc7 experiment instead freezes the convolutional prefix, replaces the fc6--ReLU--fc7--ReLU span, trains the replacement with fc8, and omits the original dropout except for the explicitly tested MLP regularization controls.  The ImageNet training set is divided into 80\% optimization and 20\% validation subsets, while the official validation set is reserved for final evaluation.  The reported comparisons use random resized $224\times224$ crops and horizontal flips during optimization, resize-to-256 followed by a $224\times224$ center crop at evaluation, and channel-wise ImageNet normalization with means $(0.485,0.456,0.406)$ and standard deviations $(0.229,0.224,0.225)$. Top-1 accuracy is the fraction of images for which the highest-scoring class is correct.

Retained replacement-core parameters count selected synaptic weights, child couplings, activation parameters and biases. The common final classifier is counted separately; nontrainable connectivity indices contribute to storage rather than trainable-parameter counts.

The original five-budget sweep, shown by connected curves in Fig.~\ref{fig:fig6}E, replaces the complete fc6--fc7 span with one of four families: learned-contact shunting, a dense bottleneck MLP, a fixed sparse MLP, or a learned-contact sparse MLP. The replacement core excludes the frozen convolutional prefix and the common trainable fc8 classifier. The reference retained-core budget is $B=1{,}376{,}320$; the downward sweep uses five caps, $B_k=98{,}320+127{,}800k$ for $k\in\{10,5,3,2,1\}$, with $B_{10}=B$. Shunting keeps 4,096 E and 820 I neurons with $[2,2]$ trees, retaining $4k$ feedforward E contacts and $k$ local-I contacts per E branch, and $k$ input contacts per I branch. All learned coupling and activation parameters remain counted. Sparse MLPs contain two affine layers separated and followed by ReLUs; their hidden width and output fan-in decrease with the cap. Dense MLP hidden width is the largest integer fitting the cap; ReLU or leaky-ReLU activation is selected on validation data. Learned sparse MLPs select connections by absolute weight; shunting selects nonnegative effective weights. Both are evaluated after exporting only retained weights and indices.

At each cap, three tuning seeds (61--63) evaluate synaptic learning rates $3\times10^{-4}$ and $10^{-3}$ for shunting, and core learning rates $10^{-4}$ and $3\times10^{-4}$ for the MLPs, for five epochs; the dense family additionally compares the two activations. Selection minimizes mean internal-validation cross-entropy on a fixed 4,096-image subset of the held-out training partition. Five fresh seeds (81--85) then train for 100 epochs with the same optimization split, augmentation, global batch size 256, frozen convolutional prefix, trainable pretrained fc8, FP16 mixed precision, and gradient-norm clipping at 5. Adam uses fixed learning rates and zero weight decay in this sweep; no active-synapse maintenance or inactive-candidate boosting is applied. For trees, coupling, activation and classifier rates are one tenth of the synaptic rate. For ReLU MLPs the core and classifier rates are equal; the leaky-ReLU alternative uses one tenth of the core rate for its classifier. All models process the same number of optimization batches. Validation loss selects checkpoints within each horizon; saved 50- and 100-epoch selections are evaluated in FP32 on all 50,000 official-validation images. Core parameter counts are checked after compact export. The common classifier contributes 4,097,000 additional parameters to every model. We use 50,000 paired seed-bootstrap resamples and report pointwise intervals. Continued validation improvement after epoch 80 is recorded separately from performance at the fixed horizon (Supplementary Note~\ref{note:downward_matched}).

The endpoint regularization controls were motivated by continued decreases in MLP training loss after validation loss began to rise, indicating overfitting. The endpoint controls in Fig.~\ref{fig:fig6}B--D repeat caps $B_{10}$ and $B_1$ and add learned-contact raw-additive trees. These trees exactly match shunting in population sizes, morphology, candidate-bank size, retained counts, classifier, and parameter-group learning-rate ratios. For each rule, three tuning seeds (91--93) test learning rates $10^{-4}$, $3\times10^{-4}$ and $10^{-3}$ under both analytical initialization and occupancy-quantile activation calibration. Calibration uses 256 optimization images and iterates to a maximum activation-parameter change below 0.001; no validation images enter initialization. For each MLP family, the earlier selected learning rate and architecture are retained while testing dropout probabilities 0 and 0.2 and Adam weight decay 0 and $10^{-4}$. Dropout follows both MLP activations and is disabled for evaluation; weight decay is the coupled L2 penalty in Adam and applies to core and classifier. These runs retain the sweep's fixed learning rates, parameter-group ratios and gradient-norm clipping at 5, without synaptic maintenance or boosting. Each candidate trains for 50 epochs. Mean minimum internal-validation loss across tuning seeds selects one protocol per family. Both policy-specific tree protocols and any distinct unregularized MLP reference are retained for confirmation.

Five new seeds (101--105) train the selected endpoint models for 100 epochs with the same data split, 4,096-image internal-validation membership, augmentations, batch size, classifier and checkpoint-selection rule as the original sweep. Within each seed, all arms receive the same augmented image batches. The tree rules both select learning rate $10^{-3}$; analytical initialization is selected for shunting and occupancy calibration for addition. Weight decay $10^{-4}$ without dropout is selected for all MLPs except the smallest learned-contact sparse MLP, which retains no regularization. All confirmation runs finish before compact 50- and 100-epoch selections are evaluated on the full official validation set in FP32. Before production, gradient and export checks pass and every family at both caps fits a fixed 256-image optimization subset within 200 updates. These diagnostics establish basic trainability, not optimal full-data fitting. Several tree models still improve late in training, so the fixed horizons do not establish asymptotic performance. This follow-up was designed after the earlier official-validation results had been inspected; protocol selection for the new runs uses only internal validation. Full selection evidence, histories and parameter inventories accompany Supplementary Note~\ref{note:alexnet_followup}.

The original fc6--fc7 confirmation, retained in Supplementary Table~\ref{tab:fig6_fc6fc7_confirmation}, uses five seeds (41--45) and a 50-epoch horizon. It compares fixed-contact shunting under global, random-local and structured-local routing; fixed-contact raw addition under global and structured-local routing; a sparse point layer; learned-contact shunting; and the retrained native dense span. Every dendritic arm contains 4,096 E and 820 I somas with $[2,2]$ trees. Each I compartment retains ten feedforward contacts; each E compartment retains 40 feedforward contacts and ten local-I contacts. Global routing samples from all 9,216 flattened convolutional features. The two local routes treat those features as 256 channels on a $6\times6$ grid. Structured-local routing places windows according to the branch path; random-local routing preserves the window-size distribution but places windows independently. All channels within a branch's window are eligible sources.

Fixed and learned-contact dendritic cores retain 1,278,000 synapses and 98,320 coupling and activation parameters. The sparse point control is a fixed-index affine $9216\mathbin{\to}4096$ map followed by ReLU, with the same total of 1,376,320 trainable core parameters. Learned-contact shunting instead stores 292,085,776 core parameters during training. The native dense control retains the pretrained fc6 and fc7 weights and retrains that span with fc8; its core contains 54,534,144 parameters. It is distinct from the untouched pretrained-AlexNet reference, which is evaluated without retraining on all 50,000 official-validation images with the pretrained center-crop transform.

These original replacements use Adam and mixed precision on NVIDIA H100 80GB GPUs, with minimum-validation-loss checkpoint reload and clipping at 5. The fc6--fc7 confirmation uses a per-process batch size of 64 across four distributed processes, giving a global batch size of 256. Synaptic parameters use learning rate $10^{-3}$; child coupling, activation-function and decoder parameters use $10^{-4}$. Weight decay, active-weight maintenance and inactive-candidate boosting are disabled. Per-run device identity is recorded in the frozen telemetry tables. Most best checkpoints occur near the 50-epoch horizon, so these original comparisons describe a common training horizon rather than asymptotic performance.

The separate conventional-network comparison uses the same fc6--fc7 boundary and a replacement-core cap $B=1{,}376{,}320$, counting every trainable scalar, including biases, branch couplings, and activation parameters. The dense bottleneck has width 103, the deeper dense MLP has two hidden layers of width 102, and the sparse MLP has width 1,024 with 256 and 270 fixed input connections per unit in its first and second layers. MLP weights are signed and unconstrained; fixed sparse weights are stored directly with their source indices. Rectification follows every core affine layer. Every condition also trains the same pretrained fc8 classifier with 4,097,000 parameters. Integer architecture sizes give slightly different core counts below the cap (Supplementary Table~\ref{tab:alexnet_matched_mlp}); neither hidden width, connectivity storage nor computational cost is matched.

MLP training settings were selected on three tuning seeds after five epochs, using the mean of each run's lowest cross-entropy on 4,096 fixed internal-validation images. The selected ReLU, deeper and sparse MLPs use output-layer initialization scaled by 0.1 and learning rate $10^{-4}$ for core and classifier. A separately retained leaky-ReLU control uses negative slope 0.1, core learning rate $3\times10^{-4}$ and classifier learning rate $3\times10^{-5}$. Five fresh seeds (71--75) train all MLPs and fixed global shunting, raw-additive and sparse-point references for 50 epochs. After inspecting the 50-epoch official-validation results, we extended these runs to 100 epochs on the same split, image augmentations and internal-validation membership. Fixed shunting and the sparse MLP were close after 50 epochs; the continuation increased the shunting advantage (Supplementary Table~\ref{tab:alexnet_matched_mlp}). Reference parameter-group learning rates retain the original protocol; hyperparameter-search budgets are therefore not matched. These single-GPU runs use global batch size 256, FP16 mixed precision and gradient-norm clipping at 5. Checkpoints minimize internal-validation cross-entropy; final evaluation uses all 50,000 official-validation images in FP32. The larger MLPs use caps $2B$ and $4B$ and retain the settings selected at $B$. Pointwise intervals use 50,000 bootstrap resamples of the five seeds; contrasts with shunting are paired by seed. These runs form a separate cohort from Figure~\ref{fig:fig6}B and are not pooled with it.

Training and deployment resources are counted separately. Figure~\ref{fig:fig6}C reports replacement-core parameters for the five fc6--fc7 families at $B$, excluding the retained AlexNet prefix and fc8. Whole-model storage additionally includes those modules and persistent connectivity buffers. A separate fc7-only compaction check uses an earlier learned-contact shunting model with $W_E=4096$, $W_I=1024$ and $[2,2]$ trees (resource definitions in Supplementary Table~\ref{tab:resources}; storage in Supplementary Table~\ref{tab:fig6_compact_storage}). During training, its candidate synaptic weights are FP32 parameters. After training, deterministic supports are frozen into fixed indexed layers: inactive candidates are discarded, retained weights are unchanged, and sorted source indices are stored as unsigned 16-bit integers. No weights are quantized. Conversion checks each replaced synapse layer at absolute and relative tolerance $10^{-5}$, reloads the compact checkpoint into a fresh model, and repeats an end-to-end deterministic-input check. This storage conversion occurs after learning and does not reduce the training candidate bank.

The representation analysis in Supplementary Fig.~\ref{fig:figS_alexnet_width_parameter_scaling}C uses a separate five-seed fc7-only cohort with 4,096 E and 820 I neurons, $[2,2]$ trees, and the same 40E/10I allocation in E branches and ten feedforward contacts in I branches. Inputs are the 4,096-dimensional outputs of frozen fc6. Five official-validation examples per class undergo a label-independent 256-dimensional CountSketch followed by three-fold out-of-fold ridge decoding and three shuffled-label fits. Accessibility is held-out accuracy minus the mean shuffled-label accuracy. Here, ``E/I population'' denotes the emitting soma population, whereas ``E/I drive'' denotes the corresponding branch-aggregated drive entering an E-population tree.

Figure~\ref{fig:fig6}F and \suppfig{figS_alexnet_corruption_decoding} evaluate the 25 endpoint models at $B$ (seeds 101--105) and untouched AlexNet without retraining.  Images are resized to 256 pixels and center-cropped to $224\times224$, and the corruption acts on the raw $[0,1]$ intensities before ImageNet normalization and the frozen convolutional prefix.  Gaussian noise adds independent $\mathcal N(0,\sigma^2)$ values to every pixel and color channel, with $\sigma\in\{0.1,0.2,0.3,0.5\}$, and clips the result to $[0,1]$.  Gain multiplies each image by one factor $g$, with $\log g\sim\mathcal U(-\log b,\log b)$ for gain bounds $b\in\{2,5,20\}$, without clipping.  Five draws, shared across models and conditions, are evaluated on all 50,000 official-validation images and averaged within model; retention, $(A-0.001)/(A_{\rm clean}-0.001)$, uses each model's own clean accuracy.  Ridge probes with standardized features, penalty 1 and balanced class weights read the 4,096 replacement outputs that feed fc8 (the E population for the trees) and the fc6 and fc7 outputs of untouched AlexNet.  They are fitted on 50 images per class from the held-out training partition, excluding the 4,096 checkpoint-selection images, and scored on the official validation images.  A probe is either refitted under each condition on equally corrupted fit images or fitted once on clean images; probes use the first corruption draw, and five label-shuffled fits give the reference.

\subsection{Statistics and reproducibility}

The \hyperref[supp:software]{supplementary software note} illustrates how to define the E and I inputs to a neuron and reproduce the branch-transmission and inhibitory-transplant analyses (\suppfig{figS_software_workflow}).
\label{sec:statistics}

The independently trained experiment seed is the replicate.  Examples, branches, somas, dendritic depths, network positions, perturbation draws, projections, probe fits, permutations, and intervention supports are technical dimensions and are reduced within trained seed before uncertainty is estimated.  Contrasts between integration rules are formed within matched seed whenever the experimental design is paired.

Unless stated otherwise, curves show means and pointwise 95\% percentile intervals from 10,000 bootstrap resamples of trained seeds~\citep{EfronTibshirani1993}.  For curves and point summaries, intervals are shown as capped error bars or bands; faint points, where shown, are individual networks. Heatmaps summarize condition means, with descriptive interval markers where specified in the caption.  Difference maps use symmetric color ranges around zero. In Fig.~\ref{fig:fig2}B, both accuracy maps share a 49.5--52\% scale, and the paired-difference scale spans $\pm1.6$ percentage points; neither clips displayed values.  Figure~\ref{fig:fig1}E--G,I are illustrative; Figure~\ref{fig:fig1}D reports ten paired training seeds.  Main Fig.~\ref{fig:fig1}H and Supplementary Fig.~\ref{fig:figS_mnist_components_source_activity} average somas within each network before bootstrapping ten networks per integration rule.  Figure~\ref{fig:fig2}B, the depth-four comparison in C, and the depth-four accuracy slices in \suppfig{figS_allocation_slices} use twenty seeds per cell; the depth-two and depth-three grids and their comparisons in C use thirteen: three original pilot seeds (53000--53002) plus ten extension seeds (53003--53012), with all thirteen retained in each allocation cell.  Figure~\ref{fig:fig2}D uses ten paired inhibitory-contact-control seeds, E ten paired seeds across readouts, and the depth-four inhibitory-tail summaries ten; F--H use twenty paired seeds, ten original and ten added after inspection.  Figures~\ref{fig:fig3} and \ref{fig:fig4} also use twenty paired seeds per integration rule and condition, ten original and ten added after inspection; Supplementary Table~\ref{tab:seed_extension} reports the original, added and pooled estimates of their main contrasts and of Fig.~\ref{fig:fig2}F--H. Comparisons with the initialization-policy controls and with the linear-readout models keep the ten original seeds, because those cohorts were not extended. Unless a caption states otherwise, Figures~\ref{fig:point_neurons} and \ref{fig:fig5} use ten seeds per integration rule and condition; Fig.~\ref{fig:point_neurons}C,D use twenty. Figure~\ref{fig:fig7} uses ten training seeds per model (62000--62009). The shared-protocol comparison was repeated by training both trees on a second GPU model.  Figure~\ref{fig:fig6}B--D and F use the five endpoint seeds (101--105), and E combines them with the five seeds of the original sweep at each retained-parameter cap (81--85); intervals in B, D and F come from 50,000 resamples of the seeds, shared across families, and F evaluates the endpoint models at $B$ without retraining. The original pretrained model is a deterministic reference.

Intervals across allocation cells, perturbation levels, depths, and network positions are pointwise and are not adjusted for simultaneous coverage.  Markers indicating exclusion of zero are descriptive rather than familywise hypothesis tests.  The allocation slices and morphology screens are structured summaries of larger factorial experiments, not independent replications of the same contrast.  In-text contrasts are computed from unrounded seed-level values and may differ from differences of displayed rounded means in the final digit.  Percentile bootstrap intervals can have below-nominal coverage with small seed samples, particularly the five-seed AlexNet comparisons (Fig.~\ref{fig:fig6}B,D--F); increasing the number of bootstrap draws does not add independent networks. Individual seed values accompany the numerical results. The learned-leak comparison illustrates interval-method sensitivity and is reported with both percentile-bootstrap and paired Student-$t$ intervals in Supplementary Note~\ref{note:contextual_details}.

\section{Data availability}
\label{sec:repro}

The per-seed measurements for the two worked examples (Figs.~\ref{fig:fig3}D
and \ref{fig:fig7}F) are available in the
\href{https://github.com/KempnerInstitute/DendriNet/releases/download/v0.5.0/empirical-example-data.zip}{DendriNet example data archive}.
The full set of numerical tables and trained checkpoints has not been deposited publicly.
MNIST, EMNIST, CIFAR-10 and ImageNet can be obtained from their cited sources.

\section{Code availability}

DendriNet v0.5.0 can be installed from
\href{https://pypi.org/project/dendritic-modeling/0.5.0/}{PyPI}.
The \href{https://github.com/KempnerInstitute/DendriNet/releases/tag/v0.5.0}{accompanying release}
includes eight representative training configurations and scripts for reproducing
the branch-transmission and inhibitory-transplant summaries
in Figs.~\ref{fig:fig3}D and \ref{fig:fig7}F. The \hyperref[supp:software]{supplementary
software note} gives installation instructions and identifies the versions used for
the original experiments.

\section*{Author contributions}
H.S., M.R. and B.L.S. developed the ideas, methods and code. H.S. and B.L.S. wrote the manuscript. R.W., A.M. and N.K. contributed to coding and software development.

\section*{Competing interests}
No competing interests are declared.

\section*{Acknowledgments}
This work has been made possible in part by a gift from the Chan Zuckerberg
Initiative Foundation to establish the Kempner Institute for the Study of
Natural and Artificial Intelligence at Harvard University.

\FloatBarrier
\clearpage
\BeginSupplementary
\begin{center}
{\LARGE\bfseries Supplementary Information\par}
\vspace{0.45em}
{\large for \textit{Trainable sparse dendritic networks reveal regime-dependent roles for shunting inhibition and morphology}\par}
\end{center}

Supplementary Notes~\ref{supp:model}--\ref{supp:vision} follow the main figures in order, from single-neuron responses to contextual classification and the AlexNet replacements. Supplementary Note~\ref{supp:protocols} gives the training protocols, cohort definitions and resource counts, with worked examples for reproducing the branch-transmission and contextual-inhibition analyses.

\section*{How to read this supplement}
\addcontentsline{sic}{section}{How to read this supplement}

An independently trained network is the experimental replicate. Paired effects are computed within training seed before resampling, and measurements over images, branches, branch selections, projections, probe fits, label shuffles and perturbation draws are averaged within seed; a cohort reused in two comparisons is counted once. Unless stated otherwise, intervals are pointwise 95\% percentile intervals from 10,000 seed-bootstrap resamples. They are not adjusted across the conditions of a grid, so counts of conditions whose interval excludes zero are descriptive. With five or ten networks per condition, percentile intervals can have below-nominal coverage; Supplementary Note~\ref{supp:contextual} gives one case in which a Student-$t$ interval includes zero. Display offsets separate overlapping symbols without changing the conditions.

Accuracy differences are in percentage points (pp), and retention is normalized to each model's own clean above-chance accuracy. Mutual information, decoding accuracy and replacement losses measure different properties. Dendritic depth $d$, population position $\ell$ and total network depth $L$ are distinct coordinates (Supplementary Table~\ref{tab:notation}). Cohort sizes, readouts and departures from the primary design are stated in each note. Recipe A, also called the common recipe or common protocol, is the training setting shared by both rules in the single-layer comparisons (Supplementary Note~\ref{note:recipe_policy}). Original seeds are the first ten seeds of a cohort later extended to twenty; the original implementation is the earlier model of the contextual task (Supplementary Note~\ref{supp:contextual}); the original AlexNet comparison is the first five-seed comparison (seeds 41--45). Endpoint refers, as stated, to the depths $L=1$ and $L=5$, to the two budgets of the AlexNet controls, or to the endpoint-gate activation.

The canonical MNIST and CIFAR-10 test sets informed earlier model development, so their intervals describe seed variability on reused benchmarks. ImageNet's official validation split was not used for checkpoint or recipe selection, although two AlexNet follow-ups were designed after earlier official-validation results had been seen (Supplementary Note~\ref{supp:vision}). No interval measures uncertainty on an unobserved data distribution.

\suppcontents

\FloatBarrier

\supppart{Synaptic inputs and single-neuron responses (Figure~\ref{fig:fig1})}
\label{supp:model}

Among retained contacts, more E sources are active for the assigned digit, whereas more I sources are active for other digits (Fig.~\ref{fig:fig1}). The analyses below distinguish where synapses are retained from how strongly their sources are activated by an image.

\suppnote{Class-related contact activity}
\label{note:mnist_components_source_activity}

\begin{figure}[tbp]
    \centering
    \IfFileExists{figures_empirical/figS_mnist_components_j1.pdf}{%
      \includegraphics[width=165mm]{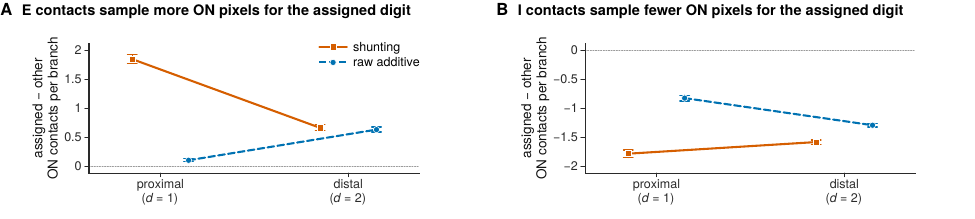}%
    }{%
      \PackageError{empirical-manuscript}%
        {Missing required figure `figures_empirical/figS_mnist_components_j1.pdf'}%
        {Restore the figure listed in the manuscript export manifest before compiling.}%
    }%
    \caption{%
\textbf{Class-related contact activity under both integration rules.}
MNIST identity classifier: one single-neuron classifier per digit with $[2,2]$ trees and 25 E and 25 I contacts per compartment, without I neurons. Assigned-minus-other numbers of ON-source contacts per branch for the E (A) and I (B) pathways at proximal and distal branches; ON means $x_p>0.1$. Squares and solid lines, shunting; circles and dashed lines, raw addition; the rules are offset horizontally. Somas and branches are averaged within network before means and 95\% seed-bootstrap intervals are computed across ten networks per rule.}
    \label{fig:figS_mnist_components_source_activity}
  \end{figure}

Across ten trained networks per rule, the identity classifier performs modestly better under shunting: test accuracy is $91.31\pm0.32\%$ versus $89.64\pm0.40\%$ (mean $\pm$ SD), with a paired advantage of 1.67 percentage points (95\% interval, 1.41--1.99). Both rules sample more ON-source E contacts and fewer ON-source I contacts for a soma's assigned digit than for pooled other digits. The between-rule difference is clearest in proximal branches (\suppfig{figS_mnist_components_source_activity}A,B).

ON-source contacts have pixel intensity $x_p>0.1$. The retained support is fixed across images, so assigned-versus-other differences describe the inputs sampled by learned contacts, rather than changing contact counts. Figure~\ref{fig:fig1}G combines mean retained weights with assigned-minus-other pixel activity to show the corresponding difference in synaptic drive.

\FloatBarrier

\supppart{Contact allocation, readout, and morphology (Figure~\ref{fig:fig2})}
\label{supp:allocation}

Relative accuracy changes with E/I allocation, and compartmentalization and local input sampling can improve classification without a consistent benefit from greater serial depth (Fig.~\ref{fig:fig2}). This note expands the allocation maps, tests how the readout, initialization and contact budget affect these conclusions, and reports the seed extension of the cohorts of Figs.~\ref{fig:fig2}--\ref{fig:fig4}.

\suppnote{Sensitivity to initialization}
\label{note:allocation_landscape}

\begin{figure}[tbp]
    \centering
    \IfFileExists{figures_empirical/figS_allocation_landscape_j1.pdf}{%
      \includegraphics[width=165mm]{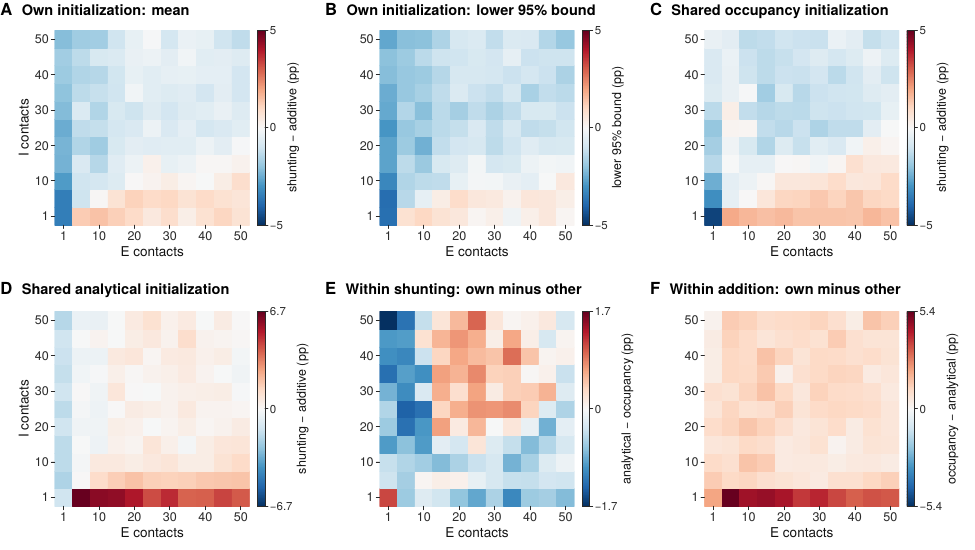}%
    }{%
      \PackageError{empirical-manuscript}%
        {Missing required figure `figures_empirical/figS_allocation_landscape_j1.pdf'}%
        {Restore the figure listed in the manuscript export manifest before compiling.}%
    }%
    \caption{%
\textbf{Initialization sensitivity across the complete allocation grid.}
CIFAR-10, 64 somas, $[3,3,3,3]$ trees and MLP readouts, at all 121 allocations with ten paired seeds per cell (seeds 0--9; the readout comparison of \suppfig{figS_readout_allocation} uses seeds 10--19, so cell values differ). (A) Shunting with analytical minus addition with occupancy initialization, each rule's own, and (B) its pointwise lower 95\% bound. (C,D) Shunting-minus-additive differences with shared occupancy or shared analytical initialization. (E,F) Within-rule effect of each rule's own initialization minus the other: analytical minus occupancy for shunting (E) and occupancy minus analytical for addition (F). A--C share one zero-centered color scale; D, E and F have their own. Intervals use 10,000 paired-seed bootstrap draws and are not adjusted across the grid; the maps of Fig.~\ref{fig:fig2}B use twenty seeds.}
    \label{fig:figS13_allocation}
  \end{figure}

The single-layer models initialize shunting analytically and addition by occupancy quantiles, under the common recipe (Methods). \suppfig{figS13_allocation} compares this pairing (A, mean; B, lower interval bound) with occupancy (C) or analytical (D) initialization for both rules across the 121 allocations, with ten paired seeds per cell (seeds 0--9; the maps of Fig.~\ref{fig:fig2}B have twenty). Averaged over the grid, shunting minus addition is $-0.6$ points with each rule's own initialization, $+0.6$ with analytical initialization for both and $-0.4$ with occupancy initialization for both, so the sign of the grid average depends on initialization. The change comes mainly from addition: analytical initialization lowers additive accuracy (49.5\% versus 50.6\% averaged over the grid), whereas shunting accuracy is similar under both (50.1\% and 50.2\%; within-rule contrasts in E,F). A shared policy does not give identical initial weights, because its calibration depends on the integration rule. Supplementary Note~\ref{note:recipe_policy} reports sensitivity to the training recipe; neither control identifies an optimizer-independent optimum.

\FloatBarrier

\suppnote{Allocation at fixed and varying contact budgets}
\label{note:allocation_slices}

\begin{figure}[tbp]
    \centering
    \IfFileExists{figures_empirical/figS_allocation_slices_j1.pdf}{%
      \includegraphics[width=165mm]{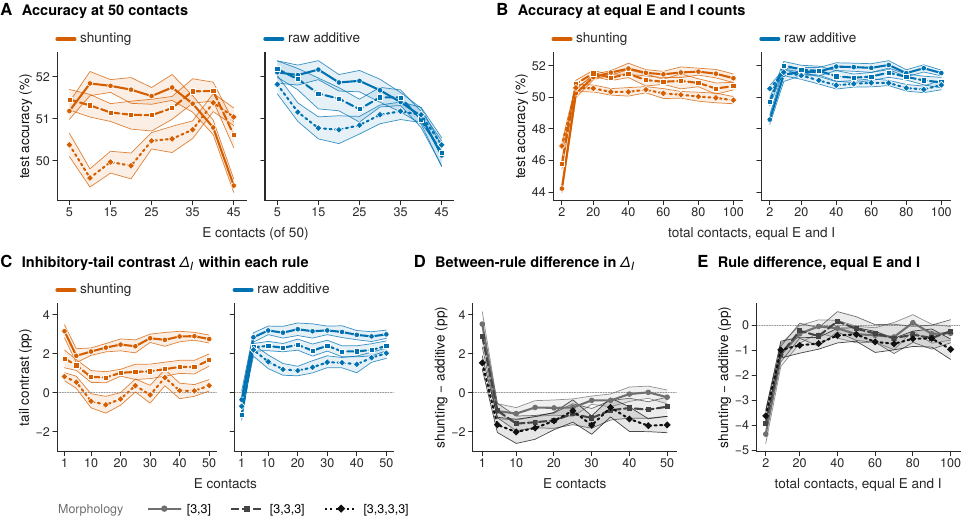}%
    }{%
      \PackageError{empirical-manuscript}%
        {Missing required figure `figures_empirical/figS_allocation_slices_j1.pdf'}%
        {Restore the figure listed in the manuscript export manifest before compiling.}%
    }%
    \caption{%
\textbf{Absolute performance and paired contrasts across E/I allocation.}
CIFAR-10, 64 somas, MLP readouts. (A) Accuracy at $N_E+N_I=50$ and (B) at equal E and I counts with increasing total contacts, each integration rule in its own labelled frame. (C) Inhibitory-tail contrast $\Delta_I$ within each rule, the mean accuracy at the two largest minus the two smallest sampled $N_I$ at each $N_E$ (Methods); (D) its shunting-minus-additive difference. (E) Shunting-minus-additive accuracy at equal E and I counts. Marker and line style identify morphology (key), and grey level identifies it in D,E. For the four-level tree, the accuracy curves (A,B,E) use all 20 seeds (0--19). The inhibitory-tail contrasts (C,D) use the ten seeds (0--9) in the initialization analysis. Both shallower trees use the same 13 seeds (53000--53012) throughout. Means and pointwise 95\% seed-bootstrap intervals are computed from the unsmoothed per-seed values; positive differences between rules favor shunting, and chance accuracy (10\%) lies below the plotted range in A,B.}
    \label{fig:figS_allocation_slices}
  \end{figure}

\suppfig{figS_allocation_slices} expands the maps of Fig.~\ref{fig:fig2}B along two slices: a fixed total $B=N_E+N_I=50$, which changes the E fraction $N_E/B$ (A), and equal E and I counts with increasing $B$, which changes capacity at a fixed ratio (B, with the difference between rules in E). Fig.~\ref{fig:fig2}C shows the difference between rules along the first slice. Neither coordinate measures the learned balance of E and I drive.

The inhibitory-tail contrast $\Delta_I$ (Methods) compares the two highest with the two lowest sampled $N_I$ at each $N_E$ within each rule (C); its difference between rules (D) is a secondary summary of allocation dependence. All models use the MLP readout; the readout effect of Fig.~\ref{fig:fig2}E is defined in the next section.

\FloatBarrier

\suppnote{Joint training with linear and nonlinear readouts}
\label{note:readout_allocation}

Both rules were trained with an MLP or a linear readout at all 121 allocations, with seeds 10--19 paired across the four arms and the common recipe (the main maps use twenty MLP seeds). The readout effect is $\Delta A_{\rm readout}=A_{\rm linear}-A_{\rm MLP}$, and the contrast between rules within a readout is $\Delta A=A_{\rm sh}-A_{\rm add}$.

\begin{figure}[tbp]
    \centering
    \IfFileExists{figures_empirical/figS_readout_allocation_j1.pdf}{%
      \includegraphics[width=165mm]{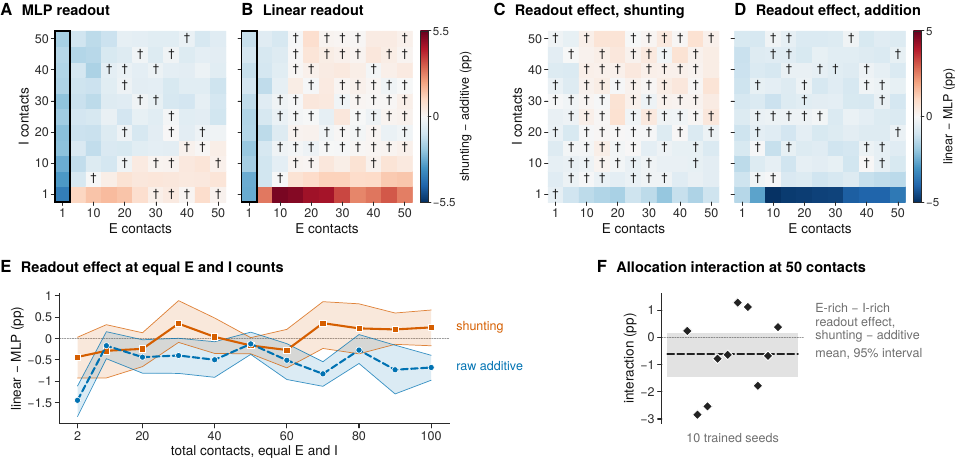}%
    }{%
      \PackageError{empirical-manuscript}%
        {Missing required figure `figures_empirical/figS_readout_allocation_j1.pdf'}%
        {Restore the figure listed in the manuscript export manifest before compiling.}%
    }%
    \caption{%
\textbf{Readout effects depend on both E/I allocation and total contact number.}
CIFAR-10, 64 somas and $[3,3,3,3]$ trees; both readouts are trained end to end, with seeds 10--19 paired across the four arms. (A,B) Paired shunting-minus-additive test accuracy under MLP and linear readouts across all 121 allocations; outlines mark the entire $N_E=1$ column, not a significance-selected subset. (C,D) Linear-minus-MLP accuracy for shunting and raw addition. Each pair of maps shares a symmetric color scale, and daggers mark allocations whose pointwise 95\% paired seed-bootstrap interval includes zero. (E) Readout effects at equal E/I counts (shunting solid, addition dashed); bands are pointwise 95\% paired seed-bootstrap intervals. (F) Fixed-total allocation interaction: for each rule, the mean readout effect at 40E/10I and 45E/5I minus that at 5E/45I and 10E/40I, then the shunting-minus-additive difference. Diamonds show the ten seeds; the dashed line and band show their mean and 95\% seed-bootstrap interval.}
    \label{fig:figS_readout_allocation}
  \end{figure}

A linear readout lowers additive accuracy by 0.9 points averaged over the grid (95\% interval, 0.9--1.0) but shunting accuracy by only 0.1 points (0.1--0.2), so the two rules are closer under the linear readout (\suppfig{figS_readout_allocation}A--D). Shunting has the higher mean accuracy at 70 of the 121 allocations with the linear readout and at 28 with the MLP; intervals exclude zero in favor of shunting at 39 (linear) and 18 (MLP) allocations and in favor of addition at 29 and 79. With a single E contact, both readouts favor addition, by 2.3 points (MLP) and 1.7 points (linear) averaged over that column. Along the equal-E/I slice, the readout effect varies with total contact number (E). For the fixed-budget interaction (F), each seed contributes its mean readout effect at 40E/10I and 45E/5I minus that at 5E/45I and 10E/40I; the shunting-minus-additive difference of these changes is $-0.6$ points ($-1.5$ to 0.2) and is negative in six of ten seeds. The data therefore do not show opposite allocation dependence of the readout effect for the two rules.

Because the representation and the readout are trained together, this comparison tests the complete classifier, not decoding of a fixed representation (next section), and it does not establish a universal allocation threshold. \suppfig{figS_readout_mechanisms}E adds conductance-normalized addition under both readouts.

\FloatBarrier

\suppnote{Decoding frozen somatic representations}
\label{note:allocation_accessibility}

\begin{figure}[tbp]
    \centering
    \IfFileExists{figures_empirical/figS_allocation_probes_j1.pdf}{%
      \includegraphics[width=165mm]{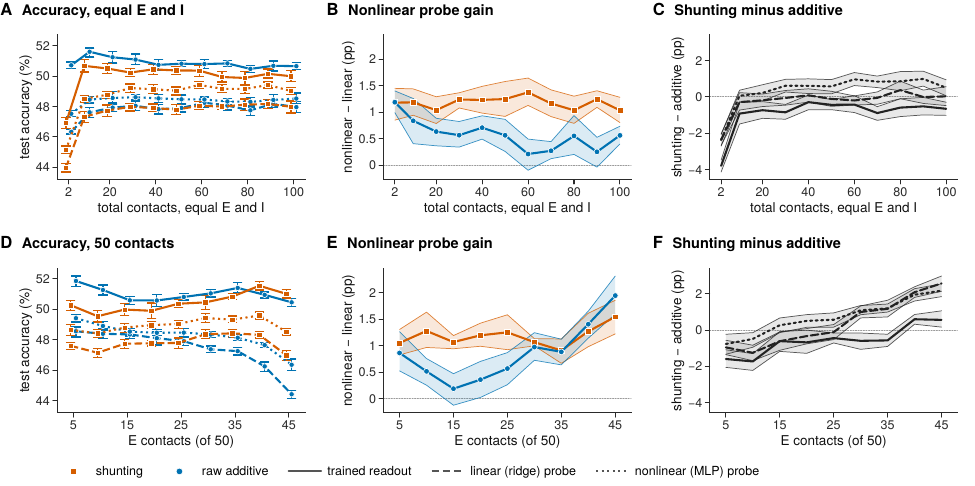}%
    }{%
      \PackageError{empirical-manuscript}%
        {Missing required figure `figures_empirical/figS_allocation_probes_j1.pdf'}%
        {Restore the figure listed in the manuscript export manifest before compiling.}%
    }%
    \caption{%
\textbf{Linear and nonlinear decoding distinguish the accessibility of frozen somatic representations.}
CIFAR-10, MLP-trained 64-soma $[3,3,3,3]$ networks, ten seeds per allocation; probe fits leave the dendritic representation fixed. (A--C) Equal E/I allocations over increasing total contacts: (A) accuracy of the trained MLP readout and of validation-fitted linear (ridge) and nonlinear (MLP) probes; (B) paired nonlinear-minus-linear probe gain; (C) paired shunting-minus-additive difference for each probe and the trained readout. (D--F) The same quantities when a fixed $N_E+N_I=50$ budget is reallocated. Squares, shunting; circles, raw addition; line style identifies the readout or probe (key). Means and pointwise 95\% seed-bootstrap intervals after probe repeats are averaged within seed: whiskers in A,D, with the rules offset horizontally, and bands in B,C,E,F. Chance accuracy lies below the range in A,D.}
    \label{fig:figS15_allocation_probes}
  \end{figure}

Ridge (linear) and MLP (nonlinear) probes were fitted on validation responses to the fixed 64-dimensional somatic outputs of MLP-trained networks (seeds 10--19, full allocation grid) and evaluated on test responses; shuffled validation labels define the null. \suppfig{figS15_allocation_probes} shows equal-allocation slices over increasing budgets (A--C) and the fixed-total slice (D--F).

At a fixed total, linear decoding increasingly favors shunting as E allocation rises, from $-1.0$ points at 5E/45I to 2.6 points at 45E/5I, whereas the trained classifier changes less (D,F). Nonlinear probes improve decoding for both rules, by 0.2 to 1.9 points along this slice, more for shunting at most allocations but not at the E-rich end (E). Across the full grid, the lowest E-contact count favors addition under the linear probe. Better linear decoding therefore coexists with class structure that only a nonlinear decoder reaches. These are performances of specified decoders, not estimates of all class information.

\FloatBarrier

\suppnote{Contact counts and learned synaptic drive}
\label{note:allocation_learned_utilization}

To separate allocated contacts from their learned use, we measured synaptic drives, retained weights and the fraction of outputs in the active range $0.05\leq a\leq0.95$ along $N_E+N_I=50$, keeping dendritic depths separate. The measurements use no labels and the same 2,000 test images for ten trained seeds per rule and allocation.

\begin{figure}[tbp]
    \centering
    \IfFileExists{figures_empirical/figS_learned_utilization_j1.pdf}{%
      \includegraphics[width=165mm]{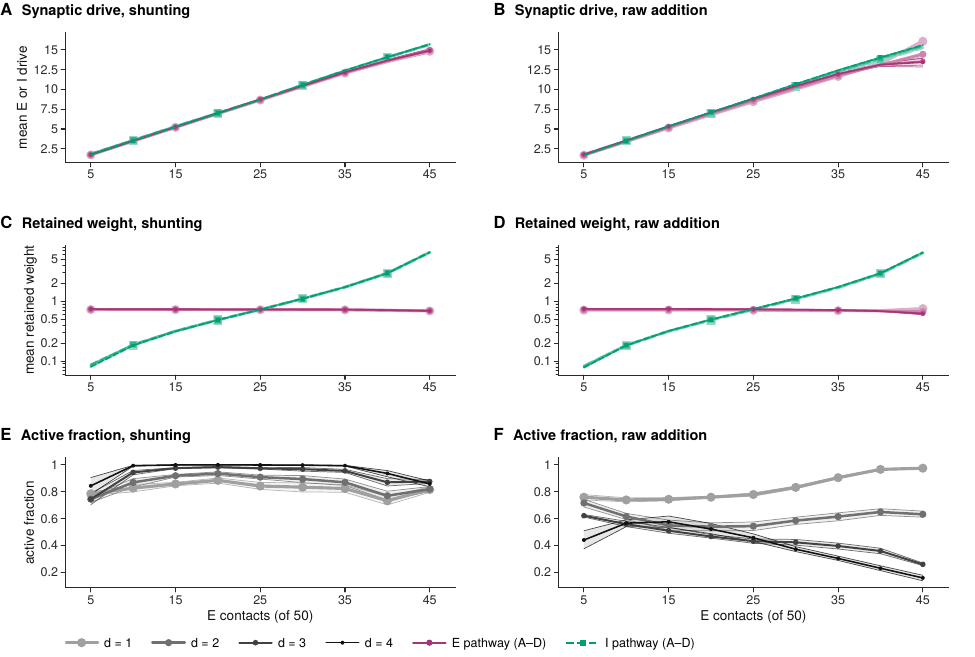}%
    }{%
      \PackageError{empirical-manuscript}%
        {Missing required figure `figures_empirical/figS_learned_utilization_j1.pdf'}%
        {Restore the figure listed in the manuscript export manifest before compiling.}%
    }%
    \caption{%
\textbf{Learned weights and activation state vary along a fixed contact budget.}
CIFAR-10, 64 somas and $[3,3,3,3]$ trees, at nine allocations with $N_E+N_I=50$; left, shunting; right, raw addition. (A,B) Mean E and I drive; (C,D) retained E and I weight; (E,F) fraction of outputs in the intermediate activation range $0.05\leq a\leq0.95$. Magenta solid lines with circles and green dashed lines with squares identify the E and I pathways in A--D; line width, marker size and shade indicate $d=1,\ldots,4$ (widest, largest and lightest for $d=1$). Coincident depths use nested markers, and E/I markers alternate between allocations while lines cover all allocations. C,D use logarithmic ordinates; each row shares one ordinate. Curves and pointwise 95\% intervals summarize ten trained seeds per cell, evaluated without labels on the first 2,000 test images.}
    \label{fig:figS28_allocation_learned_utilization}
  \end{figure}

Inhibitory weights vary strongly with allocation (\suppfig{figS28_allocation_learned_utilization}C,D), whereas mean E and I drives remain comparable (A,B). The analytical initializer anticipates part of this pattern: when both pathways are present it sets $\mathbb E[w_I]=(N_E/N_I)\mathbb E[w_E]$, which equalizes the expected summed drives under its source-activity assumption. Complete initial weights were not retained, so the change from initialization cannot be measured, and the trained profiles do not show that balance emerged through learning. The active fraction differs between rules and depths (E,F): distal additive compartments are active less often as E allocation rises, whereas the proximal profile differs. These measurements separate contact allocation from the fitted operating state; they do not identify occupancy as the cause of the accuracy difference.

\FloatBarrier

\suppnote{Inhibitory contacts and the source of inhibition}
\label{note:inhibition_presence}

The E-only versus direct-E/I control reallocates a fixed budget between the excitatory and inhibitory pathways, with fixed contact locations and matched retained contacts and trainable parameters (Fig.~\ref{fig:fig2}D). Replacing some E contacts by I contacts improves both rules, by 2.8 points (95\% interval, 1.9--3.5) under raw addition and 1.2 points (0.4--1.9) under shunting. Shunting exceeds addition by 2.7 points (2.2--3.2) with 40E/10I and by 4.4 points (3.0--5.4) with 50E/0I; in the learned-contact allocation grid, by contrast, the two rules are comparable at 40E/10I (Fig.~\ref{fig:fig2}B). With $N_I=0$ the direct inhibitory contacts are absent, but shunting still divides by $1+E+G$, so this comparison removes direct inhibition while keeping activity-dependent normalization.

\FloatBarrier

\suppnote{Compartmentalization at limited contact budgets}
\label{note:low_capacity_morphology}

\begin{figure}[tbp]
    \centering
    \IfFileExists{figures_empirical/figS_low_capacity_morphology_j1.pdf}{%
      \includegraphics[width=165mm]{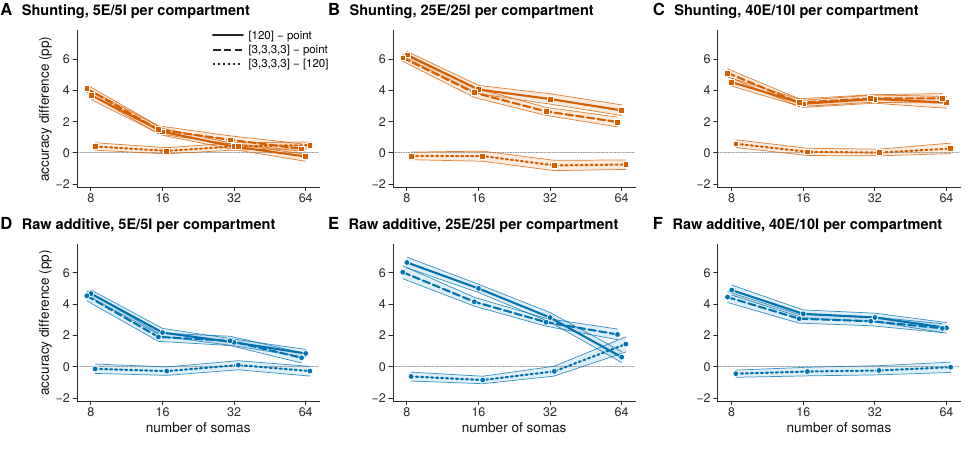}%
    }{%
      \PackageError{empirical-manuscript}%
        {Missing required figure `figures_empirical/figS_low_capacity_morphology_j1.pdf'}%
        {Restore the figure listed in the manuscript export manifest before compiling.}%
    }%
    \caption{%
\textbf{Point/tree comparisons across soma count and contact budget.}
CIFAR-10, MLP readouts. (A--C) Shunting and (D--F) raw addition, at 5E/5I, 25E/25I and 40E/10I contacts per tree compartment. Solid lines show shallow $[120]$ minus point, dashed lines deep $[3,3,3,3]$ minus point, and dotted lines deep minus shallow. Both trees have 120 nonsomatic compartments. Point/tree E and I contacts match separately at the two balanced budgets; only total contacts match at 40E/10I, and candidate parameters are not matched. Curves and pointwise 95\% seed-bootstrap intervals summarize twenty paired seeds per condition (ten original and ten added after inspection) at 8, 16, 32 and 64 somas. Small horizontal offsets separate coincident contrasts; dotted horizontal lines mark zero.}
    \label{fig:figS38_low_capacity_morphology}
  \end{figure}

This comparison varies soma number, from 8 to 64, and the contact budget for one-level $[120]$ and four-level $[3,3,3,3]$ trees and their point controls (\suppfig{figS38_low_capacity_morphology}; A--C shunting, D--F addition). At 5E/5I and 25E/25I per compartment, points match the trees' E and I contacts exactly; at 40E/10I the image size prevents matching E contacts at one unit, so only total contacts match. Candidate parameters are not matched.

Compartmentalization helps most when somas are few: with 8 somas, trees exceed points by 3.7 to 6.7 points at every budget, and the advantage shrinks as somas are added, most at the smallest budget. The two tree shapes differ little: at 5E/5I the deep tree is slightly more accurate under shunting (by 0.1 to 0.5 points), whereas at 25E/25I the shallow tree is more accurate under shunting at 32 and 64 somas. The comparison supports a benefit of distributing limited inputs across compartments, not of serial depth itself, because the two shapes also differ in branching and somatic fan-in.

\FloatBarrier

\suppnote{Resource-matched morphology and spatial routing}
\label{note:resources}

The four morphologies $[120]$, $[10,11]$, $[8,2,6]$ and $[3,3,3,3]$ each have 120 nonsomatic compartments per soma. The fixed-index comparison holds soma number and the counts of E/I contacts, stored weights, coupling parameters and activation parameters fixed. With structured-local routing, accuracy falls from the shallowest to the deepest shape under raw addition (53.4\% to 47.5\%) and changes little under shunting (52.1\% to 51.9\%). Because branching changes together with depth, this is an effect of morphology that cannot be assigned to depth alone. Fixed-index training stores only the retained weights, whereas learned top-$k$ training maintains a larger candidate bank, so resource matching distinguishes active contacts from stored parameters (Supplementary Table~\ref{tab:resources}).

Structured-local and random-local routing use the same window areas at each depth but place them differently: structured windows cover the whole image, whereas random windows leave gaps and overlap more often (Supplementary Table~\ref{tab:routing_window_coverage}). The difference is already present in the one-level tree, which has no serial ancestry. The window layout is fixed across the twenty seeds of each construction, so the intervals describe training variability for those layouts, not variability across layouts. A test of ancestry itself would keep every window and its placement and change only its assignment to the tree.

\begin{table}[htbp]
\centering
\caption{\textbf{Eligible-window coverage in the configured spatial-routing designs.} Values describe all windows at the terminal stage, independently of trained weights. Coverage is the fraction of the $32\times32$ image lying in at least one window; overlap is the mean pixel intersection of two distinct branch windows. Window areas match between routes at each depth. The same geometry is reused across training seeds, so these are design measurements without seed intervals.}
\label{tab:routing_window_coverage}
\small
\begin{tabular}{@{}lrrrr@{}}
\toprule
& \multicolumn{2}{c}{Coverage (\%)} & \multicolumn{2}{c}{Mean overlap (pixels)} \\
Morphology & Structured & Random-local & Structured & Random-local \\
\midrule
$[120]$ & 100 & 81.4 & 0.18 & 0.28 \\
$[10,11]$ & 100 & 90.7 & 0.70 & 0.90 \\
$[8,2,6]$ & 100 & 68.5 & 0.11 & 0.33 \\
$[3,3,3,3]$ & 100 & 77.8 & 0.24 & 0.40 \\
\bottomrule
\end{tabular}
\end{table}

\FloatBarrier

\suppnote{Seed extension of the Figure~\ref{fig:fig2}--\ref{fig:fig4} cohorts}
\label{note:seed_extension}

Four cohorts were extended from ten to twenty training seeds after their first ten results had been inspected: the low-capacity and resource-matched routing cohorts of Fig.~\ref{fig:fig2}F--H, the primary cohort of Figs.~\ref{fig:fig3} and \ref{fig:fig4}B,C, and the matched-morphology cohort of Fig.~\ref{fig:fig4}D--F. Every added run was completed and kept regardless of its result, and the figures show all twenty seeds wherever every compared arm was extended. Supplementary Table~\ref{tab:seed_extension} gives the main contrasts separately for the original, added and pooled seeds. Of the 33 contrasts whose original interval excludes zero, 32 keep their sign with an interval excluding zero in the added seeds, and none reverses. The shunting advantage of the one-level tree under global routing weakens, from 0.68 percentage points in the original seeds to 0.31 ($-0.18$ to 0.79) in the added seeds; pooled, it is 0.49 (0.17--0.81). Two of the three contrasts whose original interval includes zero, the branch-output/soma association at $d=1$ and the clean accuracy of the one-level matched-morphology tree, are positive in the added seeds alone, with means of opposite sign to the original ones; both remain unresolved when pooled. The initialization controls (Supplementary Figs.~\ref{fig:figS_tree_factorial_v5} and \ref{fig:figS34_tree_scalar_information}) and the linear-readout models (Supplementary Figs.~\ref{fig:figS_readout_mechanisms} and \ref{fig:figS_readout_robustness}) keep the ten original seeds, because those cohorts were not extended. For each figure, the original, added and pooled estimates use the same seed-paired summary and bootstrap random stream, with a paired seed bootstrap for differences in transmission (Fig.~\ref{fig:fig3}D); the corruption factorial (Fig.~\ref{fig:fig4}E) and the representation analyses (Fig.~\ref{fig:fig4}F, \suppfig{figS_corruption_protocols}C) were summarized separately for each seed set.

\begin{table}[!htbp]
\centering\scriptsize
\setlength{\tabcolsep}{3pt}
\caption{\textbf{Original, added and pooled estimates of the seed-extended contrasts.} Each entry is the mean seed-paired difference with its pointwise 95\% seed-bootstrap interval, in percentage points (pp) of accuracy or retention unless a unit is given; contrasts are shunting minus addition unless stated otherwise. Original: the first ten seeds; added: the ten seeds added after inspection; pooled: all twenty, as plotted. Positive: seeds with a positive difference among the original and among the added seeds.}
\label{tab:seed_extension}
\begin{tabularx}{\textwidth}{@{}lXccc@{\hspace{4pt}}c@{}}
\toprule
Fig. & Contrast & Original (10) & Added (10) & Pooled (20) & Positive\\
\midrule
\ref{fig:fig2}F & Shallow tree $-$ point, 8 neurons, shunting & $+3.58$ [$+3.19$, $+4.04$] & $+3.81$ [$+3.42$, $+4.20$] & $+3.70$ [$+3.41$, $+4.00$] & 10/10; 10/10\\
\ref{fig:fig2}F & Shallow tree $-$ point, 8 neurons, addition & $+4.73$ [$+4.39$, $+5.06$] & $+4.61$ [$+4.28$, $+4.93$] & $+4.67$ [$+4.43$, $+4.90$] & 10/10; 10/10\\
\ref{fig:fig2}F & Shallow tree $-$ point, 64 neurons, shunting & $-0.19$ [$-0.64$, $+0.28$] & $-0.27$ [$-0.67$, $+0.15$] & $-0.23$ [$-0.53$, $+0.08$] & 4/10; 3/10\\
\ref{fig:fig2}F & Shallow tree $-$ point, 64 neurons, addition & $+0.81$ [$+0.43$, $+1.23$] & $+0.89$ [$+0.59$, $+1.17$] & $+0.85$ [$+0.60$, $+1.11$] & 9/10; 9/10\\
\ref{fig:fig2}G & Shunting $-$ addition, $[3,3,3,3]$, global routing & $+2.88$ [$+2.61$, $+3.18$] & $+2.93$ [$+2.50$, $+3.39$] & $+2.91$ [$+2.65$, $+3.18$] & 10/10; 10/10\\
\ref{fig:fig2}G & Shunting $-$ addition, $[120]$, global routing & $+0.68$ [$+0.30$, $+1.08$] & $+0.31$ [$-0.18$, $+0.79$] & $+0.49$ [$+0.17$, $+0.81$] & 8/10; 7/10\\
\ref{fig:fig2}G & Shunting $-$ addition, $[120]$, structured local & $-1.31$ [$-1.64$, $-0.96$] & $-1.24$ [$-2.12$, $-0.56$] & $-1.28$ [$-1.73$, $-0.89$] & 0/10; 1/10\\
\ref{fig:fig2}H & Structured $-$ random local, $[3,3,3,3]$, shunting & $+0.91$ [$+0.50$, $+1.27$] & $+0.66$ [$+0.32$, $+1.02$] & $+0.78$ [$+0.51$, $+1.04$] & 9/10; 9/10\\
\ref{fig:fig3} & Test accuracy & $+0.48$ [$+0.01$, $+0.95$] & $+0.62$ [$+0.37$, $+0.84$] & $+0.55$ [$+0.27$, $+0.81$] & 7/10; 9/10\\
\ref{fig:fig3}B & $I(E;I\mid Y)$, $d=1$ (bits) & $+0.068$ [$+0.053$, $+0.083$] & $+0.075$ [$+0.059$, $+0.091$] & $+0.072$ [$+0.060$, $+0.083$] & 10/10; 10/10\\
\ref{fig:fig3}B & $I(E;I\mid Y)$, $d=4$ (bits) & $-0.035$ [$-0.045$, $-0.026$] & $-0.022$ [$-0.038$, $-0.006$] & $-0.028$ [$-0.038$, $-0.018$] & 0/10; 3/10\\
\ref{fig:fig3}C & $I(C;a\mid Y)$, $d=2$ (bits) & $+1.127$ [$+1.101$, $+1.150$] & $+1.088$ [$+1.055$, $+1.120$] & $+1.108$ [$+1.084$, $+1.129$] & 10/10; 10/10\\
\ref{fig:fig3}C & $I(a;S\mid Y)$, $d=1$ (bits) & $-0.004$ [$-0.044$, $+0.040$] & $+0.031$ [$+0.009$, $+0.053$] & $+0.013$ [$-0.011$, $+0.039$] & 4/10; 7/10\\
\ref{fig:fig3}C & $I(a;S\mid Y)$, $d=4$ (bits) & $+0.028$ [$+0.024$, $+0.032$] & $+0.039$ [$+0.035$, $+0.043$] & $+0.034$ [$+0.030$, $+0.038$] & 10/10; 10/10\\
\ref{fig:fig3}D & Mean transmission $\partial S/\partial a$, $d=4$ & $+0.024$ [$+0.023$, $+0.026$] & $+0.026$ [$+0.025$, $+0.027$] & $+0.025$ [$+0.024$, $+0.027$] & 10/10; 10/10\\
\ref{fig:fig3}D & Paths below $10^{-6}$, $d=4$ (pp) & $-17.5$ [$-20.0$, $-15.2$] & $-18.5$ [$-20.9$, $-16.3$] & $-18.0$ [$-19.8$, $-16.3$] & 0/10; 0/10\\
\ref{fig:fig3}E & Loss, one output per soma, $d=1$ & $+1.97$ [$+0.94$, $+3.02$] & $+1.63$ [$+1.12$, $+2.22$] & $+1.80$ [$+1.22$, $+2.43$] & 9/10; 10/10\\
\ref{fig:fig3}E & Loss, one third of outputs, $d=3$ & $+4.10$ [$+3.48$, $+4.72$] & $+4.48$ [$+3.93$, $+5.15$] & $+4.29$ [$+3.86$, $+4.77$] & 10/10; 10/10\\
\ref{fig:figS40_whole_layer_ablation}E,F & Loss, all terminal outputs ($d=4$) & $+25.0$ [$+23.4$, $+26.7$] & $+25.0$ [$+24.1$, $+26.0$] & $+25.0$ [$+24.1$, $+26.0$] & 10/10; 10/10\\
\ref{fig:fig3}F & Loss, all E drives, $d=1$ & $-14.64$ [$-15.62$, $-13.54$] & $-14.60$ [$-15.71$, $-13.38$] & $-14.62$ [$-15.38$, $-13.80$] & 0/10; 0/10\\
\ref{fig:fig3}F & Loss, all E drives, $d=4$ & $+23.7$ [$+22.0$, $+25.3$] & $+24.2$ [$+23.2$, $+25.3$] & $+23.9$ [$+22.9$, $+24.9$] & 10/10; 10/10\\
\ref{fig:fig4}B & Retention, Gaussian noise $\sigma=0.3$ & $+12.1$ [$+10.3$, $+13.9$] & $+12.8$ [$+11.7$, $+14.1$] & $+12.5$ [$+11.4$, $+13.5$] & 10/10; 10/10\\
\ref{fig:fig4}C & Retention, scaling $g=0.32$ & $-20.5$ [$-23.1$, $-17.2$] & $-23.3$ [$-25.6$, $-20.9$] & $-21.9$ [$-23.8$, $-19.7$] & 0/10; 0/10\\
\ref{fig:fig4}C & Retention, scaling $g=3.16$ & $-30.7$ [$-36.3$, $-25.2$] & $-26.6$ [$-28.4$, $-24.6$] & $-28.6$ [$-32.0$, $-25.7$] & 0/10; 0/10\\
\ref{fig:fig4}D & Retention, $\sigma=0.3$, $[3,3,3,3]$ & $+13.0$ [$+11.2$, $+14.8$] & $+11.7$ [$+9.9$, $+13.4$] & $+12.4$ [$+11.1$, $+13.6$] & 10/10; 10/10\\
\ref{fig:fig4}D & Retention, $\sigma=0.3$, $[120]$ & $-4.3$ [$-6.0$, $-2.7$] & $-4.5$ [$-6.3$, $-2.1$] & $-4.4$ [$-5.6$, $-3.0$] & 0/10; 1/10\\
\ref{fig:fig4}D & $[3,3,3,3]-[120]$ retention, $\sigma=0.3$, shunting & $+12.1$ [$+9.8$, $+14.1$] & $+11.7$ [$+9.3$, $+13.8$] & $+11.9$ [$+10.2$, $+13.4$] & 10/10; 10/10\\
\ref{fig:fig4}D & $[3,3,3,3]-[120]$ retention, $\sigma=0.3$, addition & $-5.2$ [$-6.4$, $-3.9$] & $-4.5$ [$-5.8$, $-2.7$] & $-4.9$ [$-5.8$, $-3.8$] & 0/10; 1/10\\
\ref{fig:fig4}E & Accuracy, noise $\lambda=1$, $[3,3,3,3]$ & $+4.59$ [$+4.09$, $+5.12$] & $+3.79$ [$+3.15$, $+4.58$] & $+4.19$ [$+3.73$, $+4.68$] & 10/10; 10/10\\
\ref{fig:fig4}E & Accuracy, noise $\lambda=1$, $[120]$ & $-1.65$ [$-2.52$, $-0.88$] & $-1.42$ [$-1.97$, $-0.72$] & $-1.53$ [$-2.06$, $-1.01$] & 1/10; 1/10\\
\ref{fig:fig4}E & Accuracy, gain bound $b=20$, $[3,3,3,3]$ & $-5.54$ [$-6.07$, $-4.98$] & $-7.02$ [$-7.52$, $-6.49$] & $-6.28$ [$-6.77$, $-5.78$] & 0/10; 0/10\\
\ref{fig:fig4}E & Accuracy, gain bound $b=20$, $[120]$ & $-4.79$ [$-5.39$, $-4.22$] & $-5.05$ [$-5.61$, $-4.43$] & $-4.92$ [$-5.35$, $-4.49$] & 0/10; 0/10\\
\ref{fig:fig4}D,E & Clean test accuracy, $[3,3,3,3]$ & $+0.80$ [$+0.23$, $+1.36$] & $+0.60$ [$+0.12$, $+1.05$] & $+0.70$ [$+0.32$, $+1.07$] & 7/10; 7/10\\
\ref{fig:fig4}D,E & Clean test accuracy, $[120]$ & $-0.06$ [$-0.38$, $+0.27$] & $+0.28$ [$+0.06$, $+0.52$] & $+0.11$ [$-0.10$, $+0.32$] & 3/10; 7/10\\
\ref{fig:figS_corruption_protocols}C & Class information retained, $\sigma=0.3$, $[3,3,3,3]$ & $+9.6$ [$+7.7$, $+11.5$] & $+10.2$ [$+7.9$, $+12.4$] & $+9.9$ [$+8.4$, $+11.3$] & 10/10; 10/10\\
\ref{fig:figS_corruption_protocols}C & Class information retained, $b=20$, $[3,3,3,3]$ & $-21.1$ [$-25.0$, $-18.1$] & $-20.9$ [$-22.7$, $-19.1$] & $-21.0$ [$-23.2$, $-19.2$] & 0/10; 0/10\\
\bottomrule
\end{tabularx}
\end{table}
\FloatBarrier

\supppart{Branch associations, transmission, and interventions (Figure~\ref{fig:fig3})}
\label{supp:branches}

Shunting trees transmit distal changes more strongly and depend more on image-specific distal activity, despite clean accuracy comparable to addition (Fig.~\ref{fig:fig3}). The sections below separate association, transmission and classifier dependence, and test each against the intervention, the readout, the initialization, the replacement value and the information estimator.

\suppnote{Complete-stage interventions and replacement references}
\label{note:whole_layer_ablation_reference}

Replacing a fixed number of branch outputs and replacing every branch at one depth affect different fractions of the tree. Replacing one output per soma affects one third of the proximal stage but only one of 81 terminal outputs; replacing one third affects 1, 3, 9 and 27 outputs per soma at depths 1--4 (Fig.~\ref{fig:fig3}E). Replacing a complete stage instead affects from 192 branches at $d=1$ to 5,184 at $d=4$.

\begin{figure}[tbp]
    \centering
    \IfFileExists{figures_empirical/figS_whole_layer_ablation_j1.pdf}{%
      \includegraphics[width=165mm]{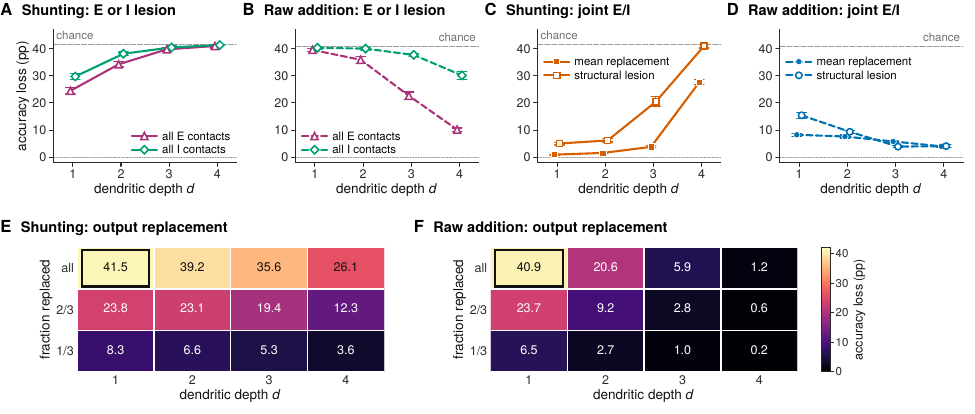}%
    }{%
      \PackageError{empirical-manuscript}%
        {Missing required figure `figures_empirical/figS_whole_layer_ablation_j1.pdf'}%
        {Restore the figure listed in the manuscript export manifest before compiling.}%
    }%
    \caption{%
\textbf{Complete-stage dependence under structural lesion and mean replacement.}
CIFAR-10 primary cohort (64 somas, $[3,3,3,3]$ trees, 40E/10I contacts per compartment, MLP readouts; seeds 0--19, twenty networks per rule, all 10,000 test images). (A,B) Accuracy loss after lesioning all E (magenta triangles) or all I (green diamonds) contacts at one branch depth, under shunting (A, solid lines) and raw addition (B, dashed lines). (C,D) Joint E/I-drive replacement by the training mean (filled) versus structural lesion (open); inherited child input is not targeted. Grey dashed lines mark the loss at chance accuracy. (E,F) Training-mean replacement of one third, two thirds or all branch outputs at each depth; cells print mean accuracy losses in percentage points on a common 0--42 color scale, and the outlined full-proximal cell gives image-independent logits by construction. A--D show pointwise 95\% seed-bootstrap intervals; E,F average three common branch-ordering draws within seed before averaging networks. Negative accuracy losses are kept. The number of affected branches grows with depth in both complete-stage and fixed-fraction interventions.}
    \label{fig:figS40_whole_layer_ablation}
  \end{figure}

\begin{figure}[tbp]
    \centering
    \IfFileExists{figures_empirical/figS_readout_mechanisms_j1.pdf}{%
      \includegraphics[width=165mm]{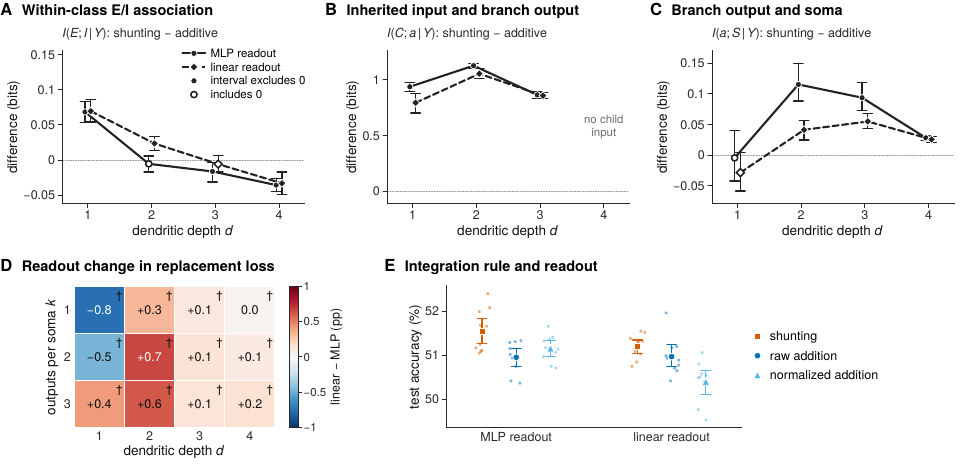}%
    }{%
      \PackageError{empirical-manuscript}%
        {Missing required figure `figures_empirical/figS_readout_mechanisms_j1.pdf'}%
        {Restore the figure listed in the manuscript export manifest before compiling.}%
    }%
    \caption{%
\textbf{Branch relationships and replacement effects recur with a linear readout.}
CIFAR-10, 64 somas, $[3,3,3,3]$ trees and 40E/10I contacts per compartment. (A--C) Shunting-minus-additive differences in corrected within-class E/I association, inherited-input/output association and branch-output/soma association under the MLP (circles, solid lines) and linear (diamonds, dashed lines) readouts; filled symbols mark pointwise 95\% intervals that exclude zero and open symbols intervals that include it. (D) Change between readouts in the shunting-minus-additive accuracy-loss contrast after replacing $k$ branch outputs per soma with their mean responses over training images; positive values mean a larger difference between rules under the linear readout, and daggers mark cells whose pointwise 95\% interval includes zero (every cell). (E) Test accuracy of shunting (squares), raw addition (circles) and conductance-normalized addition (triangles); small points show individual seeds, larger symbols and bars means and pointwise 95\% seed-bootstrap intervals. A--D use seeds 0--9 of the primary cohort under both readouts; E pairs seeds 10--19 of the 40E/10I allocation cells and the normalized-addition cohort across readouts.}
    \label{fig:figS_readout_mechanisms}
  \end{figure}

\begin{figure}[tbp]
    \centering
    \IfFileExists{figures_empirical/figS_tree_factorial_j1.pdf}{%
      \includegraphics[width=165mm]{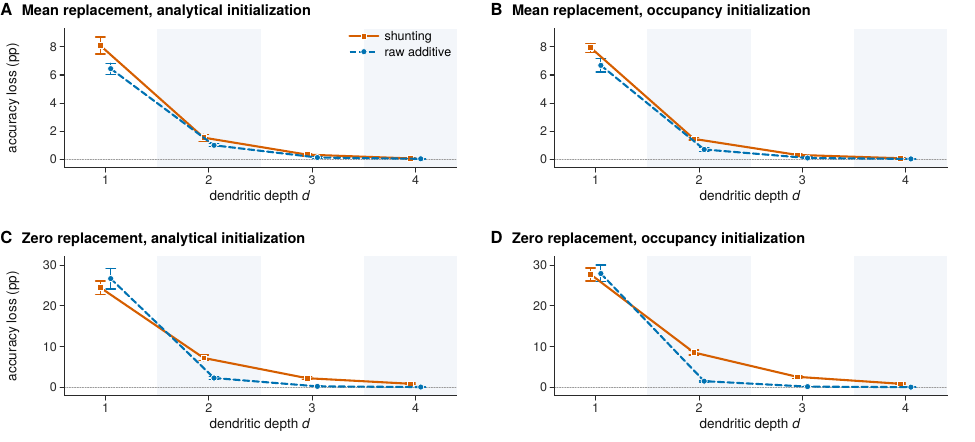}%
    }{%
      \PackageError{empirical-manuscript}%
        {Missing required figure `figures_empirical/figS_tree_factorial_j1.pdf'}%
        {Restore the figure listed in the manuscript export manifest before compiling.}%
    }%
    \caption{%
\textbf{Proximal dependence persists across initialization and replacement value.}
CIFAR-10 primary architecture (64 somas, $[3,3,3,3]$ trees, 40E/10I contacts per compartment, MLP readouts) trained with a shared initialization for both rules. (A,B) Accuracy loss after replacing selected branch outputs by their mean response over training images, under shared analytical (A) and shared occupancy (B) initialization; (C,D) the same branches replaced by zero. Five fixed selections each replace one nested soma-to-distal path per soma; interventions use all 10,000 test images. Squares and solid lines, shunting; circles and dashed lines, raw addition. Means and pointwise 95\% intervals over ten trained seeds per condition, after the selections are averaged within seed; each row has its own ordinate, and negative accuracy losses are kept.}
    \label{fig:figS_tree_factorial_v5}
  \end{figure}

\begin{figure}[tbp]
    \centering
    \IfFileExists{figures_empirical/figS_scalar_branch_information_j1.pdf}{%
      \includegraphics[width=165mm]{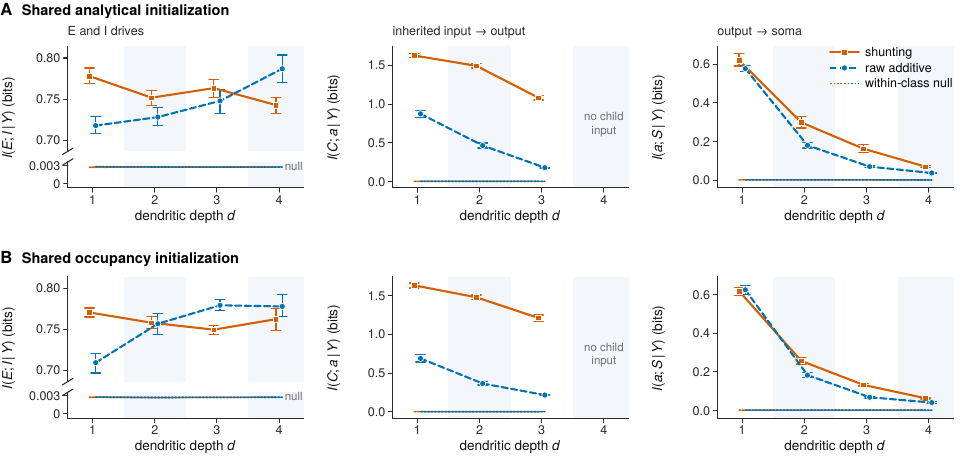}%
    }{%
      \PackageError{empirical-manuscript}%
        {Missing required figure `figures_empirical/figS_scalar_branch_information_j1.pdf'}%
        {Restore the figure listed in the manuscript export manifest before compiling.}%
    }%
    \caption{%
\textbf{Within-class branch associations persist under shared initialization.}
CIFAR-10 primary architecture (64 somas, $[3,3,3,3]$ trees, 40E/10I contacts per compartment, MLP readouts). (A) Shared analytical and (B) shared occupancy initialization. Each row shows $I(E;I\mid Y)$, $I(C;a\mid Y)$ and $I(a;S\mid Y)$, with shunting (squares, solid) and raw addition (circles, dashed) overlaid; $S$ is the owning soma's output. Dotted lines show within-class permutation-null means. Bars (bands for the nulls) are pointwise 95\% intervals over the ten original seeds after branchwise estimation and averaging; these controls were not extended. Each measure has its own ordinate, shared by A and B; the $I(E;I\mid Y)$ ordinate is broken, with its nulls in the lower strip, and inherited-input curves stop before terminal branches. The profiles of the primary cohort appear in Fig.~\ref{fig:fig3}B,C.}
    \label{fig:figS34_tree_scalar_information}
  \end{figure}

Training-mean replacement removes image-specific variation while keeping each drive's mean over optimization images; a structural lesion sets the targeted effective synaptic weights to zero (\suppfig{figS40_whole_layer_ablation}A,B). Joint E/I interventions target both local drives and leave the inherited child input intact (C,D). Both operations can produce states outside the training distribution. Complete-stage loss increases distally under shunting, whereas additive loss is smallest at the terminal stage; magnitudes depend strongly on the replacement value. These measure collective dependence, not the importance of single branches or an additive partition of accuracy.

Replacing a fixed fraction of outputs keeps the proximal concentration while revealing greater intermediate and distal dependence under shunting (E,F, all replaced fractions). Replacing all proximal outputs makes the logits constant across images, because somas receive no direct image input, so this extreme reflects the architecture as well as the fitted classifier. Replacing every output at one depth and replacing every local E/I drive at that depth target different variables and need not share a depth profile.

\FloatBarrier

\suppnote{Readout dependence of branch measurements}
\label{note:readout_mechanisms}

The main association profiles and the strong proximal replacement effect recur with a linear readout. These models retrain the primary 64-soma $[3,3,3,3]$ architecture at 40E/10I with one affine map from the somatic outputs to the class scores, on seeds 0--9 and the common recipe of the MLP-trained networks of Fig.~\ref{fig:fig3}. Under the linear readout (\suppfig{figS_readout_mechanisms}A--C), within-class E/I association is greater under shunting proximally and under addition at $d=4$, inherited-input association is greater under shunting at every depth that has an inherited input, and branch-to-soma association is greater under shunting beyond the proximal level. Information is compared between rules within each readout; replacement effects pair the readouts within seed. Replacing one proximal output per soma costs shunting more than addition under both readouts, by 1.15 percentage points with the linear readout (95\% interval, 0.28--2.04) and 1.91 with the MLP (0.96--2.90); this control keeps its nested-path branch selection (Methods), so the MLP value differs slightly from the Fig.~\ref{fig:fig3}E estimate in Supplementary Table~\ref{tab:seed_extension}. For one proximal output, the linear-minus-MLP change in this contrast is $-0.76$ ($-2.16$ to 0.71), and no change between readouts excludes zero at any depth or replacement count (D). Both readouts thus show the greater loss under shunting; intervals that include zero do not show that the readouts are equivalent.

\suppfig{figS_readout_mechanisms}E compares shunting, raw addition and conductance-normalized addition under both readouts in the 40E/10I cells of the allocation cohorts and the normalized-addition cohort, pairing seeds 10--19 across readouts. All three rules use learning rate 0.002 and at most 2,000 epochs, with patience 400 for normalized addition (which keeps learned nonlinear activations) and 150 for the other rules; settings match across readouts within each rule. Shunting has the highest mean accuracy under both readouts (51.6\% with the MLP, 51.2\% with the linear readout); normalized addition is intermediate with the MLP (51.2\%) but lowest with the linear readout (50.4\%). The comparison describes readout dependence under these optimization budgets.

\FloatBarrier

\suppnote{Initialization and replacement reference}
\label{note:tree_factorial_v5}

Branch-output replacement was repeated under shared analytical and shared occupancy initialization. Five label-independent selections each choose one nested soma-to-distal path per neuron; at a given depth, the selected outputs are replaced by their mean responses over training images (\suppfig{figS_tree_factorial_v5}A,B) or by zero (C,D). Results are averaged over the five selections within each of ten trained seeds per condition (these controls keep the original ten seeds).

Proximal replacement remains the most disruptive under both initializations, but some differences between rules change with initialization and replacement value. The proximal location of the strongest dependence is therefore robust, whereas the ordering of the two rules is not. A shared initialization policy does not give identical initial weights.

\FloatBarrier

\suppnote{Initialization and permutation controls for branch associations}
\label{note:tree_scalar_information_binned}

This control and the next test whether the branch associations of Fig.~\ref{fig:fig3}B,C depend on initialization or on the estimator. Within-class associations are estimated separately for one branch from each soma at every depth and averaged within trained seed. Shuffling the second signal within each image class keeps its class-conditional distribution while removing the pairing between examples. The primary analysis uses all 10,000 validation images, 64 branches per depth, ten quantile bins with the Miller--Madow correction, and 20 permutations.

Under both shared initializations, inherited-input/output association remains stronger under shunting at every depth, as does branch-output/soma association beyond the proximal level. Within-class E/I association depends on depth, as in the primary cohort: it is stronger under shunting proximally and under raw addition distally ($d=4$ with analytical and $d=3$ with occupancy initialization; \suppfig{figS34_tree_scalar_information}A,B). These are statistical dependencies within class, not causal contributions or a decomposition into unique, redundant and synergistic information.

\FloatBarrier

\suppnote{Sensitivity to the information estimator}
\label{note:tree_information_estimator_sensitivity}

The three association profiles and the joint E/I-to-soma association were recomputed with two estimators on identical recorded responses of the forty primary-cohort networks (twenty paired seeds, 0--19, of which 10--19 were added after inspection): ten-bin Miller--Madow estimates and nearest-neighbor estimates at $k=10$. Both use the same 3,000 validation images, 16 branches from distinct somas at each depth, label-free coordinate standardization and five within-class permutations; branch estimates are averaged within checkpoint.

\begin{figure}[tbp]
    \centering
    \IfFileExists{figures_empirical/figS_information_estimator_j1.pdf}{%
      \includegraphics[width=165mm]{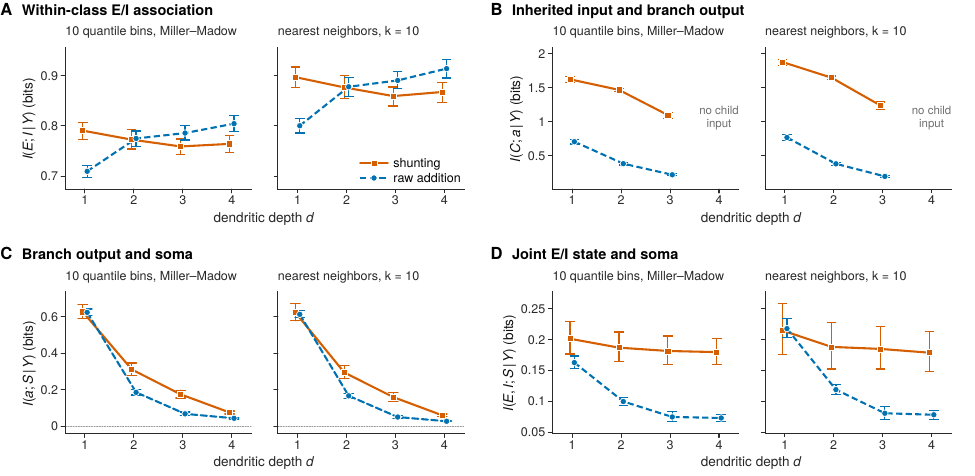}%
    }{%
      \PackageError{empirical-manuscript}%
        {Missing required figure `figures_empirical/figS_information_estimator_j1.pdf'}%
        {Restore the figure listed in the manuscript export manifest before compiling.}%
    }%
    \caption{%
\textbf{Branch-association orderings agree across estimators on identical responses.}
CIFAR-10 primary cohort (64 somas, $[3,3,3,3]$ trees, 40E/10I contacts per compartment, MLP readouts). (A) Within-class E/I association; (B) inherited-input/output association; (C) branch-output/soma association; (D) joint E/I-to-soma association. Left facets use ten quantile bins with the Miller--Madow correction~\citep{Miller1955}, right facets nearest neighbors at $k=10$~\citep{Kraskov2004}; squares and solid lines, shunting; circles and dashed lines, raw addition. Both estimators use the same 3,000 images, 16 branches per depth and five within-class permutations, and values subtract each estimator's permutation-null mean. Means and pointwise 95\% intervals over twenty trained seeds per rule after branch averaging. Terminal branches have no inherited input, so B ends at $d=3$. This common-data check is distinct from the larger binned analyses of Fig.~\ref{fig:fig3}B,C and the initialization controls of \suppfig{figS34_tree_scalar_information}.}
    \label{fig:figS41_information_estimator_sensitivity}
  \end{figure}

Both estimators give the same pattern (\suppfig{figS41_information_estimator_sensitivity}). Within-class E/I association is greater under shunting at $d=1$ and under raw addition at $d=4$, with overlapping intervals at intermediate depths (A). Inherited-input/output association is greater under shunting at every depth (B), branch-output/soma association beyond the proximal level (C), and joint E/I-to-soma association from $d=2$ onward, and at $d=1$ for the binned estimate (D). Absolute magnitudes differ, so agreement in ordering is not agreement in the amount of information. Agreement on these low-dimensional relationships neither validates high-dimensional population estimates nor decomposes information into unique, redundant and synergistic parts.

\FloatBarrier

\supppart{Input perturbations and robustness (Figure~\ref{fig:fig4})}
\label{supp:perturbations}

Shunting better tolerates pixel noise in deeper clean-trained trees but is more sensitive to strong shared input scaling, and one-level trees favor addition under noise (Fig.~\ref{fig:fig4}). The sections below separate independent pixel noise, deterministic scaling and random image gain. The primary 40E/10I cohort, the resource-matched morphology cohort and a separate 25E/25I inhibitory-source cohort (Supplementary Note~\ref{supp:networks}) are each analyzed on their own, never pooled.

\suppnote{Morphology under Gaussian noise, uniform noise, and scaling}
\label{note:morphology_noise}
\label{note:uniform_noise}

The four morphologies $[120]$, $[10,11]$, $[8,2,6]$ and $[3,3,3,3]$ each have 120 nonsomatic compartments per soma, 64 somas and 40E/10I contacts per compartment, so compartment count, active contacts and trainable parameters match while depth and branching pattern vary together. Each rule and morphology has twenty trained seeds (21000--21019, of which 21010--21019 were added after inspection).

Independent Gaussian or uniform pixel noise is added before inputs are clipped to $[0,1]$; global scaling multiplies the image by a deterministic factor without clipping. Retention $R=(A_{\rm perturbed}-0.1)/(A_{\rm clean}-0.1)$ is computed for each checkpoint, and five common noise draws are averaged within seed.

\begin{figure}[tbp]
    \centering
    \IfFileExists{figures_empirical/figS_morphology_noise_gain_j1.pdf}{%
      \includegraphics[width=165mm]{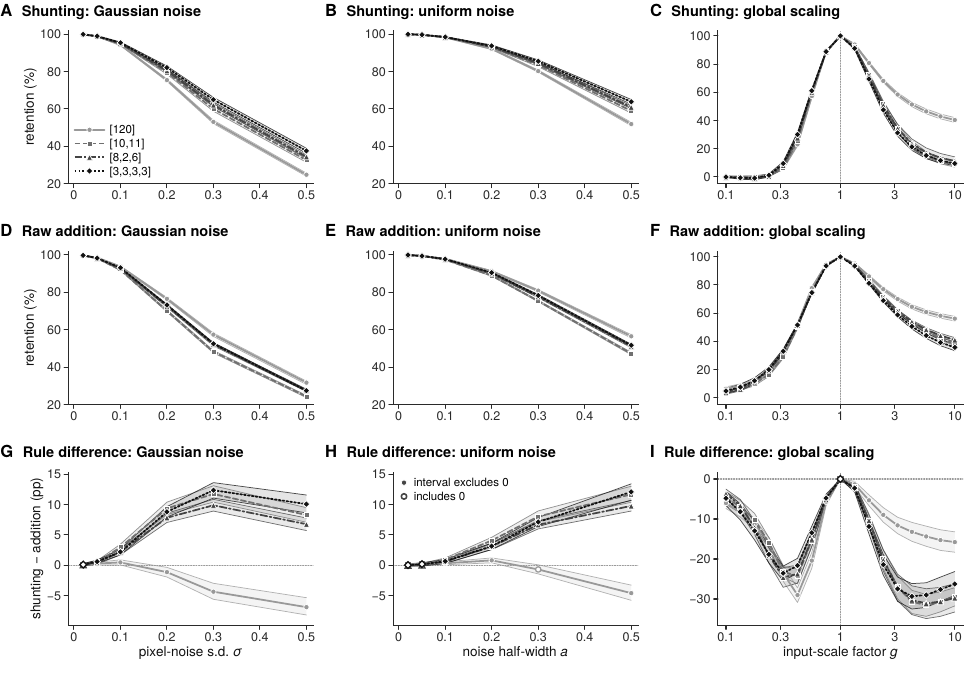}%
    }{%
      \PackageError{empirical-manuscript}%
        {Missing required figure `figures_empirical/figS_morphology_noise_gain_j1.pdf'}%
        {Restore the figure listed in the manuscript export manifest before compiling.}%
    }%
    \caption{%
\textbf{Resource-matched morphology under three input perturbations.}
CIFAR-10, MLP readouts; all four shapes have 120 nonsomatic compartments per soma, 64 somas and 40E/10I contacts per compartment, with matched trainable parameters. (A--C) Shunting retention under Gaussian noise, uniform noise $\mathcal U[-a,a]$ and global scaling $x'=gx$; (D--F) the same for raw addition; (G--I) paired shunting-minus-additive retention differences. Retention is $100R$, with $R=(A_{\rm perturbed}-0.1)/(A_{\rm clean}-0.1)$, and differences are in percentage points. Morphologies differ in grey level, marker and line style (key in A); in G--I, open symbols mark pointwise 95\% intervals that include zero. Curves and pointwise 95\% intervals summarize twenty trained seeds; Gaussian and uniform results average five common draws within seed. Dotted lines mark zero difference and the unscaled input $g=1$.}
    \label{fig:figS_morphology_noise}
  \end{figure}

\begin{figure}[tbp]
    \centering
    \IfFileExists{figures_empirical/figS_readout_robustness_j1.pdf}{%
      \includegraphics[width=165mm]{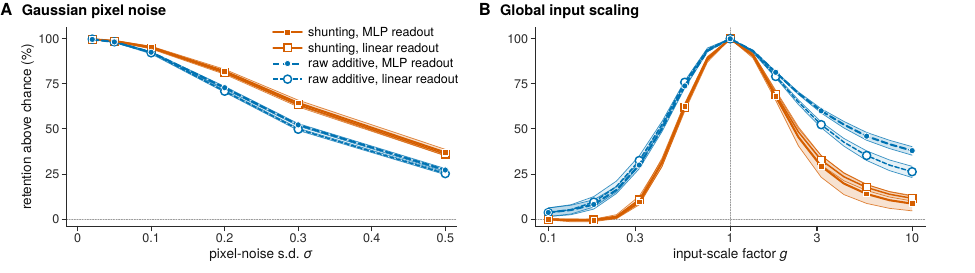}%
    }{%
      \PackageError{empirical-manuscript}%
        {Missing required figure `figures_empirical/figS_readout_robustness_j1.pdf'}%
        {Restore the figure listed in the manuscript export manifest before compiling.}%
    }%
    \caption{%
\textbf{The primary noise-versus-scaling ordering recurs with a linear readout.}
CIFAR-10 networks with 64 somas, $[3,3,3,3]$ trees and 40E/10I contacts per compartment. (A) Above-chance retention, $100R$, under Gaussian pixel noise; (B) under global input scaling. MLP and linear readouts are trained jointly with each dendritic network under the common recipe, pairing seeds 0--9. Filled markers and heavier lines show MLP readouts, open markers and lighter lines linear readouts; color, marker shape and line style (solid, shunting; dashed, raw addition) distinguish the integration rules. Curves and pointwise 95\% seed-bootstrap intervals. The full morphology grid was not repeated.}
    \label{fig:figS_readout_robustness}
  \end{figure}

\begin{figure}[tbp]
    \centering
    \IfFileExists{figures_empirical/figS_corruption_protocols_v14.pdf}{%
      \includegraphics[width=165mm]{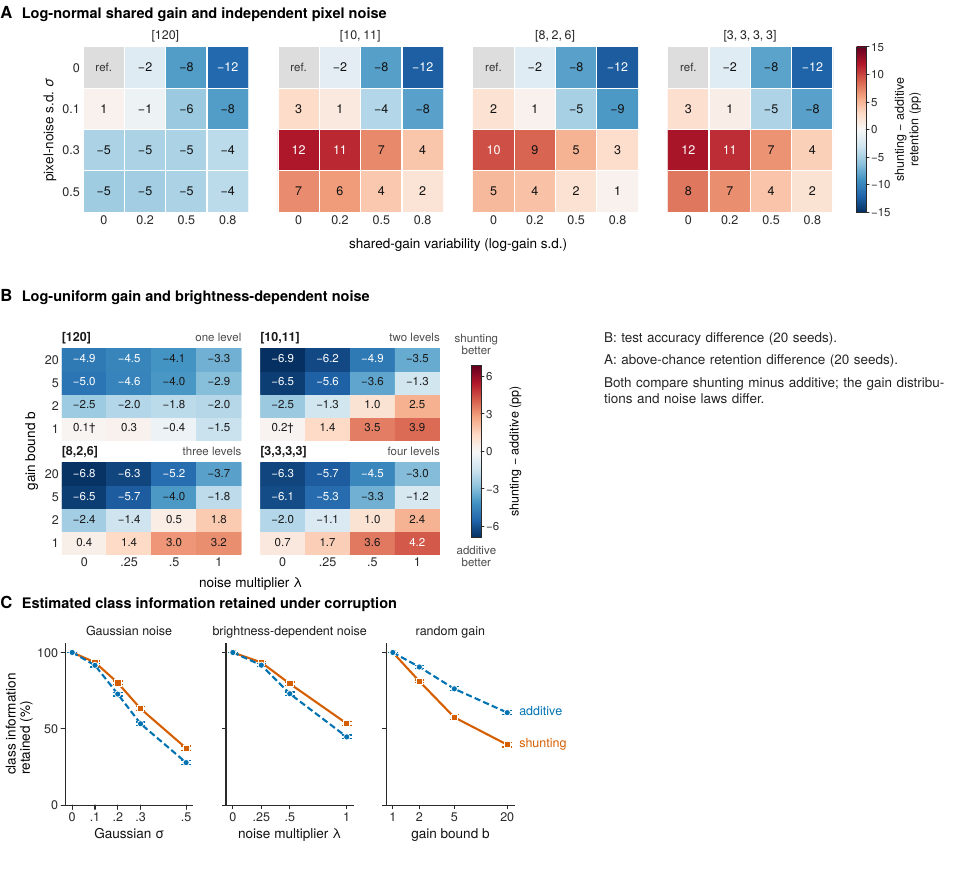}%
    }{%
      \PackageError{empirical-manuscript}%
        {Missing required figure `figures_empirical/figS_corruption_protocols_v14.pdf'}%
        {Restore the figure listed in the manuscript export manifest before compiling.}%
    }%
    \caption{%
\textbf{Full corruption surfaces and estimated class-information retention.}
(A) Shunting-minus-additive above-chance retention under mean-one log-normal gain and independent pixel noise; each map is labeled by its tree's branching factors. The four morphologies have equal resources. Each cell summarizes twenty seeds and five draws. Gray cells mark the unperturbed reference. (B) Shunting-minus-additive accuracy under log-uniform gain and brightness-dependent noise; twenty seeds, five draws. Daggers mark pointwise 95\% intervals including zero. (C) Estimated class-information retention of four-level soma outputs under Gaussian noise, brightness-dependent noise and log-uniform gain: twenty seeds, 3,000 test images and 20 label shuffles. Squares and solid lines identify shunting; circles and dashed lines addition. Seed-bootstrap intervals quantify between-network variation and do not bound estimator bias.}
    \label{fig:figS_corruption_protocols}
  \end{figure}

Relative to the one-level tree, the deepest morphology retains more performance under Gaussian and uniform noise for shunting but less for addition (\suppfig{figS_morphology_noise}A,B,D,E). Strong global scaling instead lowers retention in the deeper shape for both rules (C,F). The paired shunting-minus-additive contrasts show the noise advantage in serial trees and the scaling disadvantage (G--I); uniform noise (middle column) gives the same qualitative pattern as Gaussian noise. These findings concern the tested morphologies and input distributions, not an isolated effect of depth or general invariance to gain.

\FloatBarrier

\suppnote{Perturbation responses with a linear readout}
\label{note:readout_robustness}

The primary 64-soma $[3,3,3,3]$ architecture was retrained with a linear readout at 40E/10I under the common recipe, pairing seeds 0--9 with their MLP counterparts. Under both readouts, shunting retains more above-chance accuracy under strong independent noise (\suppfig{figS_readout_robustness}A), whereas addition retains more under strong global scaling (B). With the linear readout, the shunting-minus-additive retention difference is 13.6 percentage points at $\sigma=0.3$ (95\% interval, 12.4--14.7) and $-19.6$ at $g=3.16$ ($-22.2$ to $-17.2$). The linear readout does not change the noise difference beyond its interval and reduces the additive advantage under strong brightening (linear minus MLP at $g=3.16$: 11.1; 5.8--16.3). The perturbation ordering thus holds under both readouts; the full morphology-by-perturbation experiment was not repeated.

\FloatBarrier

\suppnote{Two protocols combining gain and noise}
\label{note:corruption_protocols}

Two protocols combine image gain with noise (Methods). The first draws a mean-one log-normal gain per image and adds independent pixel noise; the second draws a log-uniform gain and adds brightness-dependent noise. Both set negative perturbed pixels to zero without an upper cap. \suppfig{figS_corruption_protocols} shows the full shunting-minus-additive surfaces of the first protocol for the four morphologies (A; next section) and the full surface of the second (B), which complements Fig.~\ref{fig:fig4}E; together they show where the two protocols agree. Panel C estimates the class information that four-level soma outputs retain: shunting retains more under Gaussian noise ($\sigma=0.3$; +9.9 points) and less under random gain ($b=20$; $-21.0$ points; Supplementary Table~\ref{tab:seed_extension}).

\FloatBarrier

\suppnote{Shared random gain and independent pixel noise}
\label{note:gain_load_morphology}

The first protocol draws one positive, mean-one gain per image and independent Gaussian noise per pixel:
\[
\log g_n\sim\mathcal N(-\sigma_g^2/2,\sigma_g^2),\qquad
\eta_{np}\sim\mathcal N(0,\sigma_{\rm pix}^2),\qquad
x'_{np}=\max(0,g_nx_{np}+\eta_{np}).
\]
The grid crosses $\sigma_g\in\{0,0.2,0.5,0.8\}$ with $\sigma_{\rm pix}\in\{0,0.1,0.3,0.5\}$; each cell averages five common draws within each of twenty seeds per rule and morphology.

\begin{figure}[tbp]
    \centering
    \IfFileExists{figures_empirical/figS_gain_load_morphology_j1.pdf}{%
      \includegraphics[width=165mm]{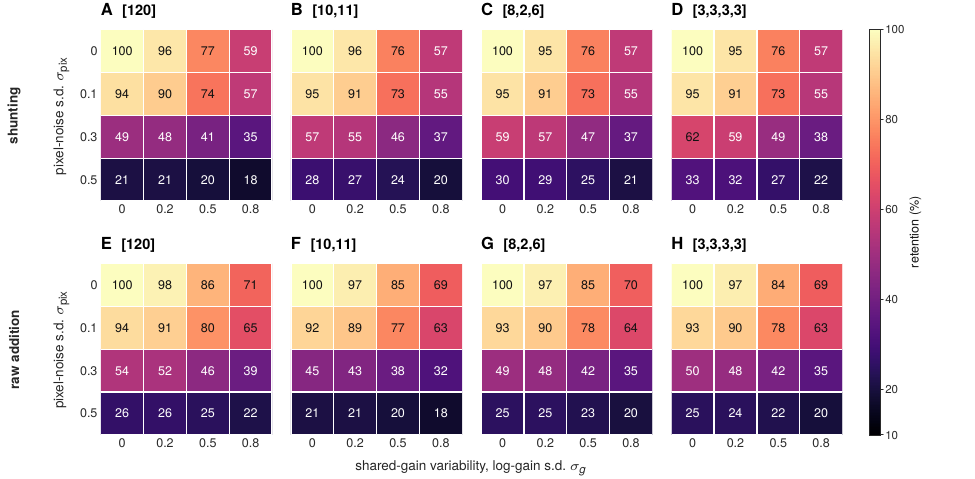}%
    }{%
      \PackageError{empirical-manuscript}%
        {Missing required figure `figures_empirical/figS_gain_load_morphology_j1.pdf'}%
        {Restore the figure listed in the manuscript export manifest before compiling.}%
    }%
    \caption{%
\textbf{Joint dependence on morphology, shared gain and pixel noise.}
CIFAR-10 resource-matched morphologies: 120 compartments per soma, 64 somas, 40E/10I contacts per compartment and equal trainable parameters. Columns show $[120]$, $[10,11]$, $[8,2,6]$ and $[3,3,3,3]$ trees; (A--D) shunting and (E--H) additive retention, $100R$. Axes vary the log-gain s.d. $\sigma_g$ (gain shared within each image, mean one) and the pixel-noise s.d. $\sigma_{\rm pix}$. Cells print means over twenty trained seeds, each averaging five common draws; intervals are not shown here, and the paired differences appear in \suppfig{figS_corruption_protocols}A.}
    \label{fig:figS_gain_load_morphology}
  \end{figure}

Shared gain alone favors addition in every tested morphology. Independent pixel noise changes the ordering in the deeper trees, whereas the one-level tree still favors addition at the stronger noise levels, $\sigma_{\rm pix}\ge 0.3$. \suppfig{figS_gain_load_morphology} shows the absolute retention behind these differences (\suppfig{figS_corruption_protocols}A) for shunting (A--D) and addition (E--H). Pixel noise alters both E and I drives; it is not an isolated injection of inhibitory conductance.

\FloatBarrier

\supppart{Point neurons and input pooling (Figure~\ref{fig:point_neurons})}
\label{supp:point_neurons}

In the primary and resource-matched morphology cohorts, compartmentalization helps shunting more than addition on clean images, while broader input pooling can improve robustness under strong corruption (Fig.~\ref{fig:point_neurons}). This note adds the point controls of the allocation grid and the accuracies of trees and points in every cohort, including multilayer and contextual networks. Points reuse the training settings of the corresponding trees and were not tuned separately, so these comparisons do not estimate the best attainable point performance.

\suppnote{Point controls of the allocation grid}
\label{note:allocation_points}

Soma-only point controls were trained for 29 depth-four allocations and five linear-readout allocations (seeds 10--19, recipe A). Two quantities must be kept apart (Supplementary Table~\ref{tab:allocation_points}): the difference between rules within trees, and the interaction, the trees' difference minus the points', which asks whether shunting gains more from compartments than addition does. Within trees, shunting is more accurate than addition only at excitation-dominated cells. The interaction is positive at 15 of the 29 cells, including some inhibition-rich cells such as 5E/25I, and negative at five, including the single-E cells. Within points, shunting is less accurate than addition at almost every cell; the exception is 50E/50I, where the additive point is least accurate (44.8\%), shunting points exceed additive points by 3.4 points and the interaction reverses. The clean-image interaction of Fig.~\ref{fig:point_neurons} is therefore not general across E/I allocations. Some points were trained on a different GPU family from their trees or from the other rule's runs: at 18 of the 34 cells, three of ten seeds mix H100 and RTX GPUs. Restricting the interaction to the hardware-matched seeds changes it by at most 0.6 points (median 0.1); at 40E/10I it is 4.37 points with all ten seeds and 4.38 with the seven matched seeds.

\begin{table}[htbp]
\centering
\caption{\textbf{Shunting-minus-additive accuracy in trees and capped point neurons at selected allocations.} Depth-four $[3,3,3,3]$ trees and their soma-only point controls, recipe A, MLP readouts, seeds 10--19. Means and pointwise 95\% seed-bootstrap intervals of the paired differences (percentage points); the interaction is the tree's difference minus the point's.}
\label{tab:allocation_points}
\small
\begin{tabular}{@{}lccc@{}}
\toprule
Allocation & Trees & Points & Interaction \\
\midrule
1E/1I & $-3.80\ [-4.14,-3.48]$ & $-0.72\ [-1.19,-0.31]$ & $-3.08\ [-3.58,-2.58]$ \\
5E/25I & $-1.01\ [-1.56,-0.49]$ & $-2.80\ [-3.14,-2.46]$ & $1.79\ [1.07,2.52]$ \\
25E/5I & $0.61\ [0.24,0.95]$ & $-3.64\ [-4.04,-3.20]$ & $4.25\ [3.70,4.76]$ \\
40E/10I & $0.59\ [0.32,0.85]$ & $-3.77\ [-4.13,-3.41]$ & $4.37\ [3.97,4.74]$ \\
45E/5I & $0.56\ [0.17,1.05]$ & $-3.51\ [-3.86,-3.19]$ & $4.07\ [3.56,4.77]$ \\
50E/50I & $-0.66\ [-1.03,-0.30]$ & $3.37\ [2.88,3.84]$ & $-4.03\ [-4.71,-3.36]$ \\
\bottomrule
\end{tabular}
\end{table}

\FloatBarrier

\suppnote{Accuracy of trees and point neurons}
\label{note:three_arms}

\begin{figure}[tbp]
    \centering
    \IfFileExists{figures_empirical/figS_three_arms_j1.pdf}{%
      \includegraphics[width=165mm]{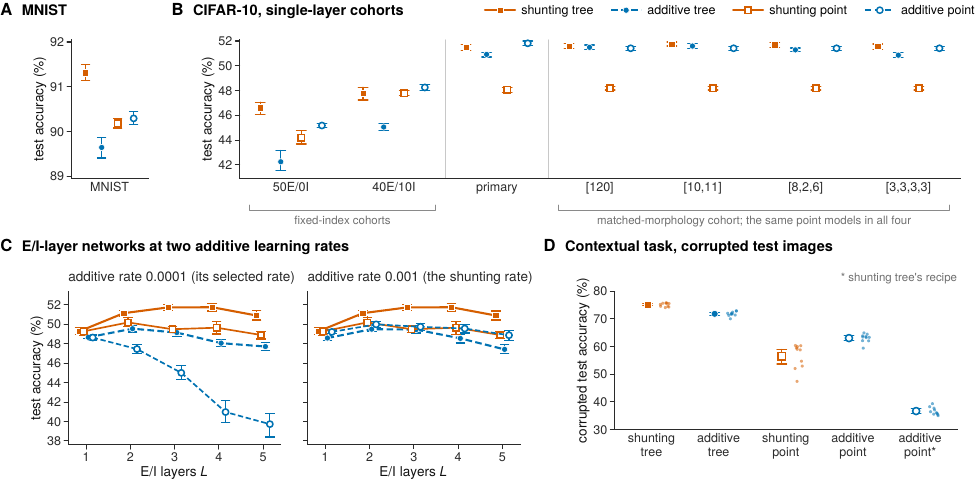}%
    }{%
      \PackageError{empirical-manuscript}%
        {Missing required figure `figures_empirical/figS_three_arms_j1.pdf'}%
        {Restore the figure listed in the manuscript export manifest before compiling.}%
    }%
    \caption{%
\textbf{Accuracy of shunting and additive trees and point neurons.}
Squares and solid lines denote shunting, circles and dashed lines addition; filled markers denote trees and open markers point neurons; arms are offset horizontally. (A) MNIST identity classifier, ten seeds. (B) Single-layer CIFAR-10 cohorts: fixed-index 50E/0I and 40E/10I, the primary cohort and the four resource-matched morphologies, all trained with the common protocol (recipe A); the points beside the four morphologies are the same models. Ten seeds for the fixed-index cohorts and 20 otherwise. (C) Networks of $L$ E/I layers with the additive arms at their selected learning rate ($10^{-4}$, left) and at the shunting rate ($10^{-3}$, right); the shunting arms are the same in both panels; ten seeds. (D) Context-dependent task, accuracy on corrupted test images; dots are the ten seeds, and the additive point marked * uses the shunting tree's recipe. Means with pointwise 95\% seed-bootstrap intervals.}
    \label{fig:figS_three_arms}
  \end{figure}

In the single-layer cohorts, the additive point was as accurate as the additive tree or more accurate (MNIST, 90.3\% against 89.6\%; primary CIFAR-10 cohort, 51.8\% against 50.9\%), and in the primary and resource-matched morphology cohorts the shunting point was the least accurate of the four arms (48.1--48.2\%; \suppfig{figS_three_arms}A,B). The interaction, the trees' shunting-minus-additive contrast minus the points', is 4.3 percentage points (95\% interval, 3.9--4.7) in the primary cohort and is positive in every seed of the resource-matched morphology family; Supplementary Table~\ref{tab:point_interactions} gives the component contrasts of every cohort. In the fixed-index cohorts, whose contacts are fixed before training, the shunting point matched the shunting tree with direct inhibition (47.8\% each) and was 2.4 points below it without.

In networks of E/I layers, additive point networks trained at the additive trees' selected learning rate ($10^{-4}$) lost accuracy with depth, from 48.6\% at $L=1$ to 39.7\% at $L=5$, whereas at $10^{-3}$ they reached 48.9\% at $L=5$ (\suppfig{figS_three_arms}C). This decline with depth depends on the learning rate and does not show an intrinsic limitation of point neurons. Shunting point networks reached 48.9--50.2\% at every depth, 2.0 to 2.2 points below the shunting tree networks at $L=3$--5.

In the context-dependent task, the additive point trained with the shunting tree's recipe performed poorly (36.7\%; \suppfig{figS_three_arms}D). Two further point arms used the rule-independent initialization and are not plotted: the shunting arm also trained poorly (42.4\%, seeds ranging from 34.0\% to 59.1\%), whereas the additive arm reached 63.2\%, similar to the plotted additive point with the tuned recipe (63.1\%). These two additive points (63.2\% and 63.1\%) remained below the tuned additive tree and the shunting tree. A separate search over training settings could improve the points.

\FloatBarrier

\begin{table}[htbp]
\centering\small
\caption{\textbf{Clean-image integration-rule contrasts in trees and points.} Shunting minus additive accuracy in percentage points, with pointwise 95\% paired seed-bootstrap intervals. Points reuse the corresponding tree recipe; E/I networks use the shared learning rate. Intervals use this table's own bootstrap draws, so endpoints can differ in the last digit from those in Supplementary Table~\ref{tab:seed_extension} and the text. Absolute accuracies are in Supplementary Fig.~\ref{fig:figS_three_arms}; the tree-minus-point interaction of the primary cohort is given in Supplementary Note~\ref{supp:point_neurons}.}
\label{tab:point_interactions}
\begin{tabular}{lrrr}
\toprule
Cohort & Seeds & Tree & Point \\
\midrule
MNIST & 10 & $1.67\;[1.41,2.00]$ & $-0.11\;[-0.28,0.03]$ \\
CIFAR-10 50E/0I & 10 & $4.36\;[2.98,5.35]$ & $-0.98\;[-1.47,-0.45]$ \\
CIFAR-10 40E/10I & 10 & $2.71\;[2.23,3.17]$ & $-0.46\;[-0.77,-0.14]$ \\
CIFAR-10 primary & 20 & $0.55\;[0.28,0.81]$ & $-3.75\;[-4.02,-3.49]$ \\
Morphology [120] & 20 & $0.11\;[-0.10,0.33]$ & $-3.23\;[-3.43,-3.03]$ \\
Morphology [10,11] & 20 & $0.17\;[-0.08,0.41]$ & $-3.23\;[-3.43,-3.02]$ \\
Morphology [8,2,6] & 20 & $0.40\;[0.12,0.67]$ & $-3.23\;[-3.43,-3.03]$ \\
Morphology [3,3,3,3] & 20 & $0.70\;[0.33,1.07]$ & $-3.23\;[-3.43,-3.03]$ \\
E/I layers L=1 & 10 & $0.69\;[0.26,1.10]$ & $0.06\;[-0.40,0.57]$ \\
E/I layers L=2 & 10 & $1.64\;[1.27,2.00]$ & $0.20\;[-0.32,0.72]$ \\
E/I layers L=3 & 10 & $2.28\;[1.68,2.84]$ & $-0.25\;[-0.64,0.17]$ \\
E/I layers L=4 & 10 & $3.20\;[2.57,3.92]$ & $0.08\;[-0.71,0.96]$ \\
E/I layers L=5 & 10 & $3.49\;[2.82,4.20]$ & $0.02\;[-0.39,0.48]$ \\
\bottomrule
\end{tabular}
\end{table}

\supppart{Multilayer E/I networks (Figure~\ref{fig:fig5})}
\label{supp:networks}

Longer dendritic E/I networks retain accuracy and decodable class structure better under shunting than under raw addition (Fig.~\ref{fig:fig5}). This note first reports a separate single-layer control that changes the source of inhibition, then separates the network result from effects of the readout, optimization and point-neuron architecture, and distinguishes population decoding from classifier dependence and from the behavior of information estimators.

\suppnote{Direct and population-mediated inhibition}
\label{note:inhibitory_source}
A separate balanced cohort asks whether the source of inhibition matters. Its single-layer networks have 25E/25I contacts per nonsomatic E compartment and receive inhibition either directly from the image or from a learned local I population, with ten paired seeds for each rule and source (forty networks, common recipe). Shunting is more accurate than addition with population-mediated inhibition (2.2 points, positive in all ten seeds), whereas the two rules are comparable with direct image-derived inhibition ($-0.3$ points; Supplementary Table~\ref{tab:inhibitory_source}). Changing the source of inhibition also changes its representation and the complete parameterized system, so this is not an isolated change of the branch integration equation.

\begin{table}[htbp]
\centering
\caption{\textbf{Clean CIFAR-10 accuracy with direct and population-mediated inhibition.} Ten paired seeds per integration rule and source of inhibition, at 25E/25I contacts per nonsomatic E compartment; mean accuracies, and the paired shunting-minus-additive difference with its pointwise 95\% seed-bootstrap interval (20,000 bootstrap draws). The local I population has its own neurons and parameters, so the two sources are not matched in source dimension or parameter count (Methods). Separate intervals are not a test of the difference between the effects of the two sources.}
\label{tab:inhibitory_source}
\small
\begin{tabular}{@{}lccc@{}}
\toprule
Inhibitory source & Shunting (\%) & Raw additive (\%) & Difference (pp) \\
\midrule
Direct image input & 50.58 & 50.88 & $-0.30\ [-0.76,0.15]$ \\
Local I population & 50.57 & 48.41 & $2.16\ [1.92,2.42]$ \\
\bottomrule
\end{tabular}
\end{table}

\suppnote{Inhibitory source and perturbation tolerance}
\label{note:balanced_routing_input_regimes}

\begin{figure}[tbp]
    \centering
    \IfFileExists{figures_empirical/figS_inhibitory_source_j1.pdf}{%
      \includegraphics[width=165mm]{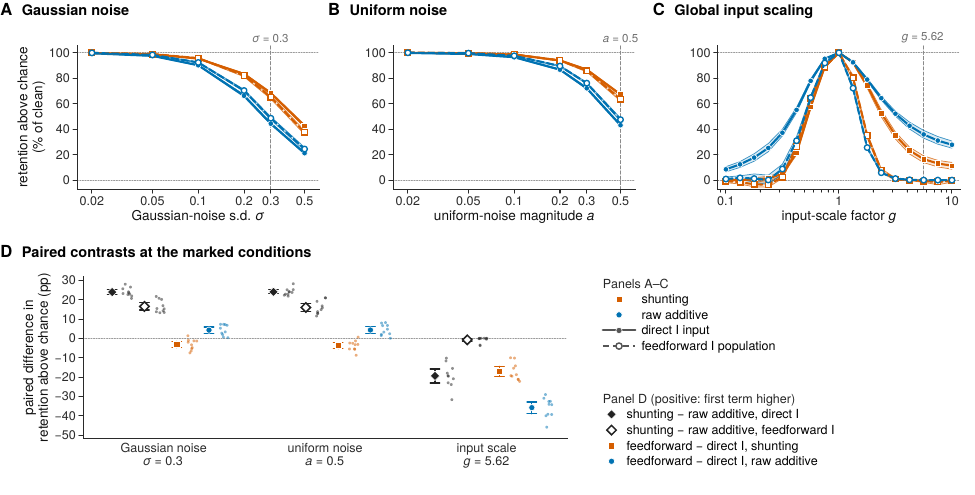}%
    }{%
      \PackageError{empirical-manuscript}%
        {Missing required figure `figures_empirical/figS_inhibitory_source_j1.pdf'}%
        {Restore the figure listed in the manuscript export manifest before compiling.}%
    }%
    \caption{%
\textbf{Perturbation tolerance depends on the source of inhibition.}
(A--C) Above-chance CIFAR-10 retention under Gaussian noise, uniform noise and global scaling, with direct image-derived inhibition (filled markers, solid lines) or a local feedforward I population (open markers, dashed lines); squares, shunting; circles, raw addition. All nonsomatic E compartments use 25E/25I contacts. Curves and pointwise 95\% intervals summarize ten trained seeds per rule and source; stochastic conditions average five common draws within seed. Labelled vertical dashed lines mark the conditions compared in D ($\sigma=0.3$, $a=0.5$ and $g=5.62$); dotted lines mark clean performance (100\%) and chance (0\%). (D) Seed-paired contrasts in percentage points of retention: diamonds, shunting minus raw addition (filled, direct I; open, feedforward I); squares and circles, feedforward minus direct I within shunting and within raw addition. Small points show seed differences, and larger symbols and bars means and 95\% seed-bootstrap intervals (20,000 draws); positive values favor the first term. The source of inhibition changes the complete trained system.}
    \label{fig:figS22_input_regime_routing}
  \end{figure}

\suppfig{figS22_input_regime_routing} tests the same forty networks under input perturbations, normalizing each model to its own clean above-chance accuracy (A--C).

Population-mediated inhibition narrows the shunting advantage under Gaussian and uniform noise (paired contrasts in D). Under strong global scaling, the direct-input contrast favors addition, whereas the contrast with an explicit I population is small, with an interval that only just excludes zero ($-0.75$ points; $-1.78$ to $-0.0004$); retention falls under that source for both rules. The result depends on the complete construction of the inhibitory source, including its learned representation and parameter count, and does not isolate a change in local inhibitory action.

\FloatBarrier

\suppnote{End-to-end readout training at the network-depth endpoints}
\label{note:readout_depth}

\begin{figure}[tbp]
    \centering
    \IfFileExists{figures_empirical/figS_readout_depth_j1.pdf}{%
      \includegraphics[width=165mm]{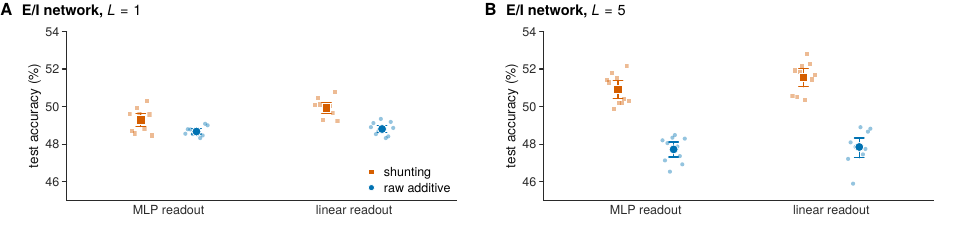}%
    }{%
      \PackageError{empirical-manuscript}%
        {Missing required figure `figures_empirical/figS_readout_depth_j1.pdf'}%
        {Restore the figure listed in the manuscript export manifest before compiling.}%
    }%
    \caption{%
\textbf{The multilayer endpoint difference persists with a linear readout.}
(A,B) CIFAR-10 test accuracy of shunting (squares) and raw-additive (circles) E/I networks at $L=1$ and $L=5$ under MLP and linear readouts. Small points show individual trained seeds; larger symbols and bars show means and pointwise 95\% seed-bootstrap intervals. Seeds 10--19 are paired across readouts, and each rule keeps its selected learning rate. The networks and readouts are trained together. Intermediate depths and representation or intervention analyses were not repeated with linear readouts.}
    \label{fig:figS_readout_depth}
  \end{figure}

Retraining the $L=1$ and $L=5$ networks with a linear readout preserves the pattern at these two depths (\suppfig{figS_readout_depth}A,B). The shunting-minus-additive difference grows from 1.1 percentage points at $L=1$ (95\% interval, 0.8--1.4) to 3.7 at $L=5$ (3.0--4.3) with the linear readout, compared with 0.6 (0.2--1.1) and 3.2 (2.6--3.8) with the MLP; under both readouts, additive accuracy is lower at $L=5$ than at $L=1$ and shunting accuracy is higher. Seeds 10--19 are paired across readouts, and each rule keeps its selected initialization and optimization settings. Intermediate depths, population probes, information estimates and interventions were not repeated with a linear readout; the analyses below use the MLP-trained networks.

\FloatBarrier

\suppnote{Network depth, optimization, and architecture controls}
\label{note:multilayer_confirmatory}

Networks with $L=1,\ldots,5$ layers have, in each layer, 64 E neurons with $[3,3,3]$ trees and 16 I neurons with $[1]$ trees; each nonsomatic E compartment receives 40 feedforward-E and ten local-I contacts (Fig.~\ref{fig:fig5}A and Methods). Changing $L$ changes the complete trained network, whereas position $\ell$ follows successive populations within one network.

\begin{figure}[tbp]
    \centering
    \IfFileExists{figures_empirical/figS_multilayer_policy_j1.pdf}{%
      \includegraphics[width=165mm]{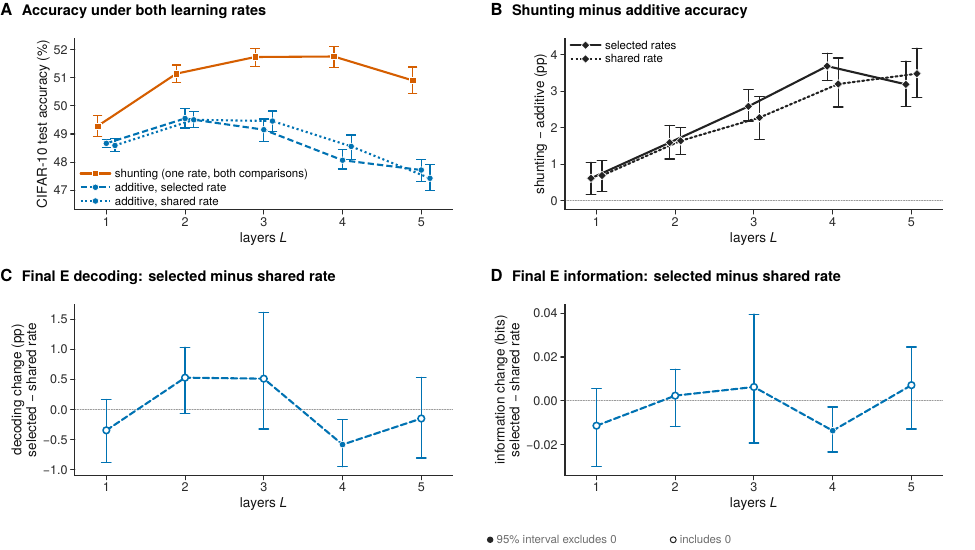}%
    }{%
      \PackageError{empirical-manuscript}%
        {Missing required figure `figures_empirical/figS_multilayer_policy_j1.pdf'}%
        {Restore the figure listed in the manuscript export manifest before compiling.}%
    }%
    \caption{%
\textbf{Multilayer performance and representation sensitivity to learning rate.}
CIFAR-10 E/I networks of $L$ layers. (A) Test accuracy under the validation-selected and shared learning rates: shunting uses $10^{-3}$ in both comparisons and is shown once, and addition uses $10^{-4}$ (selected, dashed) or $10^{-3}$ (shared, dotted). (B) Paired shunting-minus-additive accuracy differences under the selected (solid) and shared (dotted) rates. (C,D) Selected-minus-shared additive differences at the final E output: ridge decoding above shuffled labels in percentage points (C) and population class information at matched width in bits (D). Points and bars show means and pointwise 95\% seed-bootstrap intervals over ten paired seeds; in C,D, filled marks denote intervals that exclude zero. Horizontal offsets separate overlapping conditions.}
    \label{fig:figS17_multilayer_policy_sensitivity}
  \end{figure}

\label{note:multilayer_policy}
Both rules use occupancy-quantile initialization, which calibrates rule-dependent forward states and so does not give identical initial weights. Shunting uses learning rate $10^{-3}$; raw addition uses the validation-selected $10^{-4}$ or, in the shared-learning-rate control, $10^{-3}$. Every condition has ten trained seeds and a 5,000-epoch limit with patience 600, and every run stopped by the patience rule; the shunting networks serve both comparisons. The selected additive rate changes mean accuracy by less than 0.5 percentage points at every $L$, raising it at $L=1$, 2 and 5 and lowering it at $L=3$ and 4 (\suppfig{figS17_multilayer_policy_sensitivity}A). Changes in final-population decoding and estimated class information vary in sign across depths; their intervals include zero except at $L=4$, where both changes are slightly negative (C,D). The greater shunting accuracy in the longest networks holds under both learning-rate comparisons (B).

The conductance-normalized additive control of Fig.~\ref{fig:fig5}B was trained separately at $L=1,3,5$: a three-seed screen selected learning rates by validation negative log likelihood, and ten new seeds were then trained per depth. Sharing the shunting denominator nearly reaches shunting accuracy at $L=3$ (51.3\% against 51.7\%) but not at $L=5$, where eight of ten seeds reach 48.9--50.7\% and two reach only 35.0\% and 37.8\%.

\label{note:multilayer_point_reference}
Two point controls bear on this comparison. Under a shared learning rate, the point-neuron reference of Fig.~\ref{fig:fig5}B (dotted lines) performs like the dendritic networks at $L=1$ (paired differences of 0.2 and 0.1 percentage points under shunting and addition, with intervals that include zero) and is less accurate at $L=3$ and $L=5$ under both rules: by 1.7 points under shunting at both depths and by 0.9 and 1.3 points under addition, with 95\% intervals that exclude zero and nine or ten of ten seed pairs in the same direction. Under addition, the point reference also loses accuracy between $L=1$ and $L=5$ (48.7\% to 46.2\%), so in this protocol the decline with depth is not specific to dendritic trees. Its candidate parameter counts differ, so this comparison concerns complete architectures, not a resource-matched effect of dendritic depth. The source-capped point controls of Supplementary Note~\ref{supp:point_neurons} (\suppfig{figS_three_arms}C) reuse the trees' active contacts up to the number of available E or I sources. Under the shared learning rate, the shunting points remain 2.0--2.2 percentage points below the shunting trees at $L=3$--5, whereas the additive points are as accurate as or more accurate than the additive trees (by 1.5 points at $L=5$; 95\% interval, 0.7--2.3). At the additive trees' selected learning rate of $10^{-4}$, the additive points reach only 39.7\% at $L=5$, against 48.9\% at $10^{-3}$, which reflects the reused training settings, not the point construction (Supplementary Table~\ref{tab:point_interactions}).

\FloatBarrier

\suppnote{Finite-sample sensitivity of population information estimates}
\label{note:population_estimator_audit}

Population information estimates subtract a shuffled-label reference, $\widehat I_{\rm corr}(\mathbf z;Y)=\widehat I(\mathbf z;Y)-\langle\widehat I(\mathbf z;Y^\pi)\rangle_\pi$. This correction does not guarantee an unbiased estimate. The complete 64-coordinate E-soma vector $\mathbf a^E_\ell$ is the only signal passed to the next network layer. Its successor is a deterministic transform, so the data-processing inequality gives
\[
I(Y;\mathbf a^E_{\ell+1})\leq I(Y;\mathbf a^E_\ell).
\]
Estimates that increase across layers for the complete soma vector therefore call for an estimator check, not an interpretation as generated information. The projected vectors used in the E/I comparison discard coordinates and need not obey this ordering. Decoding measures accessibility to a specified readout, and seed-bootstrap intervals on information estimates do not correct estimator bias.

\begin{figure}[tbp]
    \centering
    \IfFileExists{figures_empirical/figS_network_information_audit_v10.pdf}{%
      \includegraphics[width=165mm]{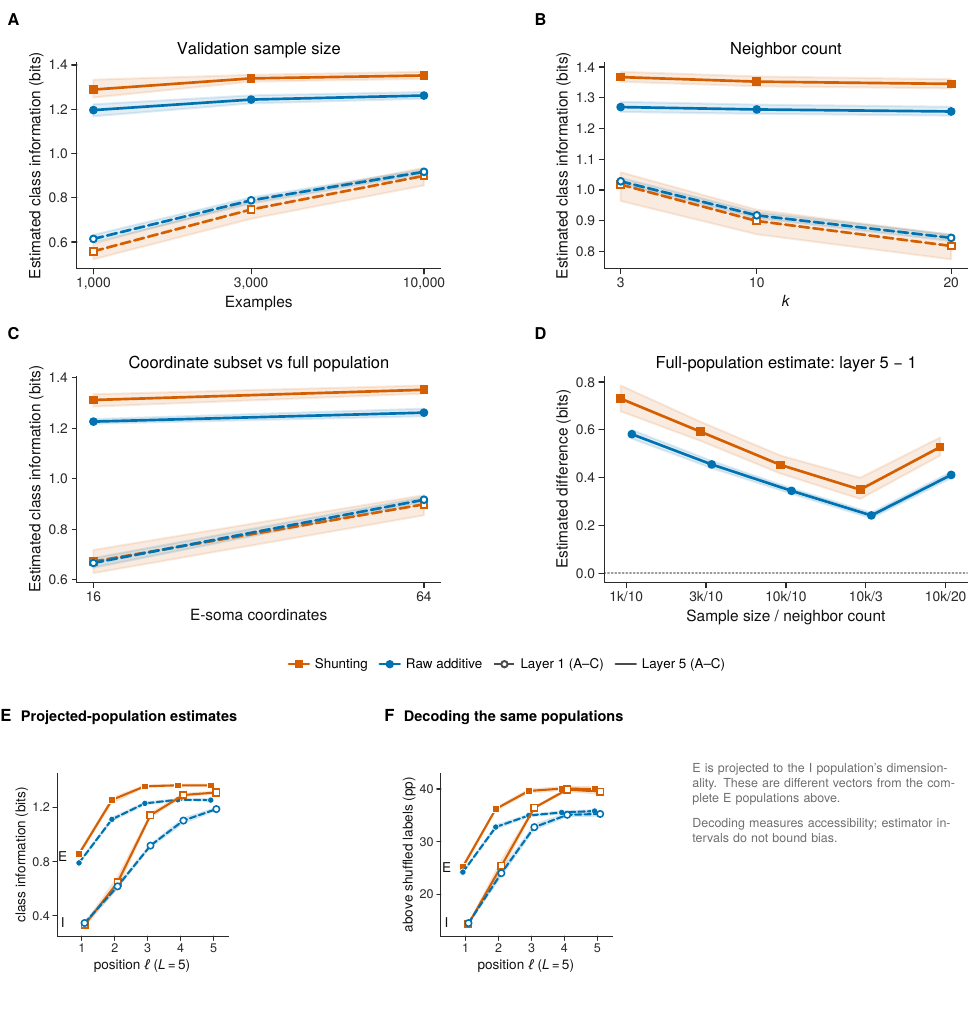}%
    }{%
      \PackageError{empirical-manuscript}%
        {Missing required figure `figures_empirical/figS_network_information_audit_v10.pdf'}%
        {Restore the figure listed in the manuscript export manifest before compiling.}%
    }%
    \caption{%
\textbf{Population-information estimates depend on sample size, neighbor count and representation.}
(A) Complete 64-coordinate estimates versus validation sample size, at $k=10$. (B) Complete-population estimates versus neighbor count, using 10,000 examples. (C) A fixed 16-coordinate subset versus the complete population, at 10,000 examples and $k=10$; these are different measured representations. A--C show layers 1 and 5 of the same fitted five-layer networks. (D) Paired layer-5 minus layer-1 estimates for the full population. Positive values cannot represent increases in true information under the data-processing inequality. All quantities subtract the mean of five label permutations. Curves and pointwise 95\% seed-bootstrap intervals summarize ten paired trained seeds per rule; uncertainty does not correct estimator bias. (E,F) Estimated information and linear decoding of E (filled) and I (open) populations at matched dimensionality, within five-layer networks. E is projected to 16 coordinates and I remains native; these are different vectors from the complete E populations in A--D. These estimates subtract twenty label shuffles and use ten seeds.}
    \label{fig:figS29_multilayer_network_tree_atlas}
  \end{figure}

All E-soma outputs were collected from the same twenty five-layer networks of the primary comparison (ten paired seeds, 10--19), standardized per coordinate without labels, and the Ross estimator~\citep{Ross2014} was applied to layers 1, 3 and 5 (the figure shows layers 1 and 5). The analysis varies the validation sample size (1,000, 3,000 and 10,000), the neighbor count ($k=3,10,20$ at 10,000 examples) and the dimensionality (all 64 coordinates or a fixed label-independent subset of 16; \suppfig{figS29_multilayer_network_tree_atlas}A--C). Image subsets are nested and shared across networks, and five label permutations define the finite-sample null. The 16-coordinate subset is a different representation, not a convergence test for the complete population. The main population-information analyses use $k=5$ (Methods); this check varies $k$ around 10 and does not include $k=5$.

The estimated layer-5 minus layer-1 difference stays positive for both rules in every full-population setting, but its magnitude decreases with more examples or fewer neighbors, a substantial finite-sample sensitivity (\suppfig{figS29_multilayer_network_tree_atlas}D). The larger final-layer estimate under shunting persists in these checks, whereas the first-layer ordering of the rules is less stable. The analysis supports a descriptive comparison of final-layer estimates; it validates neither their absolute magnitude nor an increase in true information. Decoding and interventions are the primary evidence for accessibility and classifier dependence.

\FloatBarrier

\suppnote{Native soma subsets and local inhibition}
\label{note:multilayer_subset_geometry_local_i}

\begin{figure}[tbp]
    \centering
    \IfFileExists{figures_empirical/figS_subset_local_inhibition_j1.pdf}{%
      \includegraphics[width=165mm]{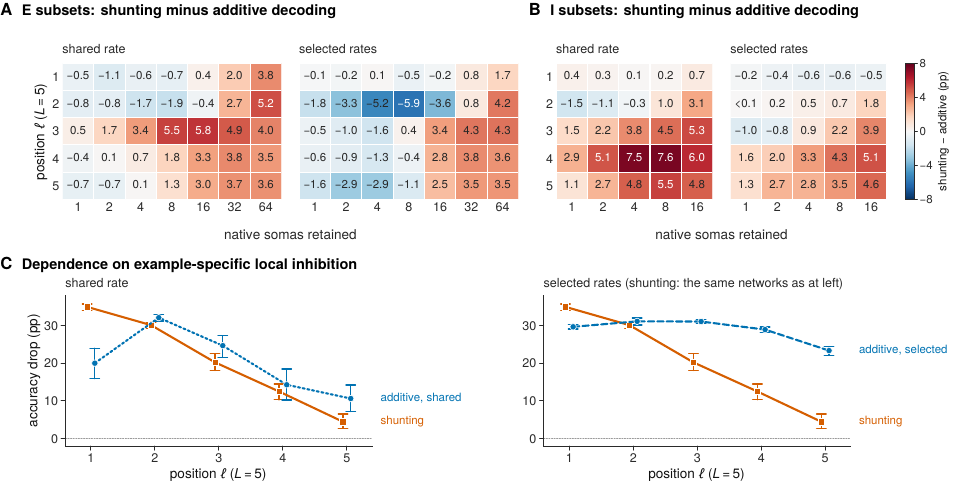}%
    }{%
      \PackageError{empirical-manuscript}%
        {Missing required figure `figures_empirical/figS_subset_local_inhibition_j1.pdf'}%
        {Restore the figure listed in the manuscript export manifest before compiling.}%
    }%
    \caption{%
\textbf{Native-population decoding and local-I dependence in five-layer networks.}
CIFAR-10 E/I networks with $L=5$. (A,B) Paired shunting-minus-additive ridge decoding, in percentage points above shuffled labels, for nested native E (A) and I (B) soma subsets under the shared and the selected learning rates, on one symmetric color scale; cells give ten-seed means, with network position 1 at the top. Probes are fitted on validation responses and evaluated on test responses. (C) Accuracy loss after class-preserving donor replacement of a complete local I population, under the shared (left) and selected (right) learning rates; both use the same shunting networks. Points and bars show means and pointwise 95\% seed-bootstrap intervals after donor draws are averaged within seed; squares and solid lines, shunting; circles with dotted or dashed lines, addition at the shared or selected rate. The donor intervention uses evaluation labels.}
    \label{fig:figS21_multilayer_subset_geometry_local_i}
  \end{figure}

Nested subsets of the unprojected soma outputs show how decoding changes as neurons are added. Ten label-independent permutations define the nested E and I subsets in the $L=5$ networks (the full population has only one subset), and ridge probes are fitted on validation and evaluated on test responses. The shunting advantage in decoding generally grows with the number of retained somas, and the full E population favors shunting under both learning-rate comparisons; very small subsets and the first I population can reverse this ordering (\suppfig{figS21_multilayer_subset_geometry_local_i}A,B).

To test example-specific inhibitory coordination, the complete local I-population response at one network position is replaced by the response to another image of the same class. This keeps the class-conditional distribution of I responses but breaks their pairing with the current E response; five donor draws are averaged within each trained seed. Under the selected learning rates, the accuracy loss is larger for shunting near the input and for addition near the output (C). The shared-rate control keeps the stronger dependence of shunting at $\ell=1$ (35.0 against 20.0 points); further on, the additive loss is larger, most clearly at $\ell=5$ (10.6 against 4.4). Replacing I outputs by zero instead leaves both rules near chance at every position under the selected learning rates, which does not discriminate between them. The class-preserving intervention uses evaluation labels and measures dependence on example-specific coordination, not total dependence on inhibition.

\FloatBarrier

\suppnote{Branch-output dependence under alternative replacement references}
\label{note:multilayer_post_gate_output_intervention}

\begin{figure}[tbp]
    \centering
    \IfFileExists{figures_empirical/figS_network_tree_intervention_j1.pdf}{%
      \includegraphics[width=165mm]{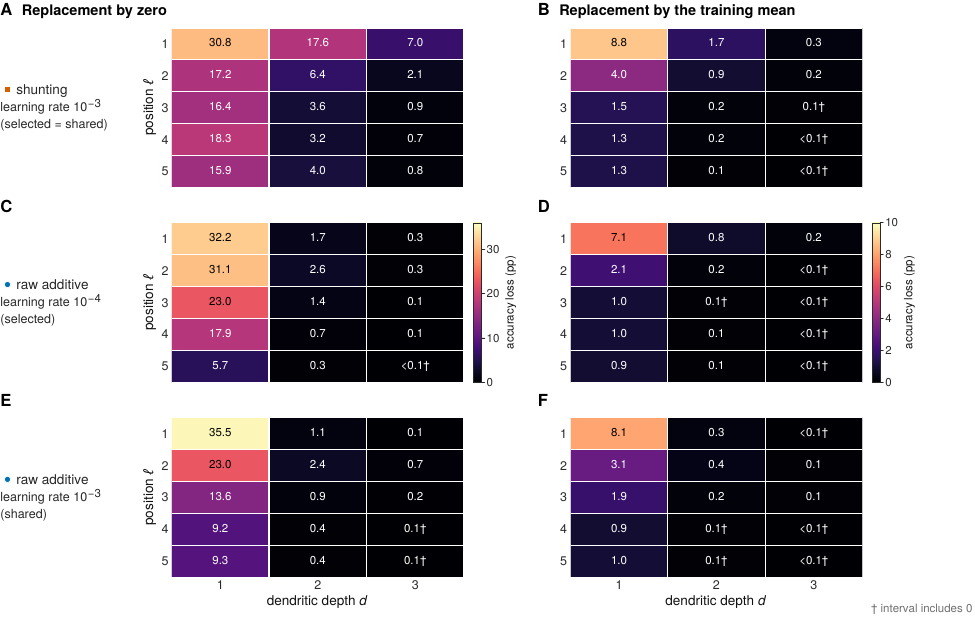}%
    }{%
      \PackageError{empirical-manuscript}%
        {Missing required figure `figures_empirical/figS_network_tree_intervention_j1.pdf'}%
        {Restore the figure listed in the manuscript export manifest before compiling.}%
    }%
    \caption{%
\textbf{Branch-output dependence remains proximal across learning rates and replacement values.}
Five-layer CIFAR-10 E/I networks. (A,B) Shunting at learning rate $10^{-3}$; (C,D) addition at the selected $10^{-4}$; (E,F) addition at the shared $10^{-3}$. Left, outputs replaced by zero; right, by their fixed mean responses over training images. Matrices show unclipped test-accuracy loss in percentage points across network position $\ell=1,\ldots,5$ and nonsomatic depth $d=1,2,3$, with one color scale per replacement value (zero, 0--36; training mean, 0--10, as in Fig.~\ref{fig:fig5}H). Printed values are ten-seed means after five fixed selections of 64 branch outputs are averaged within seed; nonzero means that round to zero are shown as bounds, and daggers mark cells whose pointwise 95\% seed-bootstrap interval includes zero. The stable result is proximal concentration; the ordering of the integration rules can depend on the replacement value and the optimization setting.}
    \label{fig:figS32_multilayer_post_gate_output_intervention}
  \end{figure}

Branch outputs were replaced along the paths of Fig.~\ref{fig:fig5}H by zero (\suppfig{figS32_multilayer_post_gate_output_intervention}A,C,E) or by their fixed mean responses over training images (B,D,F), under both learning-rate settings. At each nonsomatic depth, five fixed selections each choose one output per E soma, so every intervention replaces 64 coordinates; means come from all 40,000 optimization examples, and accuracy is evaluated on all 10,000 test examples.

Dependence remains concentrated proximally across replacement values and training settings, while effect sizes and some orderings of the rules change. Neither replacement value is a neutral biological baseline, and losses need not add across interventions. These interventions measure classifier dependence on selected branch signals; population decoding and information estimates address the separate question of which class relationships can be recovered from recorded responses.

\FloatBarrier

\FloatBarrier

\supppart{Context-dependent computation (Figure~\ref{fig:fig7})}
\label{supp:contextual}

Higher accuracy and stronger gain attenuation do not imply greater dependence on context-specific inhibition (Fig.~\ref{fig:fig7}). The sections below examine how accuracy, decoding, gain attenuation and dependence on inhibition vary across the trained networks, how conductance relative to leak affects responses to dim inputs, and how the networks select their E and I sources. Results labelled original implementation were obtained by B.L.S. in an earlier, separate implementation (unpublished data; Methods); they use a different construction and evaluation and are reported separately from the paired controls.
\label{note:contextual_details}

\suppnote{Accuracy and decoding}

\begin{figure}[tbp]
    \centering
    \IfFileExists{figures_empirical/figS_context_task_controls_v15.pdf}{%
      \includegraphics[width=165mm]{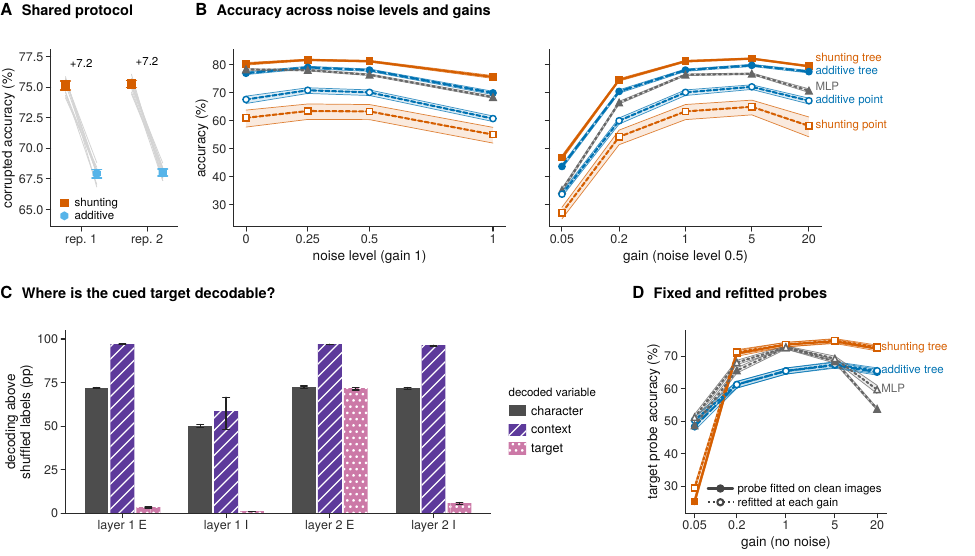}%
    }{%
      \PackageError{empirical-manuscript}%
        {Missing required figure `figures_empirical/figS_context_task_controls_v15.pdf'}%
        {Restore the figure listed in the manuscript export manifest before compiling.}%
    }%
    \caption{%
\textbf{Accuracy and decoding in the contextual task.}
(A) Corrupted accuracy of trees trained with a shared protocol in two independent trainings (rep.~1 on A100 and rep.~2 on H100 GPUs). Lines join paired seeds within each training cohort. Numbers give the paired shunting-minus-additive differences in percentage points.
(B) Accuracy across independent noise levels at unit gain and across gains at noise level 0.5. Noise level is the inverse divisor $1/q$; its training range was 0.25--1. Open markers denote point neurons.
(C) Character, cue and target decoding above shuffled-label accuracy in each population of the primary shunting tree. Probes were fitted on images held out from training and evaluated on clean test images.
(D) Target decoding on noise-free images at each gain, from second-layer E outputs or the MLP's second hidden layer. Filled markers show the probe fitted on clean images and open markers show probes refitted at each gain.
Means and pointwise 95\% seed-bootstrap intervals over ten trained networks per model.}
    \label{fig:figS_context_task_controls}
  \end{figure}

\begin{figure}[tbp]
    \centering
    \IfFileExists{figures_empirical/figS_context_conductance_v16.pdf}{%
      \includegraphics[width=165mm]{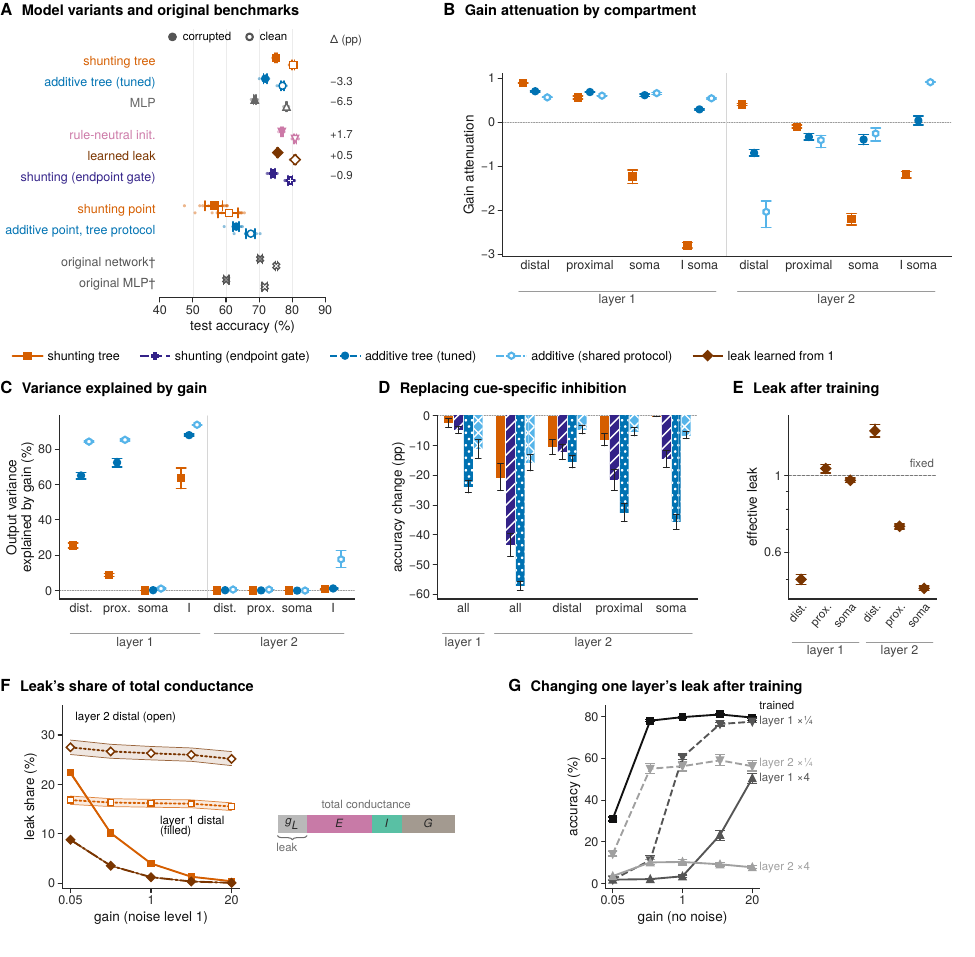}%
    }{%
      \PackageError{empirical-manuscript}%
        {Missing required figure `figures_empirical/figS_context_conductance_v16.pdf'}%
        {Restore the figure listed in the manuscript export manifest before compiling.}%
    }%
    \caption{%
\textbf{Training variants, compartmental gain sensitivity and inhibition in the contextual task.}
(A) Clean and corrupted accuracy for the tree variants, MLP and point neurons. Original-implementation values (crosses, dagger) are reported means and standard deviations over three seeds under their own protocol. Rule-neutral initialization sets initial parameters without reference to the integration rule.
(B) Attenuation across gains, measured as one minus the output-to-numerator coefficient-of-variation ratio. Negative values occur in groups with little numerator variation.
(C) Output variance explained by gain, by layer, population and compartment level.
(D) Accuracy change after replacing inhibition by its mean over all 32 cues for each corrupted image. Hatches identify primary shunting (solid), endpoint-gate shunting (stripes), tuned additive (dots) and shared-protocol additive (crosshatch).
(E) Leak conductance learned from one in E compartments; the dashed line marks unit leak.
(F) Leak's share of total conductance in distal E compartments at noise level $1/q=1$, with fixed or learned leak. Filled and open symbols denote the first and second layers, respectively. Bar widths in the schematic are illustrative.
(G) Accuracy across input gains after scaling the first or second layer's excitatory leak by one quarter or four without retraining.
Present models use ten seeds and pointwise 95\% seed-bootstrap intervals.}
    \label{fig:figS_context_controls}
  \end{figure}

Under the training corruption, the shunting tree reached 75.1\% test accuracy (Fig.~\ref{fig:fig7}B). It exceeded the separately tuned additive tree by 3.28 percentage points (95\% interval, 2.71--3.83; ten of ten seeds) and the MLP by 6.46 (5.88--7.03). On clean images the margins were 3.27 (2.58--3.91) and 2.05 (1.46--2.60) points. With the recipe held fixed, shunting was more accurate by 7.19 points (6.66--7.76), and an independent training of both arms on another GPU model gave 7.23 (6.77--7.72; \suppfig{figS_context_task_controls}A). Tuning therefore recovers about four points for the additive tree, but it does not create the advantage of shunting, which is larger when the setting is held fixed. \suppfig{figS_context_controls}A compares the activation, initialization and leak variants with the primary models. Two task-specific variants had higher mean accuracy. An initialization that ignores the integration rule (labeled ``rule-neutral init.'' in the figure) added 1.75 points (1.20--2.30; ten of ten seeds). Learning the leak from its fixed value added 0.53 (0.10--1.01; eight of ten seeds), but this interval depends on the method: a Student-$t$ interval for the same paired differences ($-0.03$--1.08) includes zero. The clearer benefit of learned leak is on dim images (below). The endpoint-gate shunting variant, which uses the endpoint-normalized tanh activation of Eq.~\ref{eq:context_activation} with different initialization and training settings, was 0.86 points less accurate than the primary shunting tree (0.49--1.25). Soma-only points also performed substantially worse under the corresponding tree protocols (Supplementary Note~\ref{note:three_arms} and \suppfig{figS_three_arms}D). For reference, the original implementation of the network that motivated the task was reported at 70.3\% corrupted and 75.2\% clean test accuracy, and its MLP at 60.1\% and 71.7\% (means of three seeds, with that model's own construction and evaluation; \suppfig{figS_context_controls}A).

Within the training distribution the shunting tree was more accurate than the separately tuned additive tree and the MLP at every gain (\suppfig{figS_context_task_controls}B); its two task-specific variants (not plotted) had higher mean accuracy still at every gain. At noise level 0.5 it exceeded the additive tree by 3.2 points (2.4--4.0) and the MLP by 11.6 (10.5--12.7) at gain 0.05, and by 1.9 (1.5--2.4) and 8.7 (7.9--9.5) at gain 20. At unit gain its advantage grew with the noise level, reaching 5.7 (4.8--6.5) and 7.2 (6.5--7.8) points at the strongest noise. Dim images without noise lie outside the training distribution: training always adds noise, whose rectification raises the mean drive of dim images. At gain 0.05 the fixed-leak shunting tree scored 30.2\%, compared with 61.1\% for the additive tree and 61.4\% for the MLP. Learning the leak raised shunting accuracy by 15.8 points (14.5--17.2; Fig.~\ref{fig:fig7}D).

Linear probes fitted on images held out from network training and scored on test images measured how readily character, context and the required target could be decoded (\suppfig{figS_context_task_controls}C). In the shunting tree, character and context were readily decoded from first-layer E outputs (71.9 and 96.9 points above shuffled labels), whereas target decoding was weak (3.3 points). Target decoding was much stronger from second-layer E outputs (71.4 points) than from second-layer I outputs (5.8 points), although character and context remained decodable from the latter. For second-layer E outputs, the accuracy of a probe fitted on clean images was within half a percentage point of the refitted probe at gains 0.2, 1, 5 and 20. At gain 0.05 without noise, the clean-fitted probe scored 25.2\% and the refitted probe 29.4\% (\suppfig{figS_context_task_controls}D). The MLP's clean-fitted probe lost about six points at gain 20, leaving the shunting tree 18.6 points (17.7--19.5) ahead there. These are decoding accuracies, not information estimates, and stable accessibility to a fixed readout does not imply an unchanged representation.

\suppnote{Gain attenuation and contextual inhibition}
We measured how compartment outputs depend on gain using training-set images held out from optimization and model selection (Fig.~\ref{fig:fig7}C; \suppfig{figS_context_controls}B,C). In first-layer distal compartments, the coefficient of variation of the numerator $E+C$ across gains was about 1.4 in every model. Shunting reduced it to 0.14 at the output, an attenuation of 0.89. The separately tuned additive tree reached 0.41 (attenuation 0.71), and the additive tree trained with the shunting recipe 0.60 (0.57). Gain explained 26\% of the shunting tree's distal output variance, against 65\% and 84\% in the two additive trees, and under one recipe shunting increased the attenuation by 0.32 (0.31--0.34). The matched-recipe contrast supports stronger distal gain attenuation in the shunting construction; the shared initialization procedure still produces rule-dependent initial parameters. Attenuation is negative in some somatic and second-layer groups, where the output varied more across gains than the numerator; this occurs only where the numerator itself varied little (coefficient of variation below 0.1), so these ratios compare small residual variations.

Dependence on context-specific inhibition varied across the trained networks (Fig.~\ref{fig:fig7}E,F). Replacing second-layer inhibition by its mean over contexts cost the tuned additive tree 57.1 points, the endpoint-gate shunting tree 43.5 and the primary shunting tree 20.8 (Fig.~\ref{fig:fig7}E). \suppfig{figS_context_controls}D also shows replacement throughout layer 1 and at individual compartment levels in layer 2. Transplanting the next context's inhibition into the second layer made the tuned additive tree report the donor context's target in 15.0\% of trials, against 1.4\% for the shunting tree and below 1\% in every intact network (Fig.~\ref{fig:fig7}F). The additive tree trained with the shunting recipe, however, behaved like the shunting tree: it lost 15.8 points under context-mean inhibition (17.0 in the second training) and reported the donor target in 1.6\% of trials (2.1\%). Under one recipe, the shunting tree lost more than the additive tree in one training (paired shunting-minus-additive difference in the accuracy change, $-5.0$ points; $-9.5$ to $-0.2$) but not clearly in the other ($-1.7$; $-6.0$ to $2.6$). A strong dependence on context-specific inhibition is therefore a property of the tuned additive recipe and of the endpoint-gate construction, not of additive integration as such. The original implementation of the task network was reported to lose 47.8 points under second-layer context-mean inhibition and to report the donor target in 13.5\% of trials after a second-layer transplant (Supplementary Table~\ref{tab:original_routing}), close to the endpoint-gate construction and the tuned additive tree rather than to the shunting tree.

\suppnote{Leak and operating point}
Leak and synaptic conductances set the operating point together (Fig.~\ref{fig:fig7}D; \suppfig{figS_context_controls}E--G). In the fixed-leak tree, scaling the first-layer excitatory leak by four lowered accuracy at unit gain from 79.7\% to 3.6\%; the learned conductances are tuned to the leak with which they were trained. Leak learned from one developed a depth profile. In the first layer it fell to 0.50 at distal compartments and stayed near one at the soma (0.97), whereas in the second layer distal leak rose to 1.35 and somatic leak fell to 0.47. To read these values as operating points, we measured the leak's share of the shunting denominator, $g_L/(g_L+E+I+G)$, on held-out training images (\suppfig{figS_context_controls}F). In most compartments the share was below 3\%, so input conductances dominate. It was largest at distal compartments. At the highest noise level, first-layer distal leak accounted for 4\% of total conductance at unit gain and 22\% at gain 0.05 in the fixed-leak tree. With learned leak its share at gain 0.05 was 9\% (\suppfig{figS_context_controls}F). The 15.8-point accuracy improvement at the same gain was measured separately on noise-free images (Fig.~\ref{fig:fig7}D). In second-layer distal compartments the share was 16\% with fixed leak and 26\% with learned leak at every gain.

\FloatBarrier

\suppnote{Source selection and E--I covariation}
\label{note:contextual_wiring}

\begin{figure}[tbp]
    \centering
    \IfFileExists{figures_empirical/figS_context_wiring_v13.pdf}{%
      \includegraphics[width=165mm]{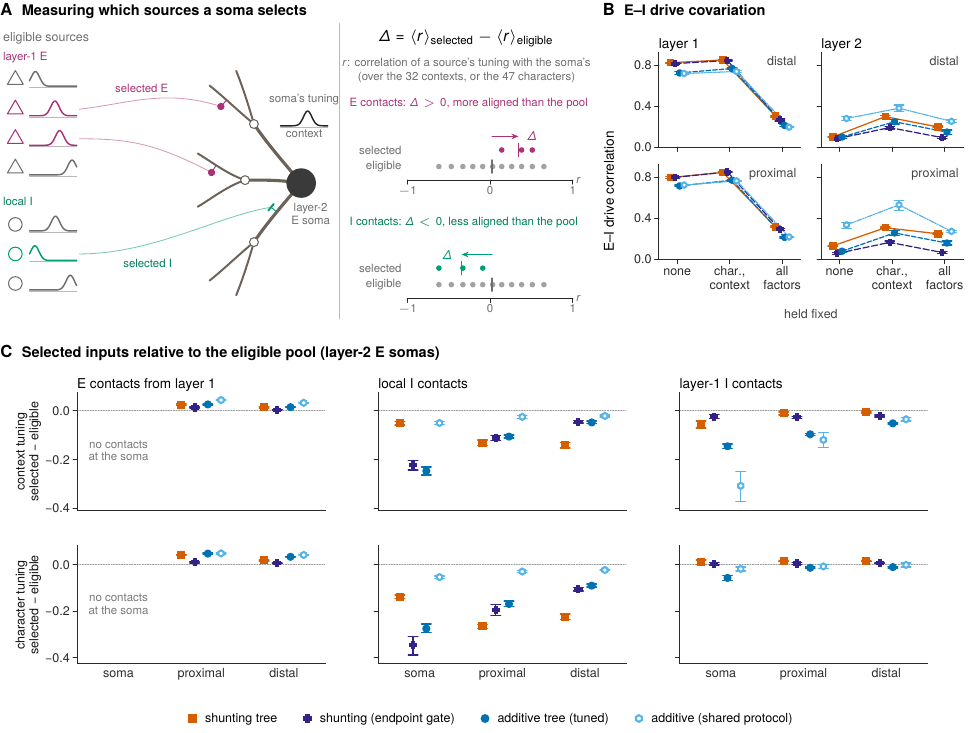}%
    }{%
      \PackageError{empirical-manuscript}%
        {Missing required figure `figures_empirical/figS_context_wiring_v13.pdf'}%
        {Restore the figure listed in the manuscript export manifest before compiling.}%
    }%
    \caption{%
\textbf{Source selection and E--I covariation in the contextual networks.}
The shunting tree, endpoint-gate construction, tuned additive tree and additive tree trained with the shunting recipe each have ten seeds. (A) Schematic of the source-selection measure, with illustrative tuning curves and correlation values, not measured data. For each second-layer E soma, source tuning over contexts or characters is correlated with somatic tuning; the difference compares selected contacts with all eligible sources. For the measurements in C, soma and source profiles use separate test-image subsets (image indices 68--71 and 72--75 within each class; four images per subset), at gain one and noise divisor two. (B) E--I drive correlations in first- and second-layer E compartments, pooled or conditioned on character and context, or on character, context, gain and noise. Correlations within exact conditions vary across images. (C) Relative context and character alignment for each input pathway and compartment level. First-layer E contacts are dendritic only. Means and 95\% seed-bootstrap intervals; source-profile and correlation controls are descriptive, whereas the main interventions test classifier dependence.}
    \label{fig:figS_contextual_wiring}
  \end{figure}

The trained networks also reproduced the E/I organization reported for the original network (\suppfig{figS_contextual_wiring}). For each second-layer E soma, we correlated the context tuning of every source it could contact with the soma's own and compared the sources of its selected contacts with that eligible pool (\suppfig{figS_contextual_wiring}A). The measure is relative: a negative selected-minus-eligible difference means lower alignment than the pool, not a negative correlation. In all four networks and all ten seeds, E contacts from the first layer came from neurons tuned more like the soma than the pool, whereas local I contacts came from neurons less aligned than the eligible pool (\suppfig{figS_contextual_wiring}C). In the shunting tree this relative alignment deficit of local inhibition was larger in dendritic compartments ($-0.13$ at proximal and $-0.14$ at distal ones) than at the soma ($-0.05$), whereas in the endpoint-gate construction and the tuned additive tree it was largest at the soma ($-0.22$ and $-0.25$). In the additive tree trained with the shunting recipe it was weak at every level ($-0.02$ to $-0.05$); its largest relative deficit was for first-layer I contacts onto the soma ($-0.31$). Character tuning showed the same opposition, with local I contacts less aligned than the pool in every seed and E contacts aligned in every seed but one (an endpoint-gate network at proximal compartments).

E and I drives covaried in a layer-specific way (\suppfig{figS_contextual_wiring}B). In first-layer E compartments of the shunting tree their correlation was 0.80--0.82 over all task conditions and 0.85 within each character--context pair, but 0.30--0.32 across images within exact conditions of character, context, gain and noise: most of the shared variation followed gain and noise. In the second layer the two drives were nearly uncorrelated over all conditions (0.10--0.13) and more correlated within character--context pairs (0.30--0.31) and within exact conditions (0.20--0.25). Every network showed both patterns, which is consistent with second-layer inhibitory sources selected for lower relative context alignment. The original implementation was reported to show 0.84 within character--context pairs and 0.26 within exact conditions in the first layer, and 0.11--0.12 over all conditions and 0.20--0.21 within exact conditions in the second. E--I correlation describes covariation; it does not show equal drive magnitudes or cancellation. These measurements describe the learned wiring; the interventions above test its consequences.

\FloatBarrier

\supppart{AlexNet replacements (Figure~\ref{fig:fig6})}
\label{supp:vision}

Shunting improves matched dendritic replacements at AlexNet width, and learned connectivity supports smaller deployed models (Fig.~\ref{fig:fig6}). The comparisons below assess conventional alternatives and distinguish retained parameter budgets from the larger candidate banks used during training. Different replacement boundaries and evaluation cohorts are reported separately.

\suppnote{Replacement protocols, performance, and resource accounting}
\label{note:scale_compatibility}

The AlexNet experiments test whether sparse dendritic populations can replace wide, fully connected transformations in a pretrained image classifier. Starting from ImageNet-pretrained AlexNet~\citep{Deng2009,Krizhevsky2012}, either fc7 and its rectifier or the complete fc6--rectifier--fc7--rectifier span is replaced; the preceding network is frozen, and the replacement and the final classifier, fc8, are trained. The original comparisons omit dropout within the replaced span; the endpoint MLP controls below test dropout and weight decay explicitly. The official training set is divided into optimization (80\%) and checkpoint-validation (20\%) subsets, and the official validation set is used only for final evaluation. The original integration-rule comparisons train with Adam~\citep{KingmaBa2015} and mixed precision for 50 epochs, the matched scaling and continuation studies for 100 epochs, and all select checkpoints by validation negative log likelihood (Methods).

Training uses random resized crops and horizontal flips, and evaluation a center crop after resizing, with standard ImageNet normalization. The replacement input differs between boundaries: fc7 receives the 4,096 features of the frozen fc6, whereas the fc6--fc7 replacement receives 9,216 flattened convolutional features. Both dendritic replacements have 4,096 E and 820 I neurons with $[2,2]$ trees; each nonsomatic I compartment receives ten input contacts, and each nonsomatic E compartment 40 input contacts and ten contacts from local I neurons. Only E outputs enter fc8. Fixed-contact dendritic and sparse point cores each have 1,376,320 trainable parameters, despite different internal computations and contact counts.

At the complete fc6--fc7 boundary, learned-top-$k$ shunting is more accurate than fixed-contact shunting with the same retained core-parameter count (Supplementary Table~\ref{tab:fig6_fc6fc7_confirmation}, seeds 41--45, 50 epochs), although learning contact locations requires a much larger bank of candidate weights during training. Among fixed-contact models, shunting is more accurate than the matched sparse point model and than raw addition, under both global and structured-local routing. Every replacement remains below untouched pretrained AlexNet, and retraining the native dense span also lowers accuracy, so these results concern performance under the specified retraining protocol, not the best attainable accuracy of each architecture. A separate cohort compares trained dense and sparse MLPs with fixed-contact references under the same core-parameter cap (Supplementary Note~\ref{note:conventional_mlp_controls}).

Training cost and deployment storage answer different questions. Supplementary Table~\ref{tab:fig6_compact_storage} gives an example: a learned-contact fc7 replacement with 1,024 I neurons. Removing its unused candidate weights after training changes no logit by more than $1.91\times10^{-6}$ (end-to-end check, Methods) and substantially reduces the stored replacement core. Whole-model savings are smaller, because the retained AlexNet layers and the connectivity indices must also be stored, and the conversion does not reduce the candidate bank needed during learning.

\suppnote{Endpoint regularization and matched learned-contact integration}
\label{note:alexnet_followup}

The endpoint controls address two comparisons. The MLP regularization screen tests whether dropout or weight decay improves the conventional replacements, and the learned-contact additive arm tests the integration rule within one tree construction. Recipes were selected on three separate 50-epoch tuning seeds, and the selected recipes were then trained on five new seeds for 100 epochs; no official-validation measurement enters recipe selection, but this follow-up was designed after earlier official-validation results had been seen (Methods). Both tree rules retain 1,376,320 parameters at the larger cap and 226,120 at the smaller cap, including all couplings and activation parameters, while storing 292,085,776 trainable core values during optimization; their 4,096 E and 820 I neurons and candidate banks are identical. MLP widths and candidate banks differ from the trees, and integer dimensions leave their retained counts slightly below the caps.

\begin{figure}[tbp]
    \centering
    \IfFileExists{figures_empirical/figS_alexnet_scaling_j1.pdf}{%
      \includegraphics[width=165mm]{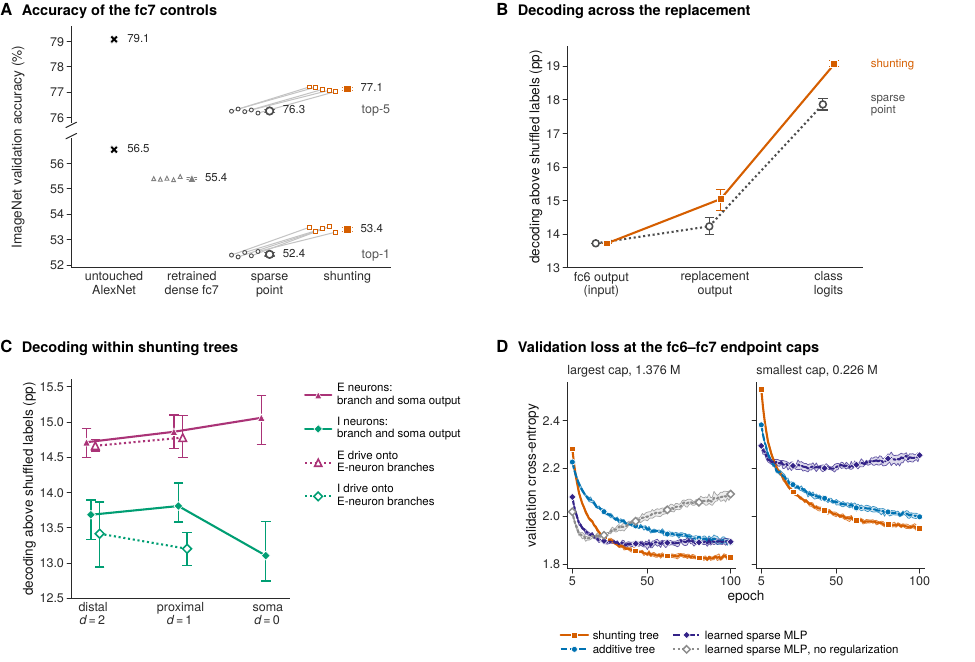}%
    }{%
      \PackageError{empirical-manuscript}%
        {Missing required figure `figures_empirical/figS_alexnet_scaling_j1.pdf'}%
        {Restore the figure listed in the manuscript export manifest before compiling.}%
    }%
    \caption{%
\textbf{Representation and training controls for AlexNet replacements.}
(A) Top-1 (lower) and top-5 (upper) ImageNet validation accuracy for untouched AlexNet, sparse point and shunting fc7 replacements; retrained dense fc7 is shown for top-1 only. Gray segments join paired seeds, the axis omits the empty interval between 57\% and 75.8\%, and printed values are the plotted means. (B) Ridge decoding above shuffled labels at the shared fc6 output (the replacement input), the replacement output and the logits. (C) Decoding within shunting trees: branch and soma outputs of E and I neurons (filled symbols, solid lines) and the E and I drives onto E-neuron branches (open symbols, dotted lines). A--C show means and pointwise 95\% seed-bootstrap intervals from the separate five-seed fc7 cohort (41--45); B,C use five images per class and label-free 256-dimensional projections, and stage differences in C are small relative to seed variability. (D) Internal-validation cross-entropy from epochs 5--100 at the two fc6--fc7 endpoint caps (1.376 M and 0.226 M retained core parameters) for the selected shunting, additive and learned sparse MLP recipes. The large-cap panel also shows the paired unregularized MLP reference; at the small cap it coincides with the selected recipe. These are validation trajectories, not training--validation gaps. Curves and bands show means and pointwise 95\% seed-bootstrap intervals over seeds 101--105, without smoothing. Initialization procedures differ between the selected tree recipes in D; Fig.~\ref{fig:fig6}D gives the matched-procedure contrasts.}
    \label{fig:figS_alexnet_width_parameter_scaling}
  \end{figure}

Regularization improves the large-cap sparse MLPs and delays their validation-loss minima (\suppfig{figS_alexnet_width_parameter_scaling}D). Weight decay without dropout is selected for every MLP except the smallest learned-contact sparse model. The selected learned and fixed sparse MLPs perform similarly at the larger cap, whereas learned connectivity helps at the smaller cap. The shunting advantage persists at both caps (Supplementary Table~\ref{tab:alexnet_followup}). These controls strengthen the baseline comparison but do not supply a retuned curve at intermediate budgets.

\begin{table}[htbp]
\centering\small
\caption{\textbf{Endpoint controls after validation selection.} Upper block: official-validation top-1 accuracy (\%) of the selected recipe in each family. Lower block: paired shunting-minus-additive differences (percentage points) with the same initialization procedure for both tree rules. Means and pointwise 95\% seed-bootstrap intervals over five new seeds (101--105), with checkpoints selected within 100 epochs. In the upper block shunting uses analytical initialization and addition occupancy calibration; the lower block matches the procedure. Caps count retained core parameters ($B=1{,}376{,}320$), excluding the common 4,097,000-parameter classifier and nontrainable indices.}
\label{tab:alexnet_followup}
\begin{tabular}{@{}lcc@{}}
\toprule
Replacement or contrast & Largest cap: $B$ & Smallest cap: 226,120 \\
\midrule
Shunting tree, learned contacts & $53.38\ [53.31,53.46]$ & $51.28\ [51.21,51.35]$ \\
Additive tree, learned contacts & $51.95\ [51.85,52.06]$ & $50.40\ [50.36,50.45]$ \\
Sparse MLP, learned contacts & $52.13\ [52.08,52.19]$ & $47.41\ [47.12,47.62]$ \\
Sparse MLP, fixed contacts & $52.14\ [51.93,52.31]$ & $45.76\ [45.59,45.92]$ \\
Dense bottleneck MLP & $50.54\ [50.37,50.71]$ & $34.30\ [34.20,34.39]$ \\
\midrule
Shunting $-$ additive, analytical & $1.87\ [1.73,2.02]$ & $1.78\ [1.65,1.96]$ \\
Shunting $-$ additive, calibrated & $0.91\ [0.73,1.06]$ & $0.93\ [0.77,1.08]$ \\
\bottomrule
\end{tabular}
\end{table}

Both matched initialization procedures favor shunting, and activation-range calibration improves addition; matching a procedure does not give identical fitted activation parameters. At the smallest cap the selected additive tree also exceeds the learned sparse MLP (50.40\% against 47.41\%), so the tree--MLP difference there cannot be assigned to shunting alone; at $B$ it is slightly below both sparse MLPs. Several additive and shunting trees still improve late in training, so these results describe a common horizon, not saturation. Every family and budget fitted a small optimization subset in a short diagnostic before the full runs, which checks basic trainability but not optimal fitting of the full data.

\FloatBarrier

\suppnote{Matched learned-connectivity scaling below the original budget}
\label{note:downward_matched}

The scaling comparison of Fig.~\ref{fig:fig6}E uses the same fc6--fc7 boundary, data split, classifier, batch size and training horizon for four replacement families. All trainable core values retained after training are counted, including couplings, activation parameters and biases; integer widths and fan-ins leave the conventional models slightly below each cap, and the figure plots their actual retained counts.

Learning connectivity improves the sparse MLP at smaller budgets, but it remains below learned-contact shunting at every tested cap (Supplementary Table~\ref{tab:downward_matched}). At $B$, the shunting-minus-learned-sparse-MLP difference is 2.32 percentage points (paired 95\% interval, 2.22--2.42); at the smallest cap it is 3.82 (3.72--3.92). These are differences between the trained constructions tested, not between their best possible performances.

\begin{table}[htbp]
\centering
\caption{\textbf{Matched downward scaling at 100 epochs.} Official-validation top-1 accuracy (\%), mean and pointwise 95\% seed-bootstrap interval over five paired seeds (81--85), of learned-contact shunting, learned-contact sparse MLP, fixed sparse MLP and dense MLP. The cap counts retained core parameters; the common classifier and nontrainable indices are excluded.}
\label{tab:downward_matched}
\scriptsize
\setlength{\tabcolsep}{3pt}
\begin{tabular}{@{}rcccc@{}}
\toprule
Cap & Shunting & Learned sparse & Fixed sparse & Dense \\
\midrule
1,376,320 & $53.41\ [53.25,53.54]$ & $51.09\ [51.00,51.22]$ & $50.96\ [50.86,51.07]$ & $50.02\ [49.93,50.10]$ \\
737,320 & $52.86\ [52.75,52.96]$ & $50.18\ [50.10,50.27]$ & $47.53\ [47.36,47.71]$ & $47.84\ [47.71,47.93]$ \\
481,720 & $52.34\ [52.23,52.45]$ & $49.93\ [49.84,50.02]$ & $47.02\ [46.95,47.08]$ & $44.74\ [44.55,44.93]$ \\
353,920 & $52.08\ [51.97,52.20]$ & $47.34\ [47.07,47.61]$ & $46.52\ [46.32,46.76]$ & $41.76\ [41.67,41.85]$ \\
226,120 & $51.23\ [51.17,51.28]$ & $47.40\ [47.30,47.52]$ & $45.63\ [45.42,45.84]$ & $34.13\ [33.54,34.43]$ \\
\bottomrule
\end{tabular}
\end{table}

The candidate banks are not matched: shunting stores 292,085,776 trainable core values at every cap, the learned sparse MLP 13,636,608 at $B$ and 2,134,176 at the smallest cap, and the fixed sparse and dense MLPs only their trained retained parameters. The comparison therefore concerns deployment parameter budgets, not equal training storage or search spaces.

Extending training from 50 to 100 epochs raises shunting accuracy at every cap, whereas the MLPs usually keep their earlier selected checkpoints (one learned-sparse run changed its selection at the second-smallest cap). A run is flagged as still improving when its validation loss improves by at least 0.01 after epoch 80 and its selected checkpoint is at epoch 90 or later: 0, 1, 3, 3 and 4 of the five shunting runs are flagged from the largest to the smallest cap, and no MLP run is. The MLP histories instead show training loss falling after validation loss reaches its minimum, a later deterioration that validation-based checkpoint selection avoids. Dropout and weight decay, not tuned in this sweep, are tested in the endpoint controls above. These trajectories qualify asymptotic interpretations but do not invalidate the comparison at a common horizon.

\begin{table}[htbp]
\centering
\caption{\textbf{Original five-seed comparison after replacing the complete AlexNet fc6--fc7 span.} Official-validation top-1 accuracy with a 95\% seed-bootstrap interval, seeds 41--45, 50 epochs. Retained core parameters include selected synaptic weights and all learned coupling, activation and bias parameters after unused candidates are discarded; training core parameters count all learned scalar values stored during optimization. Both exclude the common classifier and nontrainable connectivity indices. Learned-top-$k$ and fixed-contact dendritic cores each retain 1.278 million synaptic contacts.}
\label{tab:fig6_fc6fc7_confirmation}
\small
\begin{tabular}{@{}lrrr@{}}
\toprule
Replacement & Top-1 (\%) & Core retained (M) & Core training (M) \\
\midrule
Shunting, learned top-$k$ & $53.02\ [52.88,53.15]$ & 1.376 & 292.086 \\
Shunting, global fixed support & $51.27\ [51.20,51.32]$ & 1.376 & 1.376 \\
Shunting, structured-local windows & $51.15\ [51.07,51.23]$ & 1.376 & 1.376 \\
Shunting, random local windows & $50.37\ [49.85,50.77]$ & 1.376 & 1.376 \\
Raw additive, global fixed support & $50.28\ [50.19,50.34]$ & 1.376 & 1.376 \\
Raw additive, structured-local windows & $50.35\ [50.31,50.37]$ & 1.376 & 1.376 \\
Sparse point & $47.47\ [47.43,47.52]$ & 1.376 & 1.376 \\
Retrained native dense span & $52.64\ [52.51,52.76]$ & 54.534 & 54.534 \\
\bottomrule
\end{tabular}
\end{table}
\FloatBarrier

\suppnote{Conventional-network controls and parameter budgets}
\label{note:conventional_mlp_controls}

The matched MLP comparison replaces the complete fc6--fc7 span with randomly initialized replacement weights and the common pretrained linear classifier. Dense bottleneck, deeper dense and fixed sparse MLPs are compared with newly trained fixed-contact shunting, raw-additive and sparse-point references, all with the same optimization split, internal-validation examples, global batch size and training horizon. After the 50-epoch official-validation results had been inspected, the runs were continued to 100 epochs, with checkpoints selected by internal-validation loss over the full history (Methods). MLP settings come from a bounded validation search, whereas the references keep the original parameter-group learning rates. These five-seed results (seeds 71--75) form a cohort separate from the original comparison in Supplementary Table~\ref{tab:fig6_fc6fc7_confirmation}.

The common cap $B$ counts all trainable core parameters. The fixed-contact dendritic core has 1,278,000 synaptic weights and 98,320 coupling and activation parameters; the dense and sparse MLPs have slightly fewer parameters because widths and connection counts are integers (Supplementary Table~\ref{tab:alexnet_matched_mlp}). The fixed-contact dendritic core and the fixed sparse MLP store one weight per retained connection, without masked candidates, and their indices are nontrainable storage. The learned-contact dendritic core, by contrast, stores 292,085,776 core parameters during training, although discarding unselected candidates leaves 1,376,320 learned values. Matching a learned-contact model's retained parameters is therefore a deployment comparison, not a match to its training-time candidate bank.

\begin{table}[htbp]
\centering
\caption{\textbf{Conventional and dendritic replacements under a common parameter cap and training horizon.} Top-1 accuracy, mean and pointwise 95\% seed-bootstrap interval after 100 epochs, over five new paired seeds (71--75) with validation-loss checkpoint selection. The dendritic and sparse-point rows are fixed-contact models (global fixed support). The first block uses cap $B=1{,}376{,}320$, the next blocks $2B$ and $4B$; larger MLPs reuse the settings selected at $B$. Counts include every stored trainable core scalar; the common 4,097,000-parameter classifier, connectivity indices and optimizer state are excluded. This cohort is separate from Fig.~\ref{fig:fig6}B.}
\label{tab:alexnet_matched_mlp}
\small
\begin{tabular}{@{}lrr@{}}
\toprule
Replacement & Core parameters & Top-1 (\%) \\
\midrule
Shunting dendrites & 1,376,320 & $51.75\ [51.67,51.83]$ \\
Additive dendrites & 1,376,320 & $50.28\ [50.16,50.38]$ \\
Sparse point & 1,376,320 & $47.47\ [47.45,47.50]$ \\
Dense MLP, ReLU & 1,375,335 & $50.06\ [49.94,50.18]$ \\
Dense MLP, leaky ReLU & 1,375,335 & $50.56\ [50.40,50.72]$ \\
Deeper dense MLP & 1,372,528 & $48.36\ [48.23,48.47]$ \\
Sparse MLP & 1,373,184 & $51.01\ [50.83,51.19]$ \\
Dense MLP, $2B$ & 2,746,574 & $50.50\ [50.32,50.63]$ \\
Deeper dense MLP, $2B$ & 2,748,047 & $49.01\ [48.87,49.16]$ \\
Sparse MLP, $2B$ & 2,750,464 & $51.59\ [51.48,51.70]$ \\
Dense MLP, $4B$ & 5,502,365 & $50.96\ [50.80,51.10]$ \\
Deeper dense MLP, $4B$ & 5,503,811 & $49.61\ [49.36,49.83]$ \\
Sparse MLP, $4B$ & 5,505,024 & $51.88\ [51.80,51.98]$ \\
\bottomrule
\end{tabular}
\end{table}

At cap $B$, the sparse MLP and fixed-contact shunting were close after 50 epochs, but the continuation increases shunting accuracy while retaining the earlier selected MLP checkpoints. At 100 epochs, sparse-MLP minus shunting accuracy is $-0.74$ percentage points (paired 95\% interval, $[-0.92,-0.56]$). Larger sparse MLPs remain competitive. Two of five shunting runs and four of five additive runs still improve their best validation loss by at least 0.01 after epoch 80 and select a checkpoint at epoch 90 or later. These trajectories make the comparison sensitive to the training horizon; they do not establish asymptotic performance or an architectural capacity limit. This fixed-contact cohort is distinct from the learned-top-$k$ downward-budget comparison.

A separate comparison tests conventional compression from pretrained weights (Supplementary Table~\ref{tab:alexnet_conventional_compression}). Both fc6 and fc7 were either pruned or replaced by low-rank factors derived from their pretrained weights. Global magnitude pruning retains the largest weights across both layers. Structured pruning retains one weight per consecutive group of 43 inputs in each output row, including incomplete terminal groups. The low-rank construction uses truncated singular-value decompositions with ranks 57 and 63 for fc6 and fc7. Unlike the randomly initialized narrow MLP, it preserves two factorized affine transformations and their intervening rectifiers.

Each compression method was retrained with three seeds for 50 epochs on the same optimization partition and image augmentations as the dendritic reference, with the convolutional prefix frozen and fc8 trainable. Adam learning rates $10^{-4}$ and $3\times10^{-4}$ were applied to both replacement and classifier; the checkpoint and learning rate were selected by cross-entropy on 4,096 fixed internal-validation images. Both learning rates completed the training horizon. Evaluation used all 50,000 official-validation images. Pruning masks were reapplied after every optimizer step, and the retained weight counts or factor shapes were verified after training.

Magnitude pruning (53.26\%) reaches about the accuracy of the learned-contact shunting tree of the original 50-epoch comparison (53.02\%), one-of-43 pruning is slightly lower (52.85\%), and low-rank factors are less accurate at this budget (49.05\%). Conventional point-neuron networks can therefore implement useful compact replacements. The comparison is not matched in initialization or training storage: the conventional controls inherit pretrained weights, and the pruning implementation stores dense parameter and optimizer arrays even where weights are masked, whereas the learned-contact dendritic models learn from a large candidate bank. All three control seeds share the same pretrained starting model and data partition, so their spread measures retraining variability.

\begin{table}[htbp]
\centering
\caption{\textbf{Retrained conventional compression controls for the complete fc6--fc7 span.} Top-1 accuracy is the mean and range across three retraining seeds. Active core scalars include retained weights and biases; stored core scalars count the parameter arrays used during optimization. Both exclude the common fc8 classifier, optimizer state, and connectivity or mask buffers. The low-rank factors are stored directly, whereas pruned weights remain in dense masked arrays. These resource definitions distinguish the active-weight comparison from a training-storage match.}
\label{tab:alexnet_conventional_compression}
\small
\begin{tabular}{@{}lrrr@{}}
\toprule
Conventional core & Top-1 (\%) & Active scalars (M) & Stored scalars (M) \\
\midrule
Low-rank factors & $49.05\ [48.99,49.10]$ & 1.283 & 1.283 \\
Magnitude pruning & $53.26\ [53.12,53.41]$ & 1.286 & 54.534 \\
One-of-43 pruning & $52.85\ [52.77,52.96]$ & 1.282 & 54.534 \\
\bottomrule
\end{tabular}
\end{table}

\suppnote{Representation and training controls}
\label{note:alexnet_width_parameter_scaling}

\suppfig{figS_alexnet_width_parameter_scaling}A--C describes the separate fc7-only cohort (five seeds), and D shows internal-validation trajectories of the fc6--fc7 endpoint controls. For the representation analysis, five official-validation examples per class are projected without labels to 256 dimensions with CountSketch~\citep{Charikar2002}, and three-fold ridge decoding gives held-out accuracy above three shuffled-label fits. Measurements at the frozen fc6 input, the replacement output and the logits locate changes across a common replacement boundary (B). Panel C follows dendritic outputs toward the soma; the soma is shown once, because the population returns its root-branch output without a further transformation. E and I population outputs are those of the emitting neurons, whereas E and I drives are inputs to branches of E neurons. E-branch and soma decoding are similar relative to between-seed variability, and the I population shows no consistent rise toward the soma.

The original five-seed comparison of integration rules, fixed and learned connectivity, spatial routing and retrained native layers (Supplementary Table~\ref{tab:fig6_fc6fc7_confirmation}) provides architectural context for the separate 100-epoch learned-connectivity sweep; the cohorts are not pooled.

\FloatBarrier

\suppnote{Decoding under input corruption}
\label{note:alexnet_corruption_decoding}

\begin{figure}[tbp]
    \centering
    \IfFileExists{figures_empirical/figS_alexnet_corruption_j1.pdf}{%
      \includegraphics[width=165mm]{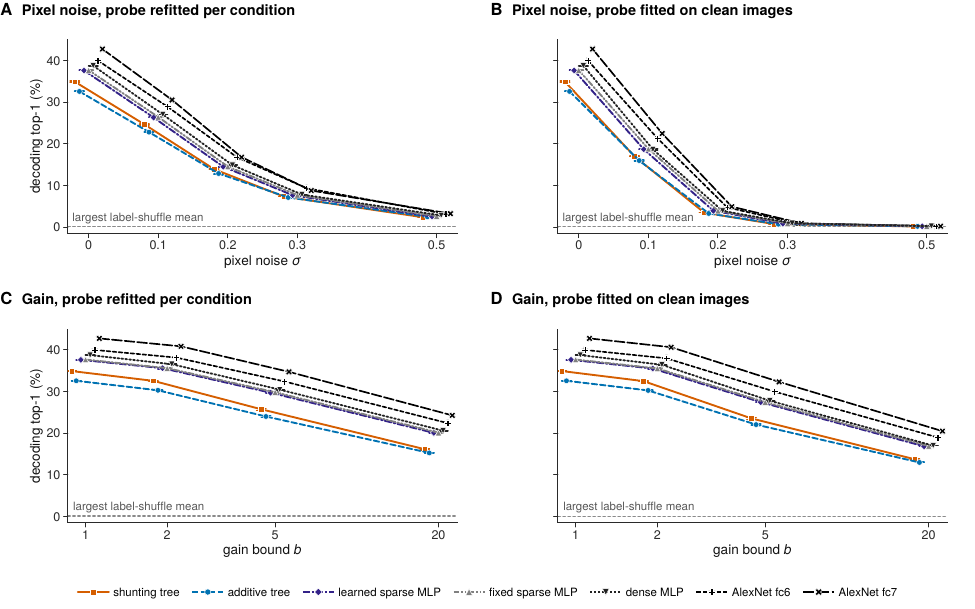}%
    }{%
      \PackageError{empirical-manuscript}%
        {Missing required figure `figures_empirical/figS_alexnet_corruption_j1.pdf'}%
        {Restore the figure listed in the manuscript export manifest before compiling.}%
    }%
    \caption{%
\textbf{Linear decoding of the AlexNet replacements under pixel noise and per-image gain.}
Top-1 accuracy of 1,000-class ridge probes on the 50,000 official-validation images. Probes read the 4,096 replacement outputs that feed fc8 in the endpoint models at the larger cap ($1{,}376{,}320$ retained core parameters; the E population for the trees) and the fc6 and fc7 outputs of untouched AlexNet. (A,B) Gaussian pixel noise $\sigma$ (0 is clean). (C,D) Gain drawn per image between $1/b$ and $b$, plotted against the gain bound $b$ (1 is clean). In A and C the probe is refitted under each condition on equally corrupted fit images; in B and D it is fitted once on clean images. Probes are fitted on 50 held-out training images per class and use the first corruption draw; the grey dashed line marks the largest mean of five label-shuffled fits. Means over five seeds (101--105) with 95\% seed-bootstrap intervals, which are narrower than the symbols; AlexNet is a single deterministic reference. Small horizontal offsets separate overlapping series.}
    \label{fig:figS_alexnet_corruption_decoding}
  \end{figure}

Fig.~\ref{fig:fig6}F reports how much accuracy each AlexNet replacement retains under pixel noise and per-image gain. \suppfig{figS_alexnet_corruption_decoding} asks whether class identity remains linearly decodable from the replacement outputs under the same corruptions, pixel noise (A,B) and per-image gain (C,D), without retraining any model (Methods). On clean images, ridge probes reach 34.9\% top-1 accuracy over the 1,000 classes from the shunting-tree outputs and 32.6\% from the additive-tree outputs, against 37.6--38.8\% from the MLP outputs and 40.0\% and 42.8\% from the fc6 and fc7 outputs of untouched AlexNet. The ordering persists under both corruptions. With probes refitted under each condition (A,C), the trees reach 24.7\% and 22.8\% at $\sigma=0.1$, against 26.3--27.0\% for the MLPs, and 16.2\% and 15.3\% at gain bound 20, against 20.1--20.6\%. Probes fitted on clean images (B,D) lose more under noise (17.0\% for the shunting tree at $\sigma=0.1$), so the corruption also moves the code relative to a fixed readout. These accuracies describe what a linear probe recovers from 50 fit images per class; they are lower than the networks' own top-1 accuracy and are not mutual-information estimates.

\FloatBarrier

\supppart{Protocols, resource accounting and software}
\label{supp:protocols}

This note collects the notation, training recipes, cohorts, resource definitions and software used across the supplement; the procedures themselves are described in the Methods.

\suppnote{Model variables and structural coordinates}
\label{note:architecture}

Dendritic depth $d$ locates a compartment within a tree, whereas network position $\ell$ and total network depth $L$ describe successive populations of a multilayer network. The integration and activation equations are those of the main text (Eqs.~\ref{eq:integration_overview},~\ref{eq:activation_overview}); Supplementary Table~\ref{tab:notation} lists the symbols used in the supplement.

\begin{table}[htbp]
\centering
\caption{\textbf{Notation used in the Supplementary Information.} Symbols with a second meaning in one note are defined there.}
\label{tab:notation}
\small
\begin{tabularx}{\textwidth}{@{}p{0.22\textwidth}X@{}}
\toprule
Symbol & Meaning \\
\midrule
$D$, $d$ & Number of dendritic levels and soma-relative compartment depth. \\
$L$, $\ell$ & Number of E/I population layers and position within one fitted network. \\
$E$, $I$ & Local aggregate excitatory and inhibitory synaptic drives. \\
$C$, $G$ & Weighted child output and summed child-coupling conductance. \\
$u$, $a$, $S$ & Activation input, branch output, and output of the branch's owning soma ($a$ is also the half-width of uniform noise in Supplementary Note~\ref{supp:perturbations}). \\
$Y$ & Image-class label. \\
$A$, $\Delta A$ & Test accuracy and its paired difference between two conditions (for example $A_{\rm sh}-A_{\rm add}$). \\
$R$ & Above-chance retention under a perturbation, $(A_{\rm perturbed}-0.1)/(A_{\rm clean}-0.1)$, shown as $100R$. \\
$\Delta_I$ & Inhibitory-tail contrast of the allocation grid (Supplementary Note~\ref{supp:allocation}). \\
$I(\cdot\,;\cdot\mid Y)$ & Mutual information conditioned on the class label, in bits. \\
$k$ & Retained contacts selected from a candidate bank (top-$k$), neighbors of a nearest-neighbor estimator, or replaced outputs per soma (\suppfig{figS_readout_mechanisms}D), as stated. \\
$N_E$, $N_I$ & Retained E and I contacts per compartment. \\
$g_L$ & Leak conductance, fixed at one unless learned from that initial value. \\
$\sigma$, $g$ & Standard deviation of Gaussian pixel noise and multiplicative image gain. \\
$\lambda$, $b$ (corruption) & Noise multiplier and bound on random gain, $g\in[1/b,b]$. \\
$q$ (contextual task) & Noise divisor; noise level is $1/q$. \\
pp & Percentage points, used for differences between percentages. \\
$B$ & Experiment-specific budget: $N_E+N_I$ contacts per compartment in allocation comparisons, or retained core parameters in AlexNet replacements. \\
$\mathcal T_{b\to s}$ & Transmission derivative $\partial S_s/\partial a_b$ from branch $b$ to its soma. \\
\bottomrule
\end{tabularx}
\end{table}

\suppnote{Training recipes and their selection}
\label{note:recipe_policy}

A validation screen trained five recipes (Supplementary Table~\ref{tab:recipe_sensitivity}) on three screening seeds at every depth-four allocation and selected, for each rule and allocation, the recipe with the highest mean validation accuracy. The linear-readout grid reused the depth-four selections after a check on ten cells showed that they carried over; the shallower grids, for which they did not, were screened at every allocation. After the complete validation screen and the per-cell test results had been inspected, the per-cell selections were replaced by a single recipe, A, at every allocation of the depth-four, shallower, linear-readout and initialization-control grids and their point controls. Runs trained for this change were evaluated after it was made. Runs already trained with recipe A were reused with their existing results only when the model rebuilt from their configuration reproduced the same computation; all others were retrained with recipe A on the same cell, rule, seed and GPU family. Recipe A gives consistency across the allocation, branch-analysis and robustness experiments, which share the same model construction; it is not claimed to be optimal. The primary cohort of Figs.~\ref{fig:fig3} and \ref{fig:fig4} is trained separately from the 40E/10I cell of the allocation grid, although both use seeds 0--19 (Supplementary Table~\ref{tab:cohort_manifest}).

With the per-cell selections (twenty seeds), shunting and addition average 51.1\% over the depth-four grid, and their difference is $-0.08$ points; addition is favored at 40 allocations and shunting at 34, and 47 intervals include zero. This near-zero average conceals allocation-dependent differences. Relative to these selections, recipe A lowers accuracy by 1.0 points on average for shunting (from $-2.7$ to $+0.6$ across cells) and by 0.5 points for addition (from $-2.1$ to $+0.7$). On the validation screen, recipe D has the highest depth-four averages, by 0.4 points for shunting and 0.15 for addition, whereas recipe A has the highest averages at depths three and two (Supplementary Table~\ref{tab:recipe_sensitivity}). The shunting-minus-additive validation map under D correlates 0.78, 0.79 and 0.68 with that under A at depths four, three and two. These are three-seed validation estimates; they are not comparable with the twenty-seed test accuracies.

\begin{table}[htbp]
\centering
\caption{\textbf{Validation accuracy of the five screened recipes, averaged over each allocation grid.} Mean validation accuracy (\%) over the 121 allocations of each grid and three screening seeds per allocation. Recipes differ in active-weight maintenance coefficient, inactive-candidate boosting, learning rate and patience (epochs): A (0.03, none, 0.002, 150), B (0.01, boosting, 0.002, 150), C (0.01, none, 0.002, 150), D (0.05, none, 0.002, 150) and P (0.01, boosting, 0.001, 400); all train for at most 2,000 epochs. Validation split only; under recipe A, the depth-four test grid uses twenty seeds per cell and the two shallower grids use thirteen.}
\label{tab:recipe_sensitivity}
\small
\begin{tabular}{@{}llrrrrr@{}}
\toprule
Grid & Rule & A & B & C & D & P \\
\midrule
$[3,3,3,3]$ & Shunting & 50.37 & 49.83 & 49.61 & 50.81 & 49.79 \\
 & Raw additive & 51.26 & 50.40 & 48.87 & 51.41 & 50.44 \\
$[3,3,3]$ & Shunting & 51.03 & 50.29 & 49.75 & 50.70 & 50.42 \\
 & Raw additive & 51.59 & 50.45 & 49.16 & 51.21 & 50.79 \\
$[3,3]$ & Shunting & 51.17 & 50.48 & 50.13 & 50.30 & 50.93 \\
 & Raw additive & 51.61 & 50.35 & 50.18 & 50.83 & 50.70 \\
\bottomrule
\end{tabular}
\end{table}

\suppnote{Training cohorts and seeds}
\label{note:cohorts}

Supplementary Table~\ref{tab:cohort_manifest} lists the training seeds for the main comparisons and identifies shared checkpoints; the seed extension is described in Supplementary Note~\ref{note:seed_extension}.

\begin{table}[!htbp]
\centering\small
\caption{\textbf{Training cohorts and checkpoint reuse.} Seed identifiers refer to training runs within a cohort. The same seed number can occur in separately trained cohorts.}
\label{tab:cohort_manifest}
\begin{tabularx}{\textwidth}{@{}p{.17\textwidth}p{.16\textwidth}X@{}}
\toprule
Panels & Training seeds & Comparison and reuse\\
\midrule
1D--I & 100--109 & Paired MNIST networks in D; H uses the shunting networks. E,F,I use the primary seed-100 checkpoint and G a separate four-level seed-100 checkpoint.\\
2B,C & 0--19; 53000--53012 & Twenty seeds for depth-four allocation; thirteen for shallower fixed-budget slices.\\
2D & 32000--32009 & Fixed-index 50E/0I versus 40E/10I inhibitory-contact control.\\
2E & 10--19 & Paired MLP/linear readout allocation; trained separately from the original 0--9 cohort.\\
2F & 52000--52019 & Low-capacity point/tree cohort; total E/I contacts matched. Seeds 52010--52019 added after inspection.\\
2G,H & 62000--62019 & Fixed routing at equal resources; each route's geometry is fixed across training seeds. Seeds 62010--62019 added after inspection.\\
3B--F, 4B,C & 0--19 & Primary MLP checkpoints, reused for association, transmission, intervention and perturbation analyses. Seeds 10--19 added after inspection; the initialization-policy and linear-readout controls use 0--9.\\
4D--F & 21000--21019 & Matched-morphology perturbation cohort with both integration rules: morphology contrasts under noise and scaling (D), the second corruption protocol (E), probes (F) and class-information estimates (\suppfig{figS_corruption_protocols}C). Seeds 21010--21019 added after inspection.\\
\ref{fig:point_neurons}C & 0--19 & Primary CIFAR-10 tree/point comparison. The wider cohort comparison is in \suppfig{figS_three_arms}.\\
\ref{fig:point_neurons}D & 21000--21019 & Matched-morphology corruption cohort; all four morphologies share one point model per rule and seed.\\
\ref{fig:point_neurons}E & 52000--52009 & 64-neuron trees and points of the contact-bottleneck cohort at exactly matched active contacts.\\
\ref{fig:point_neurons}F & 21000--21009 & Breadth control: the $[120]$ trees' active contacts on 20, 6 or 2 compartments, with the capped point as reference.\\
\ref{fig:point_neurons}G & 21000--21009; 52000--52009 & Fixed-leak shunting models reused from D--F: 80 from the matched-morphology and breadth cohorts, 50 from the 64-neuron contact-bottleneck cohort.\\
\ref{fig:fig5} & 10--19 & Multilayer cohort; selected and shared-rate comparisons reuse the shunting networks. Point references are a distinct resource condition.\\
\ref{fig:fig7}, \suppfig{figS_contextual_wiring} & 62000--62009 & Contextual remapping; matched-parameter MLP, endpoint-gate trees, leak controls and point arms.\\
\ref{fig:fig6}B--D,F; \ref{fig:fig6}E open symbols & 101--105 & Endpoint regularization and matched-tree controls; 100-epoch runs after tuning on separate seeds.\\
\ref{fig:fig6}E connected curves & 81--85 & Original fc6--fc7 downward sweep; four families, five budgets and a 100-epoch horizon.\\
\suppfig{figS_alexnet_width_parameter_scaling}A--C & 41--45 & Separate fc7-only performance and representation cohort; five trained seeds.\\
\bottomrule
\end{tabularx}
\end{table}
\FloatBarrier

\suppnote{Original implementation of the contextual task}
\label{note:original_contextual}

Supplementary Table~\ref{tab:original_routing} sets the second-layer inhibition interventions of the present networks beside the values reported for the original implementation (Supplementary Note~\ref{supp:contextual} and Methods).

\begin{table}[htbp]
\centering
\caption{\textbf{Second-layer inhibition interventions in the present networks and the original implementation.} Present networks: means over ten seeds (Fig.~\ref{fig:fig7}E,F). The original implementation's values are as reported, from its own construction, training and evaluation; they are references, not paired controls.}
\label{tab:original_routing}
\small
\begin{tabular}{@{}lcc@{}}
\toprule
Network & Context-mean inhibition (pp) & Donor target after transplant (\%) \\
\midrule
Shunting tree & $-20.8$ & 1.4 \\
Additive tree, shunting recipe & $-15.8$ & 1.6 \\
Endpoint-gate tree & $-43.5$ & 11.6 \\
Additive tree, tuned recipe & $-57.1$ & 15.0 \\
Original implementation (reported) & $-47.8$ & 13.5 \\
\bottomrule
\end{tabular}
\end{table}
\FloatBarrier

\suppnote{Resource accounting}
\label{note:resource_accounting}

Supplementary Table~\ref{tab:resources} defines the resource quantities reported by each comparison; Supplementary Table~\ref{tab:fig6_compact_storage} gives the training and deployment storage of one learned-contact AlexNet replacement.

\begin{table}[htbp]
\centering
\caption{\textbf{Resource quantities.} Each comparison states the scope (model, replacement core or whole model) at which they are counted.}
\label{tab:resources}
\small
\begin{tabularx}{\textwidth}{@{}p{0.25\textwidth}X@{}}
\toprule
Quantity & Definition \\
\midrule
Active contacts & Connections retained in each forward pass. \\
Candidate weights & Stored trainable synaptic bank used to select top-$k$ contacts. \\
Trainable parameters & Synaptic, coupling, activation-function, bias, and readout scalars in the stated model or core. \\
Persistent storage & Parameter bytes plus required nonparameter buffers, including integer indices. \\
\bottomrule
\end{tabularx}
\end{table}
\begin{table}[t]
\centering
\caption{\textbf{Training-time and deployment storage of one learned-contact fc7 replacement.} The example has 4,096 E and 1,024 I neurons with $[2,2]$ trees (one trained model). Scalar values use FP32 storage, and connectivity indices unsigned 16-bit integers; totals include both and other persistent buffers. Core scope excludes the retained AlexNet prefix and fc8; whole-model scope includes both.}
\label{tab:fig6_compact_storage}
\small
\begin{tabular}{@{}llrrrr@{}}
\toprule
Scope & Model state & Scalar values (M) & Scalar MiB & Index MiB & Total MiB \\
\midrule
Replacement core & Dense fc7 reference & 16.781 & 64.016 & 0 & 64.016 \\
Replacement core & Top-$k$, training candidates & 151.097 & 576.391 & 0 & 576.391 \\
Replacement core & Top-$k$, frozen support & 1.393 & 5.313 & 2.461 & 7.773 \\
\addlinespace
Whole model & Pretrained AlexNet & 61.101 & 233.081 & 0 & 233.081 \\
Whole model & Top-$k$, training candidates & 195.417 & 745.456 & 0 & 745.456 \\
Whole model & Top-$k$, frozen support & 45.712 & 174.378 & 2.461 & 176.839 \\
\bottomrule
\end{tabular}
\end{table}
\FloatBarrier

\suppnote{DendriNet software and reproducibility}
\label{supp:software}

\begin{figure}[tbp]
    \centering
    \IfFileExists{figures_empirical/figS_software_workflow_j1.pdf}{%
      \includegraphics[width=165mm]{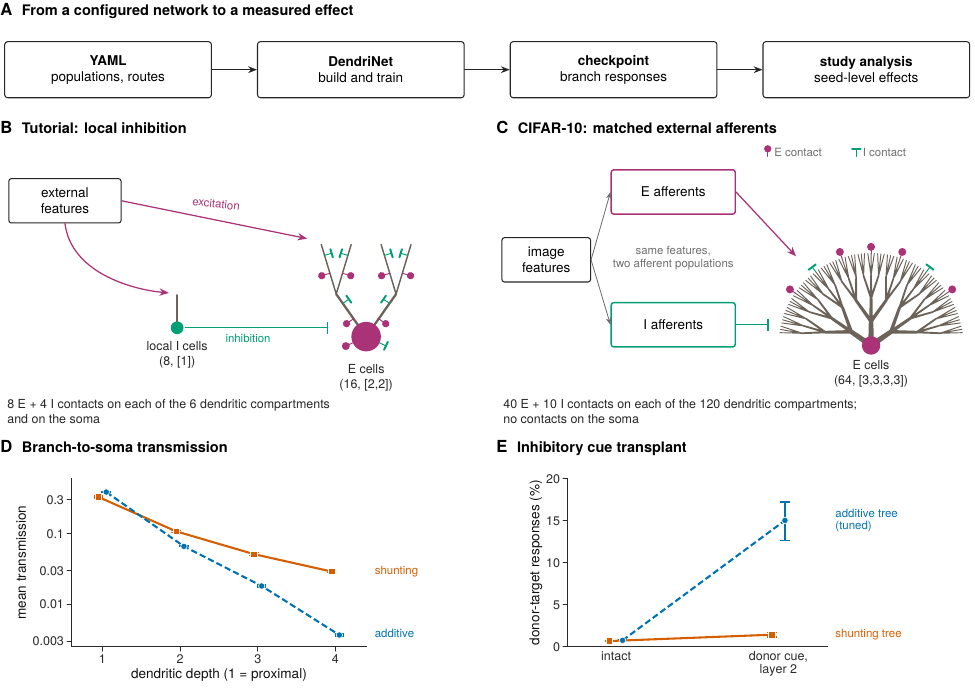}%
    }{%
      \PackageError{empirical-manuscript}%
        {Missing required figure `figures_empirical/figS_software_workflow_j1.pdf'}%
        {Restore the figure listed in the manuscript export manifest before compiling.}%
    }%
    \caption{%
\textbf{From a dendritic circuit to a reproducible measurement.}
(A) The configuration defines the network and training procedure; the study analysis evaluates the fitted network and combines measurements across training seeds. (B,C) Incoming pathways of an excitatory neuron in the synthetic tutorial and in the primary CIFAR-10 cohort, drawn with their true morphologies: arrows denote excitation and bars inhibition, and E and I contacts are drawn as dots and bars on the branches. The contact counts below each diagram apply to every receiving compartment. (D) Mean branch-to-soma transmission by dendritic depth, from Fig.~\ref{fig:fig3}D; depth one is closest to the soma, and the vertical scale is logarithmic. (E) Fraction of contextual predictions matching the next cue's target, in intact networks and after substituting that cue's layer-two inhibition, from Fig.~\ref{fig:fig7}F. Squares and solid lines denote shunting and circles and dashed lines addition in D and E. Intervals are pointwise 95\% seed-bootstrap intervals (20 networks per rule in D, ten per model in E).}
    \label{fig:figS_software_workflow}
  \end{figure}

\paragraph{From a configuration to a measurement.}
DendriNet builds a PyTorch model~\citep{Paszke2019} from a YAML file. The file specifies
which populations supply each neuron, where their contacts can terminate, the number of
contacts at each depth, and how the receiving compartment integrates them. A separate
training section sets the optimizer and stopping rule. After training, a checkpoint can
be evaluated on the same inputs while recording a branch response or replacing an
inhibitory signal. The study code specifies the intervention and the unit of analysis.
\suppfig{figS_software_workflow} illustrates this division of work
with two circuits and two measurements.

\paragraph{Where does a neuron's inhibition come from?}
In the two-class tutorial, external features excite both cell populations. Each neuron
in the excitatory population receives inhibition from the local inhibitory cells
(\suppfig{figS_software_workflow}B). Its branching factors $[2,2]$ give two primary
and four terminal compartments. Each of these compartments and the soma receives eight
excitatory and four inhibitory contacts. The primary CIFAR-10 network has a different
wiring diagram (\suppfig{figS_software_workflow}C). Two external afferent populations
carry the same image features. One supplies excitation and the other inhibition. Each
of the 120 dendritic compartments receives 40 excitatory and ten inhibitory contacts,
and the soma has no direct synapses. The upstream circuit that supplies these matched
afferents is a modeling assumption. In the configuration, the two entries under the
first layer's \texttt{connections} specify the sources and their common target:
\begin{verbatim}
- {source: input_e, target: e0, pathway: ff_excitatory}
- {source: input_i, target: e0, pathway: ff_inhibitory}
\end{verbatim}
The receiving population sets the contact budget for each pathway. When several
sources feed the same pathway, its contacts are drawn from their eligible inputs.

\paragraph{Installation and tutorial.}
In a fresh Python environment (Python 3.10 or later), install the library from
\href{https://pypi.org/project/dendritic-modeling/0.5.0/}{PyPI}.
The second command adds the optional
\href{https://pypi.org/project/dendrinet-studio/0.1.0/}{DendriNet Studio} browser interface.
\begin{verbatim}
python -m pip install dendritic-modeling==0.5.0
python -m pip install dendrinet-studio==0.1.0
\end{verbatim}
Alternatively, from the root of a checkout of the DendriNet \texttt{v0.5.0} Git tag,
run \verb|python -m pip install .| for the library and
\verb|python -m pip install ./apps/studio| for the interface.
Studio also installs a compatible core library when installed on its own.
Launch it with \verb|dendrinet-studio --workspace studio-demo| and open the
local address it prints.
DendriNet Studio, developed after the experiments, displays the population diagram,
a representative neuron's tree, and its incoming E and I pathways. Selecting the
tutorial's excitatory population shows 56 excitatory contacts from the external input
and 28 inhibitory contacts from the local population. In the first layer, separate switches
connect external E and I afferents to each receiving population. A population receiving
only external I must also receive excitation from another population in the same time step.
Readers can select matched or separate afferents, inspect contact counts, build the model,
export its YAML and train it. The same tutorial runs from the command line on generated data:
\begin{verbatim}
dendritic-example synthetic_population --output tutorial.yaml
dendritic-train tutorial.yaml --epochs 1
\end{verbatim}

\paragraph{Worked example: transmission from branch to soma (Fig.~\ref{fig:fig3}D).}
For each fitted CIFAR-10 tree we measured the derivative of the soma output with respect
to each branch output on 2,048 test images. Averaging over images and branches gives
one measurement per trained network and depth. The reader dataset contains those
measurements for 20 paired seeds. The analysis averages them across networks and
resamples the seeds to obtain the intervals. At the fourth depth the mean derivative
is 0.0290 for shunting and 0.00367 for addition (\suppfig{figS_software_workflow}D).
The fraction of derivatives below the threshold in Fig.~\ref{fig:fig3}D can be
recovered from the same table.

\paragraph{Worked example: changing the inhibitory cue (Fig.~\ref{fig:fig7}F).}
For each corrupted character image, all 32 cues were presented to the same trained
network. We supplied the inhibitory activity evoked by the next cue while retaining
the current image and cue. The source table records, for each seed and intervention,
the fraction of predictions matching the presented target and the fraction matching
the donor cue's target. Transplanting second-layer inhibition drove predictions
toward the donor target more often in the separately tuned additive network than in
the primary shunting network (\suppfig{figS_software_workflow}E). The analysis
summarizes both outcomes for the intact network and each transplanted layer,
including the two additional models shown in the main figure.

\paragraph{Running the analyses.}
Download and extract the
\href{https://github.com/KempnerInstitute/DendriNet/releases/download/v0.5.0/empirical-examples.zip}{worked examples archive}.
It contains the analysis scripts, training configurations and per-seed data.
The eight configurations cover MNIST, CIFAR-10 and
contextual EMNIST. For the contextual trees, a calibration script samples the
training presentation stream before training.
The native neuronal experiments in Figures~\ref{fig:fig1}--\ref{fig:fig7} were
trained with DendriNet v0.3.0. The AlexNet results in Figure~\ref{fig:fig6} use
historical training checkpoints and a separate revision for endpoint evaluation,
corruption and decoding. The figure index \path{reproducibility/figures.md} in the worked-examples archive
identifies these sources. For the two worked examples, we used the following numerical package
versions. From the unpacked examples directory, run
\begin{verbatim}
python -m pip install numpy==2.2.6 pandas==2.3.3 matplotlib==3.10.6
python analysis/reproduce.py empirical-example-data.zip \
  --output figure3d-results
python analysis/plot_software_guide.py --tutorial tutorial.yaml \
  --summaries figure3d-results --output software-guide.pdf
\end{verbatim}
The outputs are \path{summary.csv} (branch transmission) and \path{context_summary.csv}
(correct- and donor-target rates). The command verifies the input archive and compares
the transmission summary with the original table. Using the v0.5.0 release on
a CPU, we reproduced both reference tables byte for byte; the plotting script uses them to draw
the software figure in its original release layout. The corresponding panels in
\suppfig{figS_software_workflow} use the same measurements.
Recomputing the measurements that enter them requires the corresponding checkpoints,
input transformations and evaluation samples. MNIST, EMNIST and CIFAR-10 are obtained
from their cited sources. ImageNet inputs and pretrained weights are subject to their
respective access terms.

\FloatBarrier
\bibliographystyle{unsrtnat}
\bibliography{dendrit_empirical}

\end{document}